\documentclass[a4paper,fleqn,usenatbib]{mnras}

\usepackage{newtxtext,newtxmath}
\usepackage[T1]{fontenc}
\usepackage{ae,aecompl}

\usepackage{graphicx}	% Including figure files
\usepackage{amsmath}	% Advanced maths commands
\usepackage{amssymb}	% Extra maths symbols
\usepackage{pdflscape}	% Landscape pages
\usepackage{longtable}
\usepackage{hyperref}
\usepackage{xspace}
\usepackage{enumerate}
\usepackage{chemformula}
\usepackage{siunitx}
\usepackage{xcolor}
\newcommand{\kb}{k_\text{B}}
\newcommand{\kbmu}{\frac{\kb}{\mu m_\text{H}}}

\newcommand{\krome}{\textsc{Krome}\xspace}
\newcommand{\mpiamrvac}{\textsc{mpi-amrvac}\xspace}

\newcolumntype{h}{>{\setbox0=\hbox\bgroup}c<{\egroup}@{}}
\newcommand{\kms}{\ifmmode{~\mathrm{km}\per{s}}\else {$\mathrm{km}\per{s}$}\fi\xspace}
\newcommand{\per}[1]{\ifmmode{\mathrm{\,#1}^{-1}}\else {$\mathrm{\,#1}^{-1}$}\fi\xspace}
\newcommand{\tr}{_{\text{tr}}\xspace}
\newcommand{\rot}{_{\text{rot}}\xspace}
\newcommand{\vib}{_{\text{vib}}\xspace}
\newcommand{\elec}{_{\text{el}}\xspace}
\newcommand{\pd}{\partial\xspace}
\newcommand{\comment}[2]{#2}
\newcommand{\ado}[1]{\textcolor{black}{#1}} % to convert orange additions in black
\newcommand{\adr}{\comment} % to delete sections in red
\title[Hydrochemical AGB wind model]{Developing a self-consistent AGB wind model: \\I. Chemical, thermal, and dynamical coupling}
\author[J. Boulangier et al.]{
Jels Boulangier,$^{1}$\thanks{E-mail: jels.boulangier@kuleuven.be}
N. Clementel,$^{1}$
A. J. van Marle,$^{2}$
L. Decin,$^{1}$
and A. de Koter$^{1,3}$
\\
$^{1}$Institute of Astronomy, KU Leuven, Celestijnenlaan 200D, 3001 Leuven, Belgium\\
$^{2}$Department of Physics, School of Natural Sciences UNIST, Ulsan 44919, Korea\\
$^{3}$Anton Pannenkoek Institute for Astronomy, Universiteit van Amsterdam, Science Park 904, 1098 XH Amsterdam, The Netherlands
}

\date{Accepted 12 September 2018}

\pubyear{2018}

\begin{document}
\label{firstpage}
\pagerange{\pageref{firstpage}--\pageref{lastpage}}
\maketitle

\begin{abstract}
The material lost through stellar winds of Asymptotic Giant Branch (AGB) stars is one of the main contributors to the chemical enrichment of galaxies. The general hypothesis of the mass loss mechanism of AGB winds is a combination of stellar pulsations and radiative pressure on dust grains, yet current models still suffer from limitations. Among others, they assume chemical equilibrium of the gas, which may not be justified due to rapid local dynamical changes in the wind. This is important as it is the chemical composition that regulates the thermal structure of the wind, the creation of dust grains in the wind, and ultimately the mass loss by the wind. Using a self-consistent hydrochemical model, we investigated how non-equilibrium chemistry affects the dynamics of the wind. This paper compares a hydrodynamical and a hydrochemical dust-free wind, with focus on the chemical heating and cooling processes. No sustainable wind arises in a purely hydrodynamical model with physically reasonable pulsations. Moreover, temperatures are too high for dust formation to happen, rendering radiative pressure on grains impossible. A hydrochemical wind is even harder to initiate due to efficient chemical cooling. However, temperatures are sufficiently low in dense regions for dust formation to take place. These regions occur close to the star, which is needed for radiation pressure on dust to sufficiently aid in creating a wind. Extending this model self-consistently with dust formation and evolution, and including radiation pressure, will help to understand the mass loss by AGB winds.

% This is a simple template for authors to write new MNRAS papers.
% The abstract should briefly describe the aims, methods, and main results of the paper.
% It should be a single paragraph not more than 250 words (200 words for Letters).
% No references should appear in the abstract.
\end{abstract}


%%%%%%%%%%%%%%%%%%%%%%%%%%%%%%%%%%%%%%%%%%%%%%%%%%

%%%%%%%%%%%%%%%%% BODY OF PAPER %%%%%%%%%%%%%%%%%%

\section{Introduction}
% This is a simple template for authors to write new MNRAS papers.
% See \texttt{mnras\_sample.tex} for a more complex example, and \texttt{mnras\_guide.tex}
% for a full user guide.

% All papers should start with an Introduction section, which sets the work
% in context, cites relevant earlier studies in the field by \citet{Others2013},
% and describes the problem the authors aim to solve \citep[e.g.][]{Author2012}.
Asymptotic Giant Branch (AGB) stars are one of the main sources of chemical enrichment of galaxies, due to their massive stellar winds with mass loss rates up to $10^{-4}M_{\sun}$\,yr$^{-1}$ \citep[e.g.][]{Schoier2001, Olofsson2002,Groenewegen2002}. The general hypothesis is that the mass loss mechanism of these AGB winds is a combination of pulsations and radiative pressure on dust grains, whose mutual interaction leads to a sustainable stellar wind. These slow winds (5--20\,\si{\km\per\s}, e.g. \citealt{Schoier2001, Olofsson2002,Groenewegen2002}) collide with the interstellar medium, enriching it with a chemically diverse gas-dust mixture.\\\\
Modelling an AGB wind consists of three core disciplines: gas and dust dynamics, chemistry, and radiative transfer. All three are closely intertwined and therefore need to be modelled simultaneously. Each physical mechanism requires sufficient rigour and detail that modelling them separately constitutes an active field in AGB research in its own right. \\\\
Considerable effort has been made in modelling the dynamics of a pulsation-induced wind, either semi-analytically \citep{Willson1984,Bertschinger1985} or numerically \citep{Wood1979a}. The former is less reliable because it prescribes the wind in the ballistic limit, which is a crude approximation. The latter does not suffer from this approximation as the dynamical evolution is determined by the hydrodynamical conservation laws. Such hydrodynamical wind models have later been improved by including parametric non local thermodynamic equilibrium (non-LTE) radiative cooling \citep{Bowen1988}. However, this cooling prescription is simplified, suffering from inefficient cooling below  $\sim$8000\,K, resulting in a warm pressure-driven wind and a so-called `calorisphere' \citep{Willson2000}. Such high wind temperatures are inconsistent with molecular observations \citep[e.g.][]{Fonfria2008,Cernicharo2010,Decin2010}. This unrealistic behaviour disappears when including simple, parametrised radiation pressure on dust grains. This extra force causes larger velocities, thus stronger adiabatic cooling \citep{Wood1979a,Bowen1988}. The addition of this dust force can drive the wind, but its prescription does not consider detailed dust formation. Later on,  \citet{Fleischer1992,Fleischer1995,Winters2000} introduced an improved dust prescription by including time-dependent dust evolution and modified classical dust nucleation theory, yet, under the assumption of gas-phase chemical equilibrium. Their wind model also includes a prescription for local radiative equilibrium. The introduction of such detailed dust physics results in AGB winds that are mainly dust-driven where pulsations play a vital role by temporarily creating favourable dust condensation regions of high density. Meanwhile, \citet{Feuchtinger1993} improved the prescription of radiation by developing a self-consistent radiation hydrodynamics (RHD) wind model. Later, \citet{Hofner1995} extended this RHD model by including the above-mentioned dust prescription. They initially used grey radiative transfer, but later improved this with a frequency-dependent radiative transfer scheme \citep{Hofner2003}. All aforementioned models assume spherical symmetry, and a simplified mechanism to induce pulsations. However, \citet{Woitke2006} showed the spontaneous formation of dust-gas instabilities when extending to 2D. Additionally, 3D RHD models with focus on the star itself, reveal pulsations and shocks by self-excited large scale convective bubbles \citep{Freytag2008,Freytag2017}.\\\\
Radiative heating/cooling is a key process in the driving mechanism of the wind. Directly due to energy loss, and indirectly since nucleation and dust formation critically depend on temperature. Due to the constant compression and expansion by shocks, strong deviations from radiative equilibrium may occur and the gas temperature structure cannot be obtained from radiative transfer calculations alone. Therefore, \citet{Woitke1996a} have calculated detailed non-LTE radiative heating and cooling rates. Next, \citet {Schirrmacher2003a} used an updated and enlarged version of these rates to include in a self-consistent dust-(grey-)RHD wind model.\\\\
Note that all above-mentioned studies assume chemical equilibrium. Yet, due to the complex dynamical behaviour of the wind, this is not necessarily true. Shocks pass through the gas on typical time scales of one year, compressing and heating up the gas, hence affecting its composition. However, equilibrium time scales for chemical reactions can be longer than the dynamical time scale, preventing the composition to be chemically stable. Such non-equilibrium chemical evolution has been modelled in the dynamically complex AGB wind \citep{Cherchneff1992,Cherchneff2006a,Marigo2016}, including detailed dust formation and gas-grain chemistry \citep{Cherchneff2012,Gobrecht2016}. However, these non-equilibrium chemical evolution studies are post-processed on the wind structure derived by a semi-analytic approximation for dynamics. As mentioned above, the wind structure of an AGB star is far from what this approximation predicts. Additionally, since chemical reactions critically depend on the local temperature, it is crucial to determine a detailed temperature structure before solving the chemical evolution. Moreover, this is a self-dependent problem and one needs to solve the chemical and thermal evolution simultaneously as heating/cooling is regulated by the presence/absence of its heating/cooling species. In conclusion, the chemical evolution will indirectly affect the dynamical behaviour of the system since it can regulate the heating and cooling of the gas plus steer dust evolution.\\\\
We have taken a next step in unravelling the AGB wind driving mechanism by self-consistently evolving its gas-phase chemical, thermal and dynamical behaviour. We modelled the wind as a multi-fluid, where each chemical species follows the hydrodynamical equations. Additionally, a time-dependent chemical evolution is traced using a reduced chemical network. We determined this network by a flux-reduction and validation algorithm to ensure it contains all relevant reactions but is still computationally feasible for on the fly calculations. Furthermore, the temperature is determined by time-dependent non-LTE heating/cooling rates, depending on the local non-equilibrium chemical composition.\\\\
This work, foremost, serves as a proof-of-concept, since it lacks critical physics needed to represent a realistic AGB wind. Firstly, it does not consider dust formation and evolution. Secondly, it neglects radiation pressure on dust grains. Thirdly, the reduced chemical network is not exhaustive as it is the first of its kind for an AGB wind. Fourthly, the heating/cooling rates are limited to the mircophysical processes provided by \citet{Grassi2014}. The hydrodynamical part is limited to 1D to keep the computations feasible. This work provides a basis for more complex chemistry, thermal physics, dynamics, and inclusion of dust. It is the first in a series of upcoming papers in which we strive towards a more self-consistent AGB wind. The next paper will focus on rendering the current nucleation theory in AGB wind models more self-consistently. This will allow for a more correct prediction of seed particles abundances which are needed to initiate dust formation. A third paper will then introduce size dependent dust evolution including gas-grain and grain-grain processes. It will combine dynamics, chemistry, nucleation, dust evolution, and a radiation field, representing an improved and more self-consistent AGB wind model. We opt for this bottom-up approach to more easily disentangle effects of different processes since we suspect the introduction of coupled time-dependent chemical and thermal evolution to already have large repercussions for the driving mechanism of the wind.\\\\
Section~\ref{sec:method} describes the equations governing the dynamical, chemical, and thermal evolution, as well as the computational framework of the self-consistent model. Section~\ref{sec:model} elaborates on the hydrodynamical setup, the construction of the reduced chemical network, and the details of the heating/cooling processes involved. Section~\ref{sec:results} presents the results of a purely hydrodynamical model and a hydrochemical model. Section~\ref{sec:discussion} discusses the limitation of the current model and interprets its results. Section~\ref{sec:summary} ends with a summary and future perspectives.

\section{Methods}\label{sec:method}
This work consists of two main aspects, namely hydrodynamics and chemistry. This section will discuss what both aspects represent, by which equations they are dictated, and how to solve these equations numerically. The section will end with a description of how to self-consistently combine hydrodynamics and chemistry, some caveats and how to address them. 

\subsection{Hydrodynamics}
\label{sec:hydro} 
To model the hydrodynamical evolution, we use the \mpiamrvac\footnote{This code has been developed with the emphasis on shock-dominated problems and has been widely used in high performance computing simulations, making it ideal for our purposes (\url{http://amrvac.org/}).} hydrodynamics code \citep{Keppens2012}, which solves the conservation of mass:
\begin{equation}\label{eq:cons_mass}
\frac{\partial \rho}{\partial t} + \nabla \cdot (\rho \mathbfit{v}) = 0,
\end{equation}
momentum:
\begin{equation}
\label{momentum}
\rho \left( \frac{\partial \mathbfit{v}}{\partial t} + \mathbfit{v} \cdot \nabla \mathbfit{v}\right) + \nabla p = \rho \mathbfit{g},
\end{equation}
and energy:
\begin{equation}
\label{energy}
\frac{\partial e}{\partial t} + \nabla \cdot (e \mathbfit{v}) + \nabla \cdot (p \mathbfit{v}) = \rho (\mathbfit{g} \cdot \mathbfit{v}),
\end{equation}
with $\mathbfit{v}$ the velocity vector, $\rho$ the mass density, $p$ the internal pressure, and $\mathbfit{g}$ the gravitational acceleration. The right hand sides of Eq. \eqref{momentum} and \eqref{energy} represent a gravitational point source of mass $M_\star$ as extra force term, with
\begin{equation}
\mathbfit{g} = \frac{-GM_\star}{r^2}\hat{\mathbfit{r}},
\end{equation}
where $\mathbfit{r}$ is the radial unit vector. This set of equations is closed by the equation of state, which we assume to be of an ideal gas, 
\begin{equation}\label{eq:IG}
p=T \rho \kbmu,
\end{equation}
relating the pressure and temperature\footnote{In a purely dynamical framework, the temperature of the gas is solely regulated by compression and expansion of the gas (i.e. adiabatic heating and cooling).} to the internal energy of the gas. The energy density $e$, defined as the sum of internal and kinetic energy density, is then:
\begin{equation}\label{eq:energy_density}
e = \frac{p}{\gamma - 1} +\frac{\rho v^2}{2},
\end{equation}
with $\gamma$ the adiabatic index. The adiabatic index is a local variable and depends on the temperature and composition of the gas (see Appendix~\ref{app:gamma} for detailed calculations).

\subsection{Chemistry}
\label{sec:chem}
To solve the chemical and thermal\footnote{\label{note:temp}Form this point onward, we use the term `thermal' to describe non-adiabatic behaviour, unless stated otherwise.} evolution of a gas mixture, we use the \krome\footnote{\url{http://kromepackage.org/}} package \citep{Grassi2014}. The evolution of an initial set of chemical species is prescribed by a set of formation and destruction reactions of that species. This translates into a set of ordinary differential equations (ODEs). The change in number density of the \textit{i}th species is given by:
\begin{equation}
\label{ODEsys}
\frac{\text{d}n_i}{\text{d}t} = \sum_{j\in F_i} \left( k_j \prod_{r \in R_j} n_r \right)- \sum_{j\in D_i} \left( k_j \prod_{r \in R_j} n_r \right),
\end{equation}
where the first term represents the rate of formation of the \textit{i}th species by a reaction $j$ of a set of formation reactions $F_i$, while the second term is the analogous for a set of destruction reactions $D_i$. Each reaction $j$ has a set of reactants $R_j$, where $n_r$ is the number density of each reactant. The rate coefficient of this reaction is represented by $k_j$ and has units m$^{3(N-1)}$ s$^{-1}$ where $N$ is the number of reactants involved.\\\\
The change in temperature\textsuperscript{\ref{note:temp}}, is solely due to loss and gain of energy and is given by:
\begin{equation}\label{eq:temp_change}
\frac{\text{d}T}{\text{d}t} = \frac{\partial T}{\partial E} \frac{\text{d}E}{\text{d}t} = \frac{\gamma-1}{\kb}\frac{\Gamma(T,\mathbf{n})-\Lambda(T,\mathbf{n})}{\sum_i n_i},
\end{equation}
where the first factor of the last equality uses Eq. \eqref{eq:CV} and the second factor represents the change in energy by heating, $\Gamma$, and cooling, $\Lambda$, in units of J m$^{-3}$ s$^{-1}$. Both are a function of temperature, $T$, and of the vector $\mathbf{n}$, containing the number densities of all species. The sum in the denominator represents the total gas number density in m$^{-3}$.\\\\
From a mathematical point of view, a chemical network is represented by a set of ODEs (Eq. \ref{ODEsys}) which, in an astrochemical framework, often constitutes of a system of stiff, coupled equations. Very often astrochemical networks present a sparse or very sparse Jacobian matrix associated with their ODE system. \krome uses the DLSODES solver \citep{Hindmarsh1983} which takes advantage of the sparsity structure of the Jacobian matrix. Hence, using this solver can lead to a large speed-up compared to other widely-used solvers like DVODE/CVODE in the SUNDIALS package \citep{Hindmarsh}, which use the same integration scheme but do not exploit the sparsity of the Jacobian \citep{Nejad2005}. \citet{Grassi2013} have shown that using DLSODES over DVODE can produce a speed-up of a factor $\sim$100 for a Jacobian matrix of sparsity $\sim$94\%, which is quite common for astrochemical networks.\\\\
For more details on available microphysics and numerical implementation we refer the reader to the \krome paper \citep{Grassi2014}.

\subsection{Hydrochemistry}
\label{sec:hydrochem}
\begin{figure}
	\includegraphics[width=\columnwidth,height=5cm]{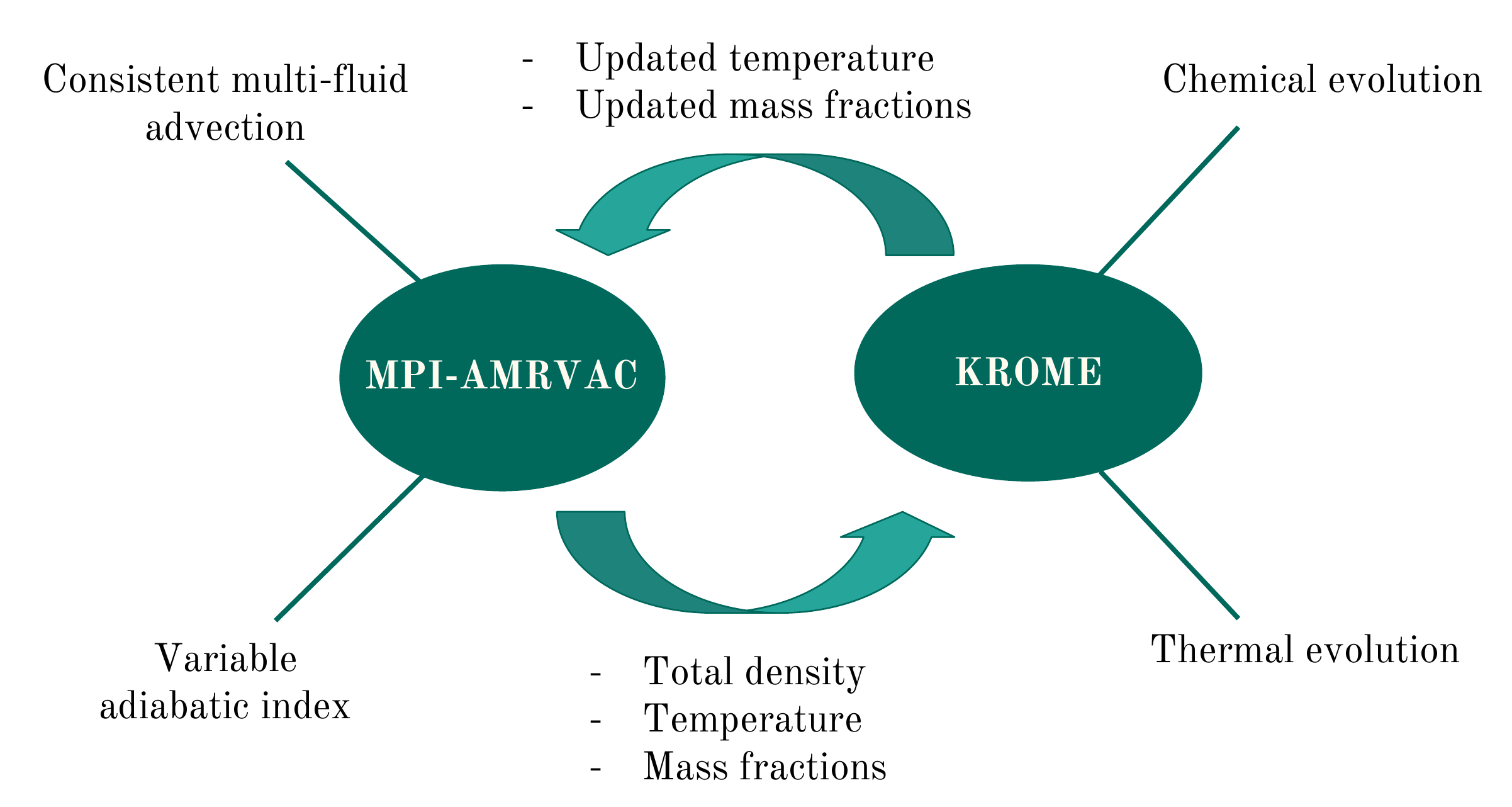}
    \caption{Schematic overview of the hydrochemical coupling cycle between \mpiamrvac and \krome. This cycle occurs every hydrodynamical time step in every hydrodynamical grid cell.}
    \label{fig:amrvackrome}
\end{figure}

\krome is a standalone 0D code that can be incorporated into an external framework code, in this case \mpiamrvac. \krome independently evolves chemistry and temperature for a specific time, given the necessary input. Here, \mpiamrvac provides \krome with local values of the total gas density, gas temperature, and mass fractions of all chemical species. On its turn, \krome returns updated values of temperature and mass fractions to \mpiamrvac, which it uses to update its hydrodynamical parameters \ado{(see Appendix~\ref{appAsect:e-gamma} for details on updating the energy density)}. This cycle is executed each hydrodynamical time step in every hydrodynamical grid cell, evolving chemistry and temperature over that same hydrodynamical time step (Fig.~\ref{fig:amrvackrome}).\\\\
We have extended \mpiamrvac with a new module to make this hydrochemical coupling possible. This includes the option to have an arbitrary number of chemical species, represented by a passive scalar field that advects as the density (i.e. multi-fluid). Additionally, to ensure that elemental abundances are conserved locally as well as globally, we have implemented a Modified Consistent Multi-fluid Advection (MCMA) routine described by \citet[][appendix A]{Glover2010}, based on the original version of \citet{Plewa1999}. For simplicity of implementation, we use the similar procedure offered by \krome \citep[][appendix E]{Grassi2017}. This extra conservation routine is crucial because, when advecting hydrodynamic quantities or locally refining/coarsening grid cells, the chemical abundances are interpolated using a slope limiter, and the interpolation weight can be different for different chemical abundances. If this interpolation scheme were applied naively, the chemical abundances and chemical composition would therefore not always be conserved;  the error might be small initially but could grow dramatically over time. Lastly, \mpiamrvac is extended with a local and variable adiabatic index $\gamma$, in contrast to a global and fixed value that was adopted before. Its calculation is done by \krome, as it depends on the local temperature and composition of the gas.

\section{Model setup}\label{sec:model}
The hydrodynamical and chemical aspects of our model will be discussed in detail. The first part will describe the hydrodynamical setup and justify the choice of physical parameters values. The second part will elaborate on the construction of the chemical network and the included microphysical processes regulating the temperature.

\subsection{Hydrodynamics}
\label{sec:hydromod} 
This section will elaborate on the hydrodynamical model setup. The first part will describe the numerical grid and how the hydrodynamical equations are solved. The second part will expand on the chosen boundary conditions. The third part will clarify the choice of initial conditions.

\subsubsection{Numerical setup}
Fully coupling chemical evolution with a hydrodynamical model drastically increases the computational cost. To keep this cost low for this proof-of-concept, we opt for a 1D model of the stellar wind\footnote{Extension to higher dimensions is straightforward using \mpiamrvac, but the computational time will drastically increase.}. The radial grid of this model extends from 1~au to 10~au, with 1~au being a typical stellar radius\footnote{It is difficult to speak of a stellar radius as AGB stars are dynamically highly active and non-spherical. When referring to the stellar radius $R_\star$ or stellar surface at $R_\star$, we imply the start of the numerical grid at 1~au.} of an AGB star and 10~au an expected transition point to a constant radial outflow due to dust acceleration \citep[][and references thererin]{Habing2003}. Further out, the wind will also be affected by external radiation (e.g. photochemistry) which is beyond the scope of this paper. The adaptive mesh refinement (AMR) is turned off because we currently only implemented MCMA when advecting species and not when switching AMR level. The latter is less trivial and will be implemented in a completely refurbished version, \mpiamrvac~2.0, that is current under development. Without the AMR option, we made sure our model setup has enough spatial resolution so the result stays unaffected. We determined that 2000 grid cells is sufficient. When later coupled with chemistry, this setup is viable to run on a small computer cluster.\\\\
We solve the hydrodynamics equations using a second order total variation diminishing Lax-Friederich TVDLF scheme (\citealt{Yee1989}, improved version in \citealt{Toth1996}) combined with a Koren flux limiter \citep{Kuzmin2006}. A 4-stage, 3rd order strong-stability-preserving Runge-Kutta method is used as time discretisation scheme \citep{Ruuth2002}. The time step is determined by the Courant-Friedrichs-Lewy CFL condition using a Courant number of 0.8.

\subsubsection{Boundary conditions}
Our model grid is a fixed slice in space, therefore we can use the output of simulations modelling the star itself as an input for our wind model, which is further extended in space. According to 3D~RHD models, the mean radial velocity at the stellar radius is predicted to vary with a few \kms, and an averaged period of a few 100 days (\citealt{Freytag2008}, fig.~5; \citealt{Freytag2017}, fig.~6). Since we are only interested in the effect of pulsations on the chemical-dynamical interaction further down in the wind, rather than in the driving mechanism itself, we use a simplified version of this velocity variation:
\begin{equation}
v(t,r=R_\star)=\Delta v \sin\left(\frac{2\pi t}{P}\right)
\end{equation}
at the inner boundary. Here, $\Delta v$ is the velocity amplitude and $P$ the pulsation period where we choose values of $\Delta v=2.5$ \kms and $P=300$~d, consistent with the 3D~RHD simulation results. Note that, because we specify the velocity field of the pulsation, the boundary is open and mass can flow through, which is in line with using the 3D~RHD model output as our input. This prescription is \ado{different from} the widely used piston approximation, which is a solid wall pushing the material in the numerical box \citep[e.g.][]{Bowen1988,Fleischer1992,Winters2000,Schirrmacher2003a,Freytag2008,Liljegren2016,Hofner2016}. The latter approach was used to induce pulsations in the wind by lack of a proper interior pulsation model.
\ado{We opt for the former approach because it relies on less assumptions on the pulsation mechanism by using the output of simulations specifically trying to model that mechanism. Additionally, the solid piston approximation has no material replenishment from the star to the wind as the lower boundary is a numerical wall. This allows for the total mass in the numerical box to decreases over time due to wind mass loss at the outer boundary, which is most likely not what happens in reality.}
In our model, we do keep the temperature and density at the inner boundary fixed. This is not the most realistic approach and preferably 3D~RHD variations should be used, but it limits the degrees of freedom of the pulsation model, allowing us to have more control on what exactly affects the chemical-dynamical interaction. We choose typical values of temperature and density at the stellar surface with $T_\star=2500$~K \citep[][table~4]{Menshchikov2001} and  $\rho_\star = 10^{-6}$~kg\,m$^{-3}$. We opt for this density value as it is consistent with the 3D~RHD model results (Table~\ref{tab:rhoLit} gives an overview of values of the surface density typically used for AGB winds). \ado{Note that both values need to be chosen and one cannot be inferred from the other on grounds of equilibrium (either hydrostatic or a more complex radiative equilibrium as in model atmosphere calculations)}.\\\\
The outer boundary of the grid is an open boundary, where material can flow in and out. \ado{This means that the gradient (of the conservative variables density, momentum, and energy) is kept zero by copying the variable values from the edge of the mesh into the ghost cells.}

\begin{table*}
	\centering
	\caption{Literature values of the stellar surface density.}
	\label{tab:rhoLit}
	\begin{tabular}{lll}
		$\rho_\star$ (\si{\kg\per\m\cubed}) & Method & Reference \\
		\hline
		$10^{-8}-10^{-5}$ & Calculated (with chosen outer boundary pressure)  & \citet{Marigo2013} -- \citet[fig. 1]{Marigo2016}  \\
		$6.7\cdot10^{-6}$ & Calculated (not explained)   & \citet{Bowen1988} \\
		$6.03\cdot10^{-4}$ & Calculated (not explained)  & \citet{Cherchneff1992} \\
		$\sim 10^{-6}$ & Calculated (with chosen inner boundary opacity and optical depth)  &\citet[fig. 3]{Fleischer1992} -- \citet[fig. 5]{Winters2000} \\
		$\sim 5 \cdot10^{-6}$ & Model results  & \citet[fig. 4]{Freytag2008} \\
		$\sim 5 \cdot10^{-6} - 10^{-5}$ & Model results  & \citet[fig. 4]{Freytag2017} \\
	\end{tabular}
\end{table*}

\subsubsection{Initial conditions}
Using a hydrostatic equilibrium solution for the density as initial condition is less meaningful than suggested by intuition. This is due to the still existing parameter freedom in the choice of chemical composition and temperature profile of the gas. Even within realistic choices of both, the density profile will result in vastly different hydrostatic equilibrium solutions (Appendix~\ref{app:HE}). As there exists no single equilibrium solution, we take a density power law as initial condition,

\begin{equation}\label{rhoPow}
\rho(r) = \rho_\star \left( \frac{r}{R_\star} \right)^{-\alpha},
\end{equation}
with $\alpha=10$. This value of $\alpha$ is consistent with model results of \citet{Hofner2016} and \citet{Freytag2017}. Regarding the temperature profile, we opt for a power law,

\begin{equation}
T(r)= T_\star \left( \frac{r}{R_\star} \right)^{-\beta},
\end{equation}
with $\beta=0.5$. In detailed models $\beta$ typically has a value between $0.4$ and $0.8$ but varies radially, with a steeper temperature profile closer to the star \citep[][fig. 10]{Freytag2017}. The equation of state (Eq.~\ref{eq:IG}) is then used to convert to initial hydrodynamical profiles where the mean molecular weight will be calculated based on the assumed chemical composition (discussed in Section~\ref{sec:network}). Note that the choice of initial profile, within reason, is not that important as the hydrodynamical evolution will erase any preset information. The used parameter values are summarised in Table~\ref{tab:param}.

\begin{table}
	\centering
	\caption{Model parameter values.}
	\label{tab:param}
	\begin{tabular}{ll}
		Parameter & Value \\
		\hline
		$M_\star (M_{\sun})$        & 1.0 \\
		$R_\star$ (au)               & 1.0 \\
		$T_\star$ (K)               & 2500 \\
		$\rho_\star$ (kg m$^{-3}$)  & $1.0\cdot 10^{-6}$ \\\\
		$\Delta v$ (km s$^{-1}$)    & 2.5 -- 20\\
		$P$ (days)                  & 300\\
		$\alpha$                    & 10 \\
		$\beta$                     & 0.5 \\
	\end{tabular}
\end{table}

\subsection{Chemistry}
\label{sec:chemmod}
Chemical evolution is controlled by a complex of chemical reactions. However, including all possible reactions will results in an impractically huge chemical network that is nearly impossible to combine with dynamical evolution, due to the extreme computational costs of the chemical computations. A feasible alternative is to use a reduced network.\\\\
Additionally, one has to include all microphysical heating and cooling processes since chemical reaction rates are mainly governed by the surrounding's temperature. However, due to the vast number and complexity of these processes, one has to restrict themselves to the most relevant ones. Hence, the choice of a set of reactions and thermal processes needs to be carefully considered.

\subsubsection{Reduced chemical network}\label{sec:network}
\begin{figure}
	\includegraphics[width=\columnwidth]{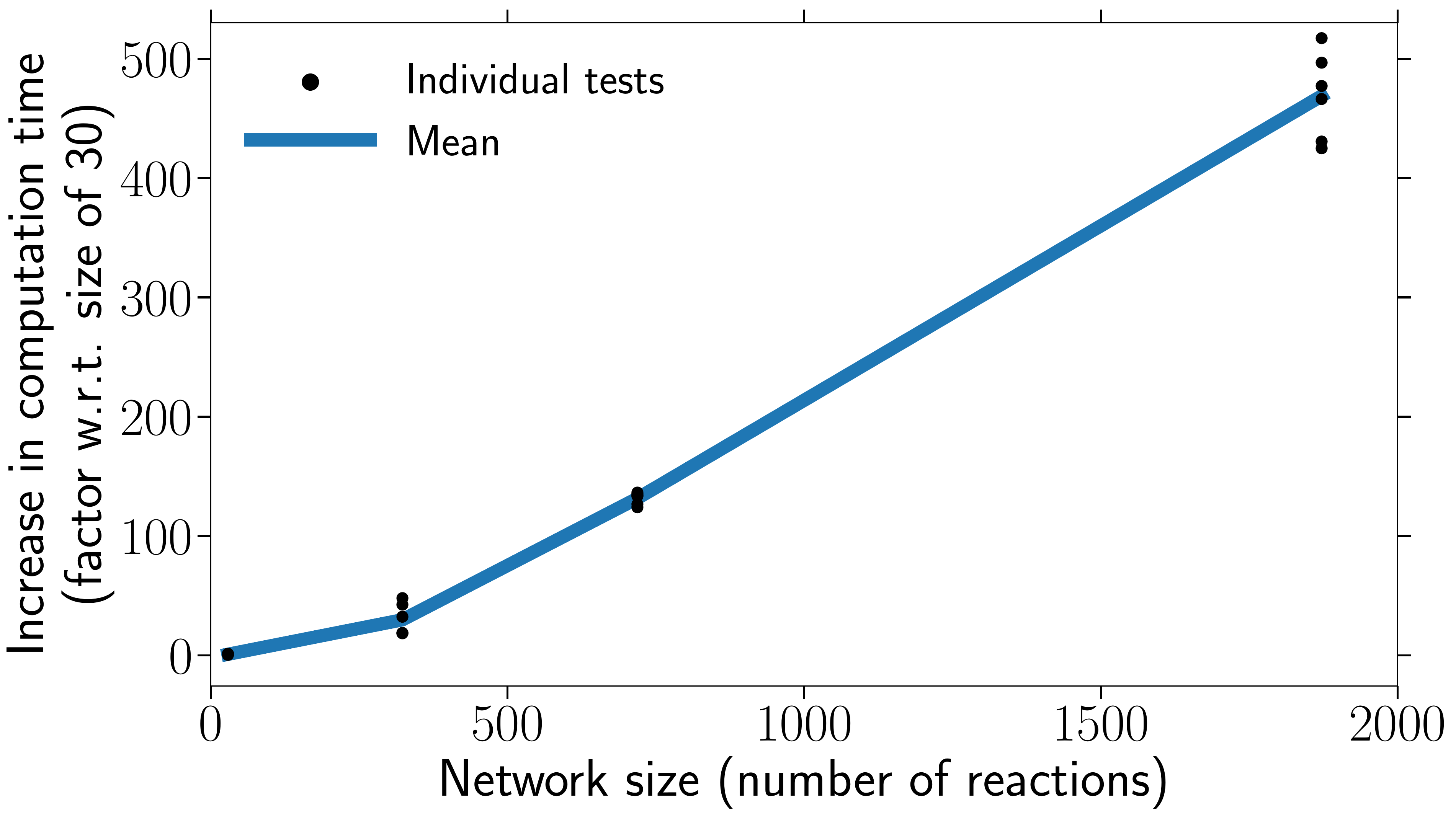}
    \caption{The size of the chemical network is the main increase factor for computation time, rather than different local conditions. To compute this factor, we ran tests for four different network sizes, where each test consists of chemically evolving initial conditions at a fixed temperature and density, for a specific time. For each network, we performed a test in different temperature density regimes. The mean computational time of these tests, normalised to the smallest network, is shown in the figure.}
    \label{fig:ntwSize}
\end{figure}

\begin{figure*}
\centering
\begin{flushright}
    \includegraphics[width=0.32\textwidth]{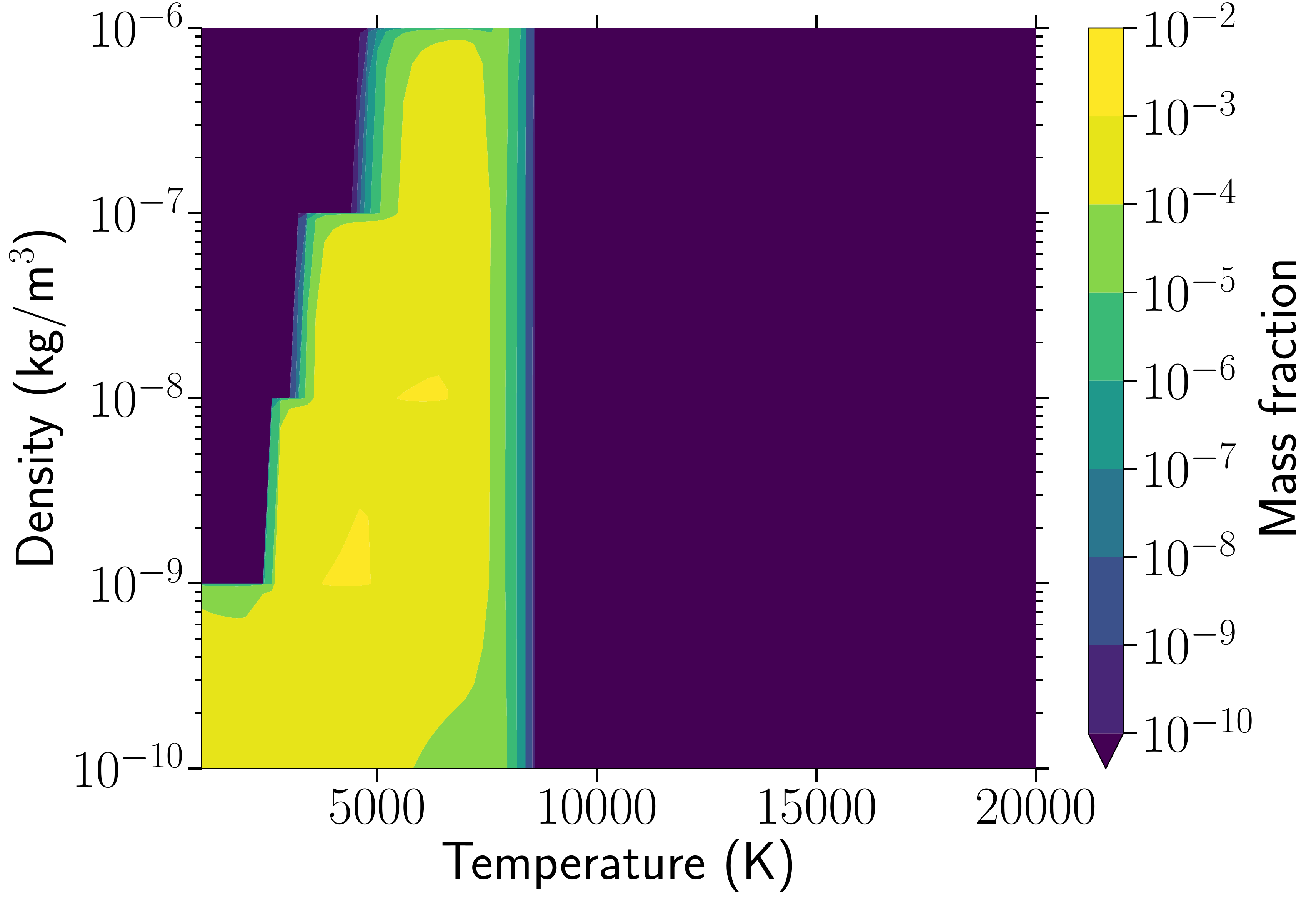}
    \includegraphics[width=0.32\textwidth]{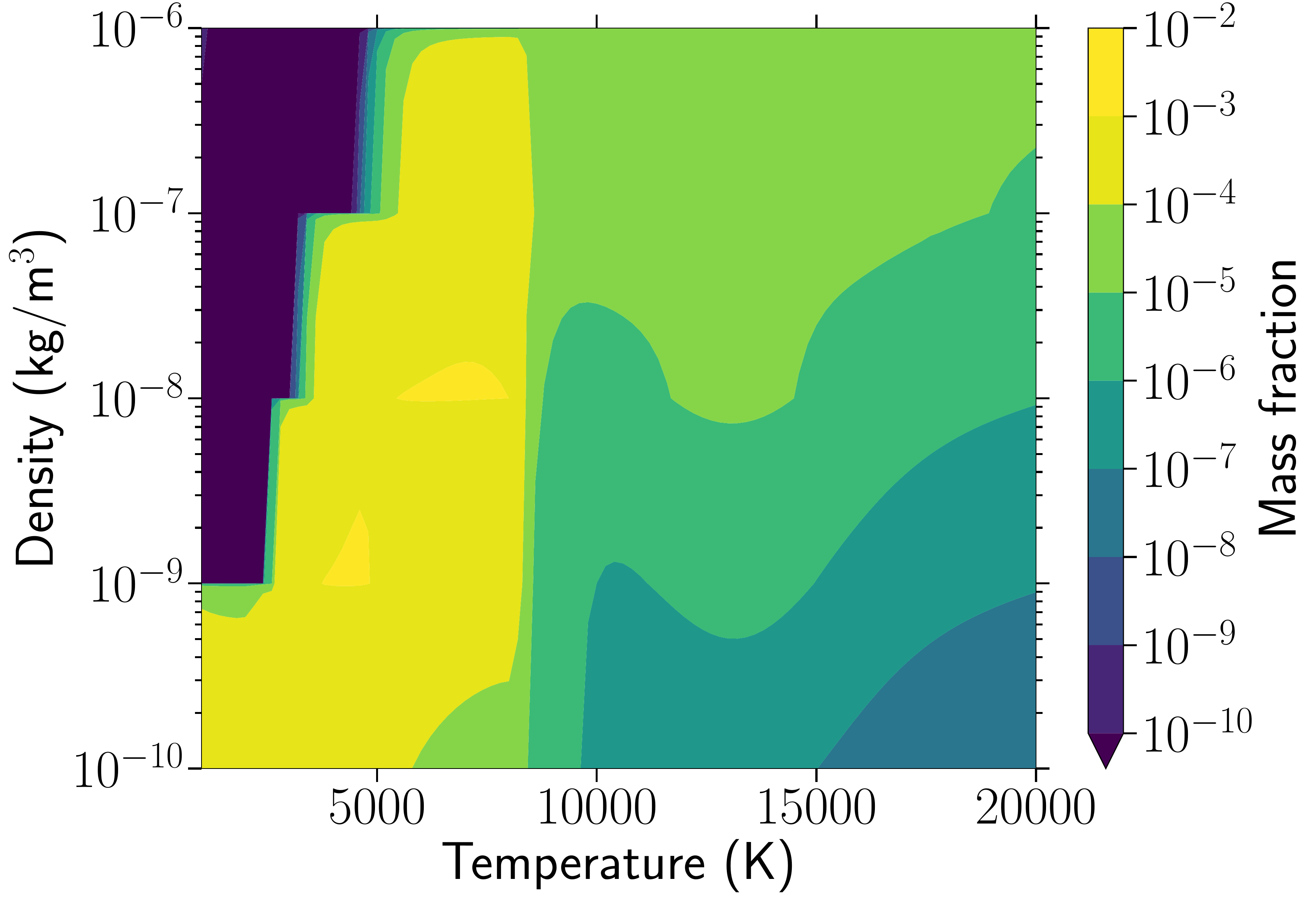}
    \includegraphics[width=0.32\textwidth]{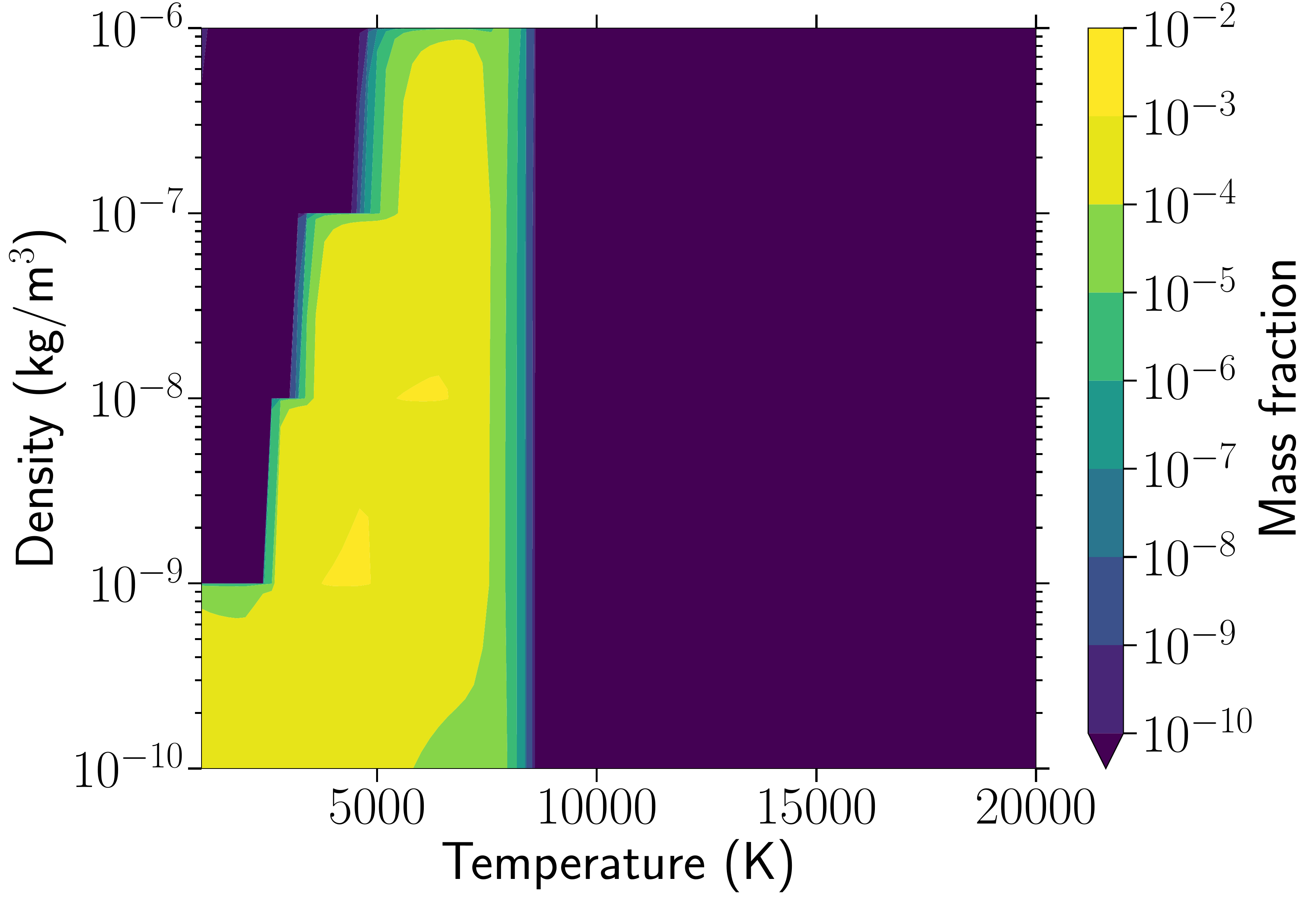}
    \includegraphics[width=0.32\textwidth]{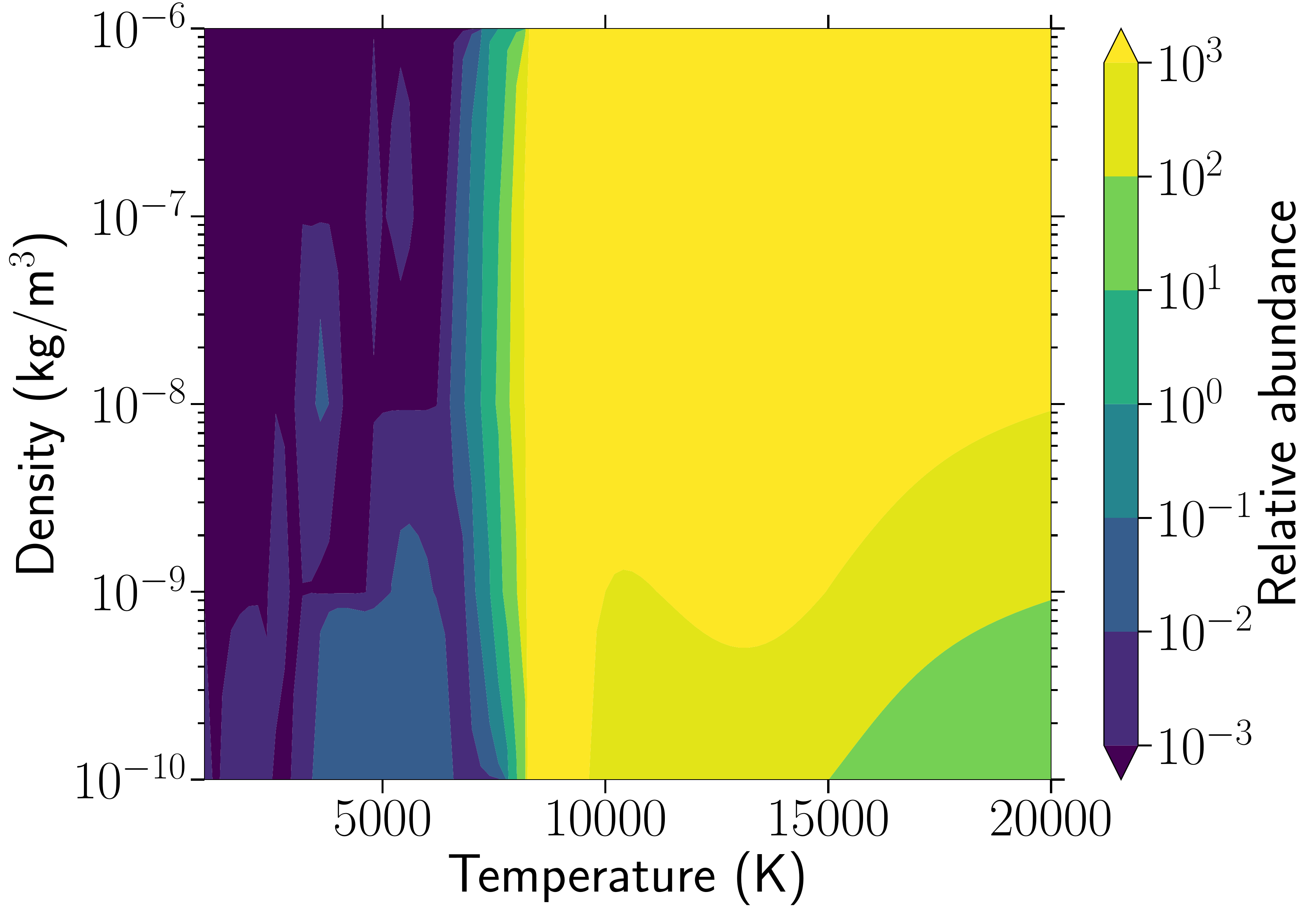}
    \includegraphics[width=0.32\textwidth]{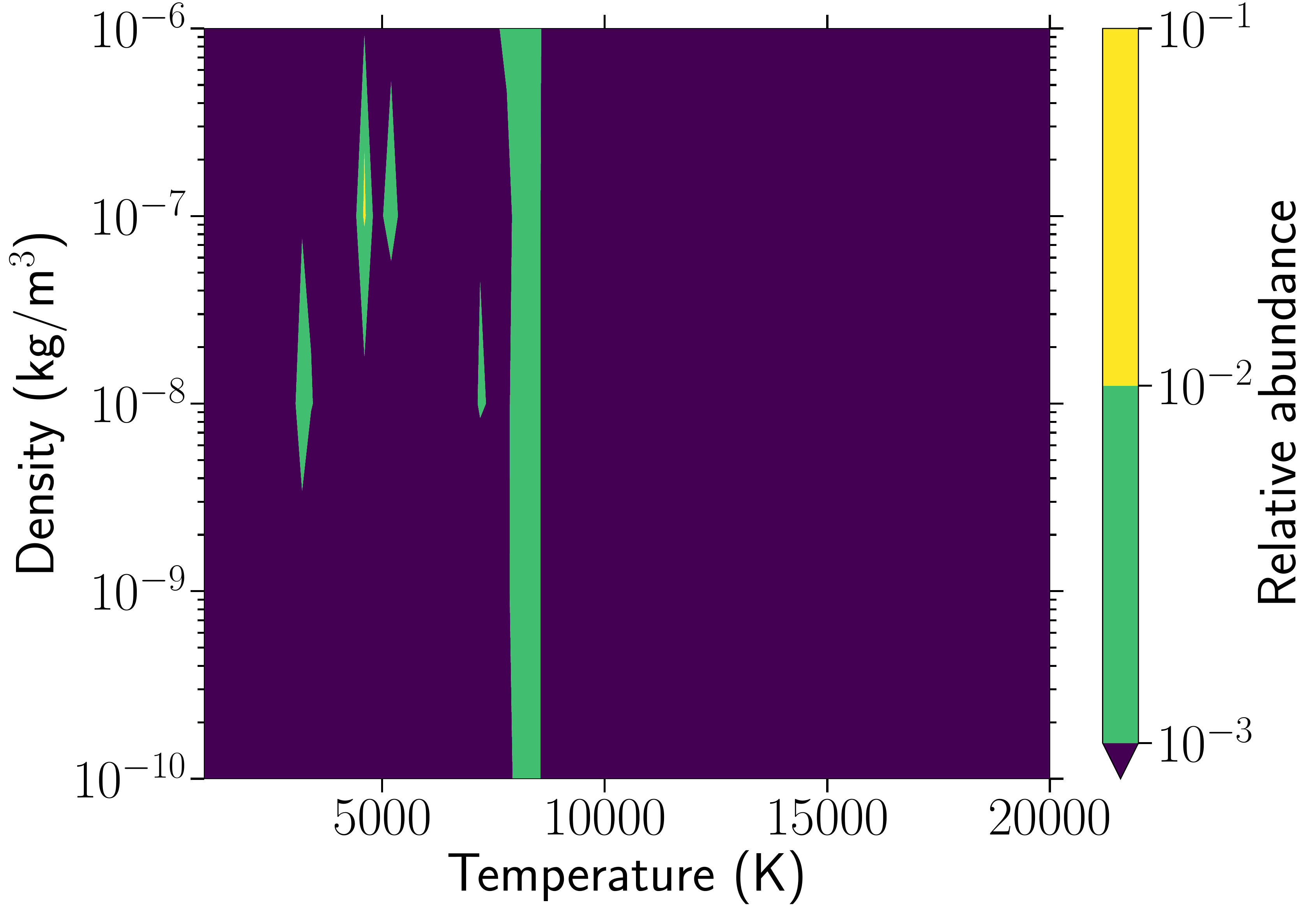}
    \end{flushright}
    \caption{Reducing a network with user defined threshold $\varepsilon= 10^{-7}$ , Eq.~\eqref{eq:reduc}, is sufficient to reproduce all chemical abundances of the comprehensive network, over the desired density--temperature range (Table \ref{tab:reducParam}). As an example, the final abundances of \ch{SiO} in the comprehensive (left), $\varepsilon= 10^{-4}$ (middle), and $\varepsilon= 10^{-7}$ (right) case are shown, where the second fails to reproduce the abundances at high temperatures. The upper row represents the absolute abundance, whereas the lower row depicts the relative difference $\left(\frac{|A_{\text{comp}}-A_{\text{red}}|}{A_{\text{comp}}}\right)$, each with a lower cut-off at $10^{-10}$ and $10^{-3}$, respectively.}
\label{fig:reducSiO}
\end{figure*}

A reduced chemical network prioritise most important species and reactions. Such a network can become several times smaller than the original one, hence reducing the computational cost tremendously (Fig.~\ref{fig:ntwSize}). The established method of reduction is to start with a comprehensive network and systematically remove components, checking that it does not alter the outcome significantly.\\\\
Because this is a proof-of-concept paper, we limit the complexity of our comprehensive network, as the construction of such is already a delicate task. Yet, we make sure it is sufficient for our purpose. For constructing the comprehensive network, we use the UMIST Database for Astrochemistry\footnote{\url{http://udfa.ajmarkwick.net}}\citep{McElroy2013} as a primary source of reactions, with the addition of several reactions from the Kinetic Database for Astrochemistry\footnote{\url{http://kida.obs.u-bordeaux1.fr}} \citep[KIDA;][]{Wakelam2012}, and a few from independent studies. Our construction method (see below) yields a chemical network of 1684 reactions and 163 different species.\\\\
As there is no unique method to determine the important reactions of a network, \citet{Grassi2013} compared different techniques and concluded that a reaction flux-reduction scheme \citep{Grassi2012,Tupper2002} proves most adequate. It is an on the fly procedure, aimed at determining the less active reactions and excluding them from the network. However, such on the fly calculation is not practical when coupling chemistry with a hydrodynamical framework. We therefore determine a reduced network which is valid during the entire hydrochemical simulation, using the algorithm presented below, which is based on the flux-reduction scheme. After executing the algorithm, the network is reduced to 255 reactions and 70 species, which is significantly less than before (Appendix \ref{app:network}).\\\\
\textbf{Comprehensive network comprises}
\begin{enumerate}[(1)]
\item All triatomic molecule reactions\\\\
This serves as a limited basis network yet also includes molecules of interest in AGB winds, e.g. \ch{HCN}, \ch{SO2}, \ch{H2O} \citep[e.g.][]{Schoier2013,Danilovich2016,Maercker2016,Lombaert2016}. Any photon reaction is  excluded because the stellar photons are energetically too weak, and the inner wind is self-shielded from any external intense radiation field due to its high density.\\\\
\item All direct cosmic ray reactions\\\\
No secondary cosmic rays ionization reactions are included because these rates are specifically calculated for dense molecular clouds and therefore not treated self-consistent in the chemical evolution. In this specific case of molecular clouds, the secondary photon reaction are calculated with an \ch{H2} line emission spectrum that arises after excitation by secondary electrons created by direct cosmic ray ionization of \ch{H2} \citep{Gredel1987,Gredel1989}.\\\\
\item Critical collisional H, He, and \ch{H2} reactions\\\\
Such high density, high temperature reactions are highly efficient in our region of interest, yet lacking in astrochemical databases that focus on low temperature, low density regimes like the interstellar medium.
\\\\
\item Three-body \ch{H2} formation reactions\\\\
Besides \ch{H2} formation on dust grains \citep{Gould1963,Cazaux2009}, which is beyond the scope of this work, three-body reactions become an important mechanism for \ch{H2} formation in the high density regime close to the stellar surface.
\\\\
\item Extra reactions to avoid sinks/sources\\\\
It is artificial to have sink or source molecules in the network. We can only resolve \ch{CNO} as a source molecule by adding one extra formation reaction. Other source molecules (\ch{PH2}, \ch{C2P}, \ch{HS2}, \ch{SiH2}, \ch{HCP}, and \ch{SiNC}) lack simple formation reactions. For these molecules, astrochemical databases only contain formation reactions by destruction of more complex molecules, which is beyond the scope of this network. We therefore ignore this issue for these source molecules.\\\\
\end{enumerate}
\textbf{Network reduction algorithm}
\begin{enumerate}[(1)]
\item Evolve chemistry of the comprehensive network in an appropriate temperature-density grid.\\\\
\item Determine the flux\footnote{Defined as the number of reactions per unit volume per unit time.} of each reaction at different evolutionary time steps.\label{reduc2}\\\\
Some reactions might only be important for a short period of time but will affect the overall outcome. To make sure we do no exclude these, we sample the reaction fluxes in time as well.\\\\
\item Deem a reaction important if:\\\\
in any given (temperature, density, time step)-grid point

\begin{equation}\label{eq:reduc}
\frac{\text{Flux of reaction}}{\text{Sum of fluxes of all reactions}} > \varepsilon,
\end{equation}

with $\varepsilon$ a user defined threshold.\\\\
\item Verify reduced network.\label{reduc4}\\\\
Verify that the chemical network does not include source or sink species and that recombination reactions are included. If this is not the case, then add relevant reactions to make the network internally consistent. Note that cosmic ray reactions might not be deemed important by the reduction process, but they are necessary for the electron production and thus electron reactions.\\\\
\item Evaluate reduced network.\label{reduc5}\\\\
Because $\varepsilon$ is user defined, one has to compare the reduced output abundances with the original network and make sure they do not differ significantly. If this is not the case, $\varepsilon$ needs to be lowered.\\
\end{enumerate}
We executed this reduction process in a temperature-density grid (Table~\ref{tab:reducParam}) applicable to our hydrodynamic results (Section~\ref{sec:hydrores}). We decided to evolve the comprehensive chemical network over a typical pulsation period. Locally, this corresponds to the longest dynamically stable period (between consecutive shocks), resulting in a roughly constant temperature and density. According to reduction step \ref{reduc2}, the temporal evolution also needs to be sampled. This is sampled exponentially to capture the very fast reactions in the reduction process. For this reduction process, we opted for the results of AGB evolution models of \citet{Karakas2010} as initial chemical composition  (Table~\ref{tab:chemInit}). These models comprise post-processed nucleosynthetic evolution on stellar structure evolution starting from the zero main sequence to near the end of the thermally pulsating AGB phase \citep{Karakas2007,Karakas2010}. We adopt the time-averaged elemental mass fractions in the wind as our initial chemical composition (defined as $\left<X(i)\right>$ in \citet{Karakas2007}). Their output tables only consider elements with a mass fraction higher than $10^{-10}$, limiting the elements we take into account. Additionally, elements of which no reactions are present in our comprehensive network are also excluded from the initial composition.  Executing the verification and evaluation steps \ref{reduc4}-\ref{reduc5}, we empirically determined that a threshold of $\varepsilon = 10^{-7}$ is sufficient to reproduce the original abundance covering the entire temperature-density grid. All species' abundances of the reduced network do not differ significantly from the comprehensive network results. We refrain from showing all comparisons and provide one example species (Fig.~\ref{fig:reducSiO}).

\begin{table}
    \center
    \caption{Parameter space for reduction scheme.}\label{tab:reducParam}
    \begin{tabular}{lllr}
        Parameter& Range & Iteration  & Grid points \\ \hline
        Temperature & 1000--\num{20000} K & T$_{i+1} = $ T$_i + 200$ K & 100 \\ 
        Density & 10$^{-6} - 10^{-10}$ kg/m$^3$ & $\rho_{i+1} = \rho_i \cdot 10$ & 5 \\ 
        Time & 1h $-$ 1yr & t$_{i+1} = $ t$_i \cdot 1.5$  & 22 \\ 
    \end{tabular}
\end{table}

\begin{table}
    \center
    \caption{Initial chemical composition. This is equal to the time-averaged mass fractions in the wind for a nucleosynthetic AGB evolutionary model  with initial solar mass and solar metalicity of \citet{Karakas2010}.}\label{tab:chemInit}
    \begin{tabular}{ll}
        Element $i$ & Mass fraction $X_i$ \\ \hline
        He & $3.11\cdot 10^{-1}$ \\
        C & $2.63\cdot 10^{-3} $\\
        N & $1.52\cdot 10^{-3} $\\
        O & $ 9.60\cdot 10^{-3} $\\\\
        S &  $3.97\cdot 10^{-4} $\\
        Fe &  $1.17\cdot 10^{-3} $\\
        Si &  $6.54\cdot 10^{-4} $\\
        Mg &  $5.16\cdot 10^{-4} $\\\\
        Na & $3.38\cdot 10^{-5} $\\
        P & $8.17\cdot 10^{-6} $\\
        F &  $4.06\cdot 10^{-7}$ \\
        H & $1 - \sum_i^N X_i$ \\ 
        \ch{e-} & 0 \\
    \end{tabular}
\end{table}

\subsubsection{Thermal processes}\label{sec:thermalproc}
The thermal state of the gas is regulated by a number of microphysical processes where different processes are relevant in different regimes. Their relevance and efficiency mainly depends on the local gas temperature. Due to the large range in temperatures covered in an AGB wind, this implies a vast number of processes. In line with the proof-of-concept layout, we limit ourselves to the most prominent processes available in \krome (Table~\ref{tab:thermalProc})\\\\
\begin{enumerate}[(1)]
    \item{\ch{H} and \ch{He} line cooling}\\\\
    This includes  the collisional ionization of H, He, and \ch{He+} by electrons; \ch{H+} and \ch{He+} recombination; He dielectric recombination; and H (all levels), He (n=2, 3, 4 triplets), and \ch{He+} (n=2) collisional excitation by electrons. All cooling functions are as described in \citet{Cen1992}, which are taken or adapted from \citet{Black1981} and \citet{Spitzer1978}.\\\\
    \item{\ch{H2} line cooling}\\\\
    This comprises \ch{H2} rovibrational line cooling which in the low density regime uses a cooling function of \citet{Glover2008} whereas in the high density regime an LTE cooling function of \citet{Glover2015}. The low density limit considers collision by \ch{H}, \ch{H+}, \ch{H2}, He, and \ch{e-}, and is an improvement over the cooling function of \citet{Galli1998} that only considers collisions between \ch{H} and \ch{H2} \citep[][fig.~1]{Glover2008}. The high density limit LTE cooling function is an improvement over the widely used expression of \citet{Galli1998} which is based on an analytic LTE approach by \citet{Hollenbach1979}. \citet{Glover2015} showed that this latter differs up to a factor of 2 for temperatures above $\sim$2000~K. Both low and high density cooling rates are only valid in the optically thin limit. However, at number densities $n_\text{\ch{H}} \ge 10^{14}$~m$^{-3}$ ($n_\text{\ch{H}}\equiv \rho/m_\text{\ch{H}}$) atomic hydrogen quickly gets turned into molecular form via three-body reactions, which become important \citep{Palla1983}, making the gas optically thick for \ch{H2} line radiation. In this regime, \krome follows the model of \citet{Ripamonti2004} where the optically thin cooling is scaled with the total number density \citep[][eq. 13]{Grassi2014}.\\\\% \ch{H2} cooling is defined as:
    %\begin{equation}
    %\Lambda_{\text{\ch{H2},thick}} = \Lambda_{\text{\ch{H2},thin}} \cdot \text{min} \left[ 1,\left(\frac{n_\text{tot}}{8\cdot 10^9 \si{\per\cm\cubed}}\right)^{-0.45} \right].\\\\
    %\end{equation}
    \item{Chemical \ch{H2} cooling}\\\\
    The gas can also cool via collisional dissociation of \ch{H2} molecules. According to \citet{Omukai2000}, this process absorbs the same amount of energy as the binding energy, that is 4.48~eV per \ch{H2} molecule. Currently, \krome supports only a fixed number of dissociation reactions, of which only two are applicable to our reduced network (Table~\ref{tab:H2diss}).\\\\
    \item{Chemical \ch{H2} heating}\\\\
    Heating by formation of \ch{H2} is the only relevant chemical heating source available in \krome. Only formation via \ch{H}, \ch{H-}, and three-body reactions are currently supported\footnote{Although our chemical network considers more \ch{H2} formation reactions. E.g., \ch{H + H + He -> H2 + He} might be an important heating reaction.}  (Table~\ref{tab:H2diss}), following the approach of \citet{Omukai2000}. \krome follows the prescription of \citet[][eq. 6.45]{Hollenbach1979}, who state that the heat released per formed \ch{H2} molecule is weighted by a critical density factor, which depends on the fractional abundances of \ch{H} and \ch{H2}.\\\\
    %\begin{equation}\label{eq:fcrit}
    %f_{\text{crit}}\equiv \left (1+\frac{n_{\text{crit}}}{n_\text{tot}} \right )^{-1},
    %\end{equation}
    %with $n_{\text{crit}}$ (in cm$^{-3}$) prescribed by:
    %\begin{align}
    %n_{\text{crit}} = 10^6 T^{-1/2}&\left\{1.6~n_\text{H}~ %\text{exp}\left[-\left(\frac{400}{T}\right)^2\right] \right. \nonumber\\
    %                &+ \left. 1.4 ~n_\text{\ch{H2}} \text{exp}\left[ \frac{-12000}{T+1200}\right] \right\}^{-1}.\\\nonumber\\\nonumber
    %\end{align}
    \item{\ch{CO} line cooling}\\\\
    Cooling by CO rotational lines is incorporated as well. The cooling table includes \ch{H2} and H as collision partner, provided by \citet{Omukai2010} and private communication between T. Grassi and K. Omukai, respectively. Calculation of the cooling rates is based on the method of \citet[eq. 5]{Neufeld1993}.\\\\ 
  \item{Collisionally induced emission (CIE)}\\\\
  At densities higher than $n_\text{\ch{H}} \sim 10^{20}$~m$^{-3}$, hydrogen molecules collide so frequently to form pairs of atoms/molecules (\ch{H2-H2}, \ch{H2-He}, \ch{H2-H}) that the collision pair temporarily induces a non-zero electric dipole making either molecule emit a photon. In the same way, a H-He collision perturbs both atoms resulting in a high probability of emitting a photon through a dipole interaction. Because of the very short collision times, CIE lines become very broad and essentially appear as continuum radiation. At densities above $n_\text{\ch{H}} \sim 10^{22}$~m$^{-3}$, the gas becomes optically thick to this continuum radiation by absorption of a photon (CIA) instead of emission. Currently \krome supports CIE of \ch{H2}-\ch{H2} and \ch{H2}-He pairs with original data taken from \citet{Borysow2001,Borysow2002} and \citet{Jørgensen2000}, respectively. \citet{Grassi2014} have extended this data to be valid to lower and higher temperature (100--10$^6$~K instead of 400--7000~K). Again, these cooling functions are only valid in the optically thin limit, yet \krome provides a optically thick option based on a fit of \citet{Ripamonti2004}. The validity of this high density fit must be carefully checked by the user to judge if it is suitable for their specific problem \citep{Hirano2013}. Fortunately, our density regime is low enough for the optically thin limit to be valid.\\\\
    \item{Metal line cooling}\\\\
    \krome, on the fly, solves the linear system of fine-structure metal transitions in a time dependent way. The method is described in \citet{Grassi2014} which is based on \citet{Glover2007} and \citet{Maio2007}. It includes transitions for the most important atoms and ions (Tabel~\ref{tab:Zcool}) with data taken from \citet{Hollenbach1989,Santoro2006,Glover2007,Maio2007}.\\\\
    \item{Cosmic ray heating}\\\\
    The heat contribution of incoming cosmic rays can be described by:
    \begin{equation}
        \Gamma_\text{CR} = \sum_{r\in R_{\text{CR}}} k_r n_r \Delta E_r,
    \end{equation}
    where the sum goes over all comic ray reactions $R_\text{CR}$, each with their reaction rate $k_r$ and collision partner number density $n_r$, releasing of amount of energy $\Delta E_r$. Following \citet{Goldsmith1978}, \krome attributes a mean $\Delta E=20$~eV to each cosmic ray reaction. Except for \ch{H} and \ch{He} ionization which have a $\Delta E = 4.3$~eV \citep{Dalgarno1999, Glassgold2012}, and \ch{H2} where $\Delta E_{\ch{H2}} = f(T,n_{\ch{H2}} )$ (\citealt{Galli2015}, fig.~2; \citealt{Glassgold2012}).
\end{enumerate}

\begin{table}
    \center
    \caption{Included microphysical heating and cooling processes (adapted from \citet{Grassi2014}).}\label{tab:thermalProc}
    \begin{tabular}{lr}
        Process & Reference\\\hline
        \ch{H} and \ch{He} line cooling &  \\
         \quad \quad H, He, \ch{He+} collisional ionization by \ch{e-}            & 1, 2\\
        \quad \quad \ch{H+} and \ch{He+} recombination                      & 1, 2, 3 \\
        \quad \quad He dielectric recombination                                 & 1, 2 \\
        \quad \quad H (all levels) collisional excitation by \ch{e-}               & 1, 2 \\
        \quad \quad He (n=2,3,4 triplets) collisional excitation by \ch{e-}        & 1, 2 \\
        \quad \quad \ch{He+} (n=2) collisional excitation by \ch{e-}                   & 1, 2\vspace{3pt}\\
        %bremsstrahlung (all ions) & \citet{Cen1992} \\  not used probably not that important (only very high T)
        \ch{H2} rovibrational lines cooling             &  \\ 
        \quad \quad Low density: collision by \ch{H}, \ch{H+}, \ch{H2}, \ch{He}, \ch{e-} & 4\\
        \quad \quad High density: LTE & 5\vspace{3pt}\\
        \ch{H2} chemical cooling & \\
        \quad \quad See Table~\ref{tab:H2diss} &  6\vspace{3pt}\\
        \ch{H2} chemical heating & \\
        \quad \quad  See Table~\ref{tab:H2diss} & 6, 7\vspace{3pt}\\
        CO rotational lines & \\
        \quad \quad Collisions by H and \ch{H2} & 8, 9, 21\vspace{3pt}\\
        Collisionally induced emission cooling           & \\
        \quad \quad \ch{H2}-\ch{H2} and \ch{H2}-\ch{He} pairs & 10, 11, 12, 13\vspace{3pt}\\
        Metal fine-structure line cooling & \\
        \quad \quad See Table~\ref{tab:Zcool}             & 14, 15, 16, 17\vspace{3pt}\\
        Cosmic ray heating & \\
        \quad \quad Cosmic ray reactions in Appendix~\ref{app:network}  & 18, 19, 20 \\\\\hline
        \multicolumn{2}{p{\columnwidth}}{1 \citet{Cen1992} - 2 \citet{Black1981} - 3  \citet{Spitzer1978} - 4 \citet{Glover2008} - 5 \citet{Glover2015}  - 6 \citet{Omukai2000} - 7 \citet{Hollenbach1979} - 8 \citet{Omukai2010} -
        9 \citet{Neufeld1993} - 10 \citet{Grassi2014} - 11 \citet{Borysow2002} - 12 \citet{Borysow2001} - 
        13 \citet{Jørgensen2000} - 14 \citet{Glover2007} - 15 \citet{Maio2007} -
        16 \citet{Santoro2006} - 17 \citet{Hollenbach1989} - 18 \citet{Dalgarno1999} - 19 \citet{Glassgold2012} - 20 \citet{Galli2015} - 21 Priv. comm. Grassi - Omukai}
    \end{tabular}
\end{table}

\section{Model results}\label{sec:results}
This section will present results of two simulations, a purely hydrodynamical one, and a hydrochemical one using our reduced chemical network. The hydrodynamical model uses a fixed chemical composition, equal to the initial one described in Section~\ref{sec:network}. This composition is used to calculate the mean molecular weight and the adiabatic index, which are needed in the hydrodynamical calculations. The hydrochemical simulation consists of switching on chemical and thermal evolution after an empirically determined `burn-in' phase of the hydrodynamical simulation. The hydrochemical simulation takes about 12h on 24 cores for a simulation of 10 pulsation periods, which is roughly a factor 10 longer than the purely hydrodynamical simulation.

\subsection{Hydrodynamical simulation}
\label{sec:hydrores} 
\begin{figure}
	\includegraphics[width=\columnwidth]{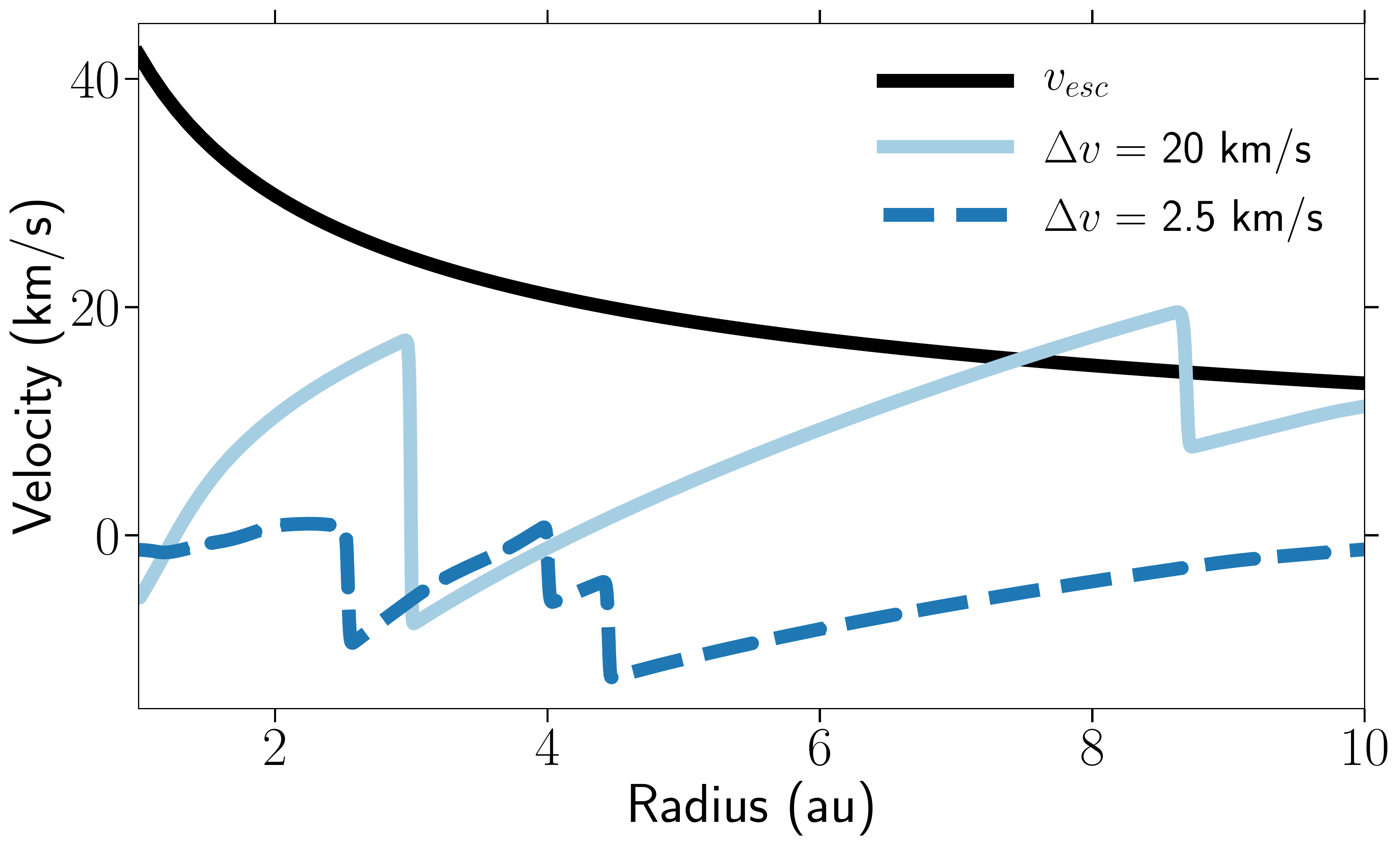}
    \caption{In a purely hydrodynamical framework an inner velocity amplitude $\Delta v = 2.5$ \kms does not lead to a sustainable stellar wind as the gas velocities do not exceed the local escape velocity. A velocity amplitude $\Delta v = 20$ \kms is needed to surpass the local escape velocity and eventually lead to a sustainable wind. This figure depicts a snapshot after 7.5 pulsation periods.}
    \label{fig:escapeVelocity}
\end{figure}

\begin{figure}
	\includegraphics[width=\columnwidth]{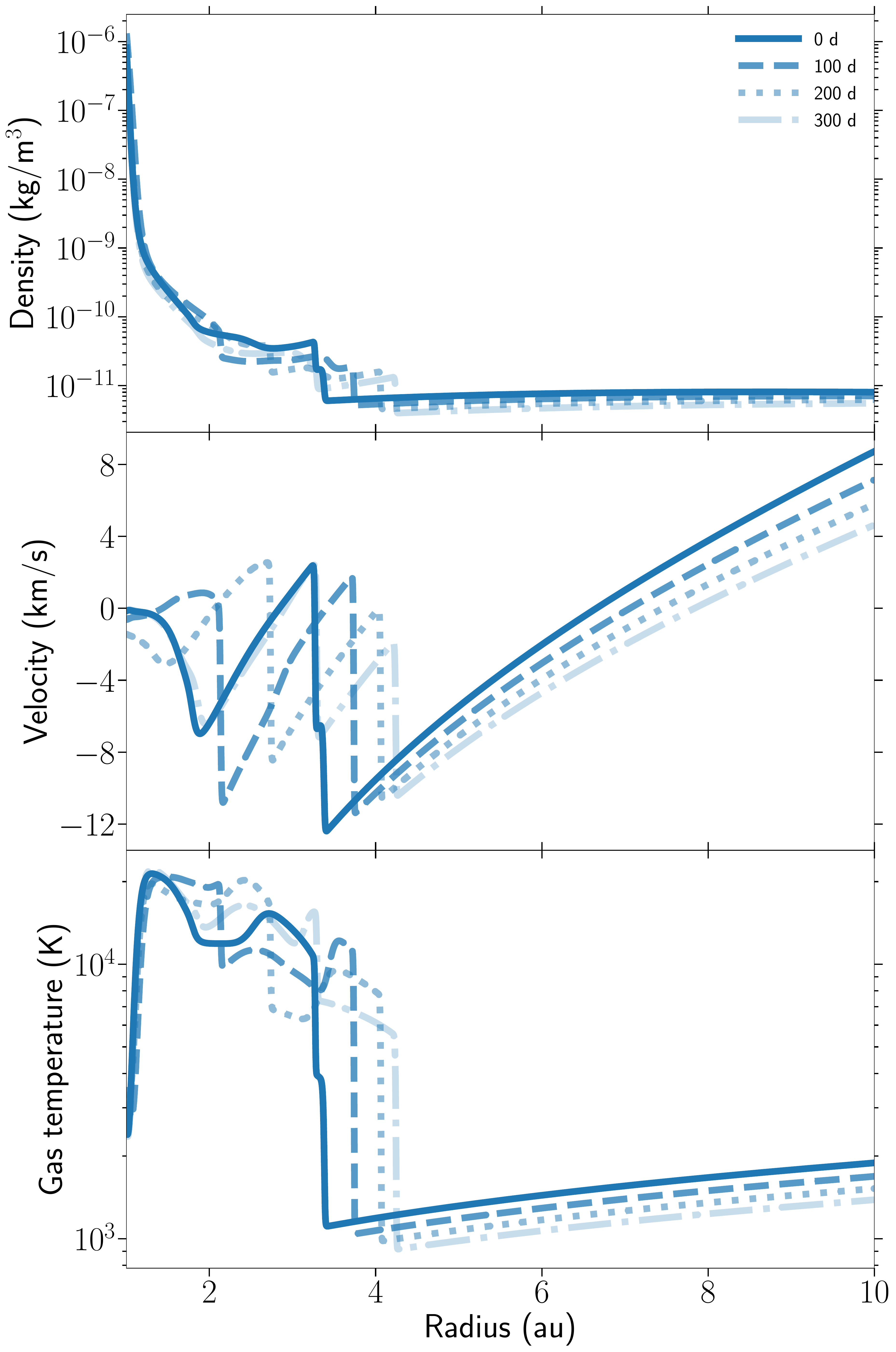}
    \caption{In a purely hydrodynamical framework with $\Delta v = 2.5$ \kms, no sustainable stellar wind is created. The gravitational pull of the star is too strong to overcome, so material starts to fall back around 4~au. Note that the small amount of outgoing gas beyond this point is a leftover from the `burn-in' phase and will gradually become fall back to the star. All three panels depict the temporal evolution over one period after a `burn-in' phase of four pulsation periods.}
    \label{fig:structure2k5}
\end{figure}

\begin{figure}
	\includegraphics[width=\columnwidth]{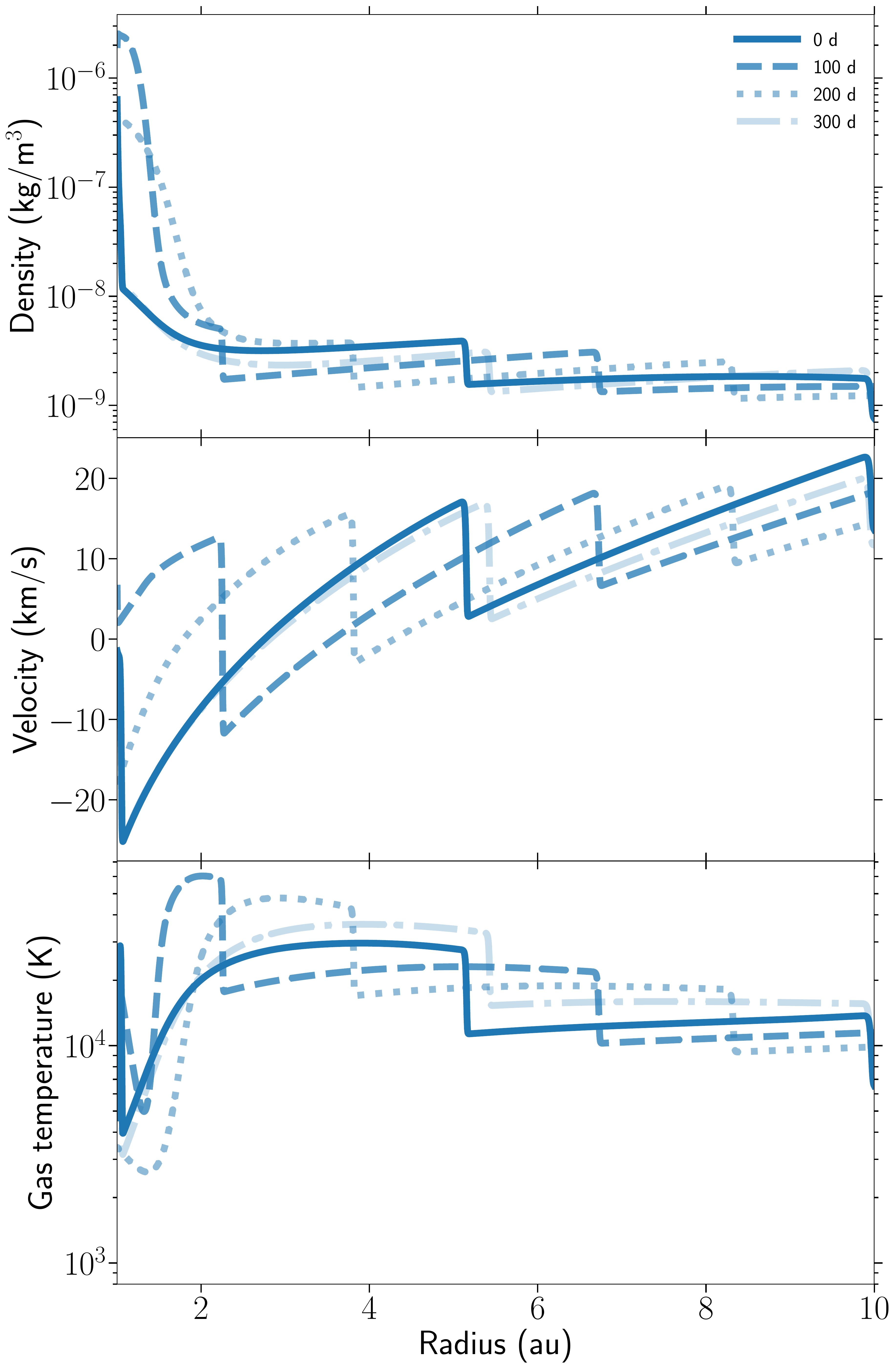}
    \caption{In a purely hydrodynamical framework with $\Delta v = 20$ \kms, a quasi-steady state of a sustainable stellar wind in reached. All three panels depict the temporal evolution over one period after a `burn-in' phase of four pulsation periods.}
    \label{fig:structure20k}
\end{figure}

The result of the purely hydrodynamic model reveals that there is no sustainable stellar wind (Fig.~\ref{fig:escapeVelocity}). This is because the gravitational pull of the star is stronger than the outwards acceleration triggered by the pulsations. This imbalance eventually leads to a fallback of all wind material. However, we can create a continuous stellar wind when increasing the pulsation velocity amplitude $\Delta v$ by roughly an order of magnitude (Fig.~\ref{fig:escapeVelocity}). This is a logical consequence of injecting more energy at the bottom of the wind, enough to overcome the gravitational pull. Even though this higher velocity amplitude can create a wind, it is most likely not a realistic value, since it is roughly an order of magnitude larger than predicted by 3D RHD simulations \citep[][fig.~6]{Freytag2017}, and half an order of magnitude larger than derived from line observations of Mira stars \citep[][fig.~14, note the reversed y-axis]{Nowotny2010}\\\\
When starting the simulation, the inner velocity perturbation leads to an unphysically fast shock propagating through the wind material. This happens because of the steep initial density profile and therefore a steep initial pressure profile. It is this large pressure gradient combined with the sudden kick from the input velocity that leads to a tremendous acceleration of the gas. Because the velocities quickly reach values larger than the local sound speed, the pulsations turn into shock waves. The recovery of this unphysically fast initial shock lasts a few pulsation periods, which we call a `burn-in' phase. After this `burn-in' phase, the simulation arrives at a more quiet stage where the gas reaches maximal velocities of several \kms. For $\Delta v = 2.5 $ \kms, these `post-burn-in' velocities in the wind stay below the local escape velocity, hence the material eventually falls back to the star and no sustainable wind is achieved (Fig.~\ref{fig:escapeVelocity}). The wind velocities in the  $\Delta v = 20 $ \kms case do exceed the local escape velocity, therefore the material escapes from the star, and a sustainable wind is created (Fig.~\ref{fig:escapeVelocity}).\\\\
After a reasonable `burn-in' phase of four pulsation periods, the physical wind structure in both cases is similar, other than that the $\Delta v = 2.5 $ \kms structure ceases at $\sim4$ au (Figs.~\ref{fig:structure2k5}--\ref{fig:structure20k}). The outward moving shocks sweep up the pre-shock gas, giving rise to higher densities once they have passed through (the absolute density differs in both cases but qualitatively they are identical). The inner shock heats the gas up to $\sim$\num{60000}~K. While moving outwards, it cools down by adiabatic expansion, reaching averaged temperatures of roughly \num{10000}~K. The lowest temperatures, $\sim$2000~K, occur close to the star where the infall velocities are high, thus bringing about efficient adiabatic cooling. Note that in the $\Delta v = 20 $ \kms model, the shocked material in the wind reaches a quasi-steady state but is not perfectly periodic, even though we have a periodic pulsation mechanism.

\begin{figure}
	\includegraphics[width=\columnwidth]{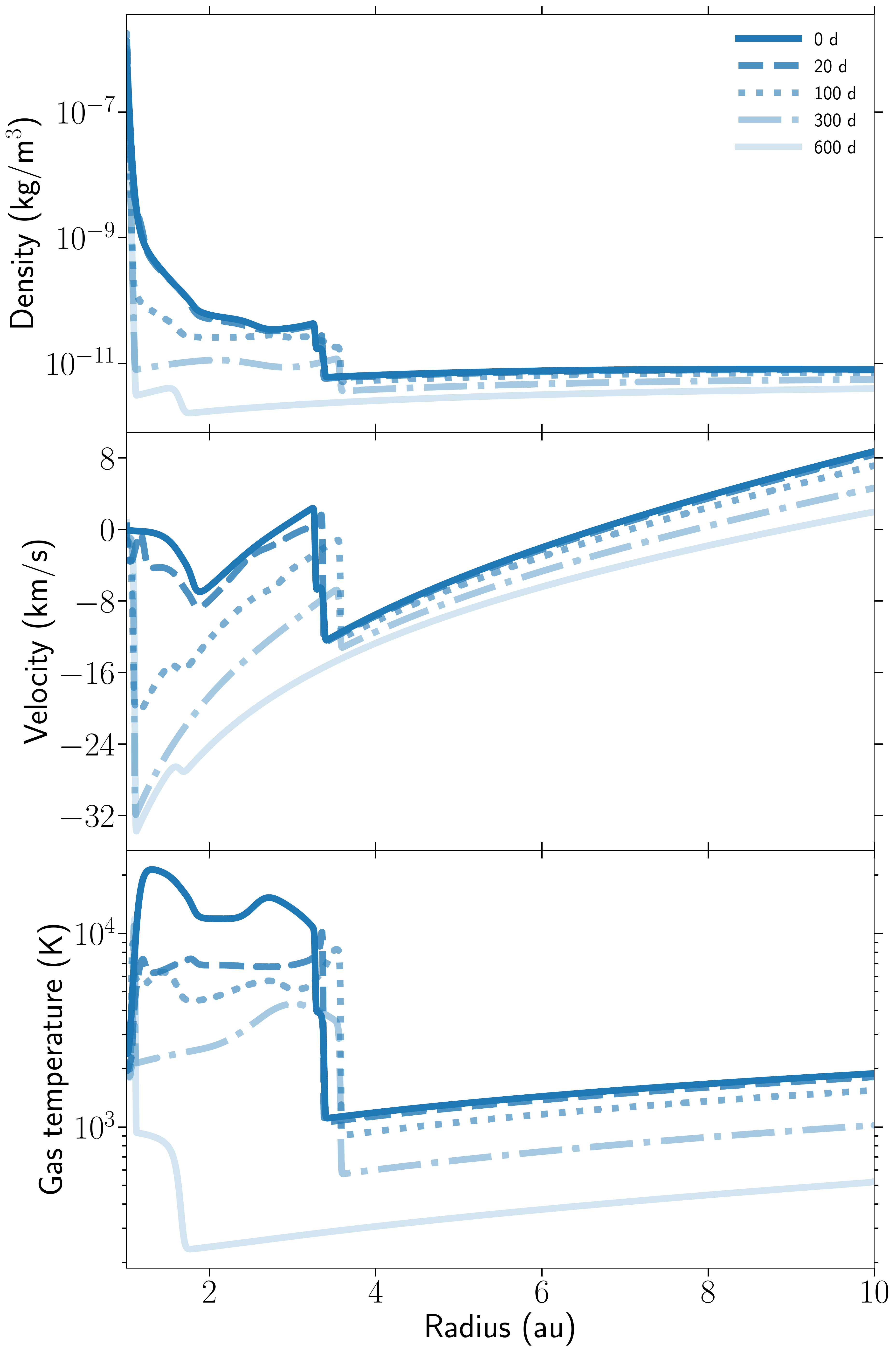}
    \caption{In a hydrochemical framework with $\Delta v = 2.5$ \kms, the high temperature, pulsating inner region disappears when switching on the chemical and thermal evolution. This happens because of the immense loss of internal energy due to efficient cooling. After roughly two pulsation periods, there is only the incoming shock which dissipates too quickly to persist throughout the wind. Eventually all gas in turns cold and falls back onto the star. However, close to the star, the conditions are ideal for dust formation to occur (i.e. cold and dense). All three panels depict the temporal evolution when switching on the chemical and thermal evolution after a `burn-in' phase of four pulsation periods.}
    \label{fig:structureCooling2k5}
\end{figure}

\begin{figure}
	\includegraphics[width=\columnwidth]{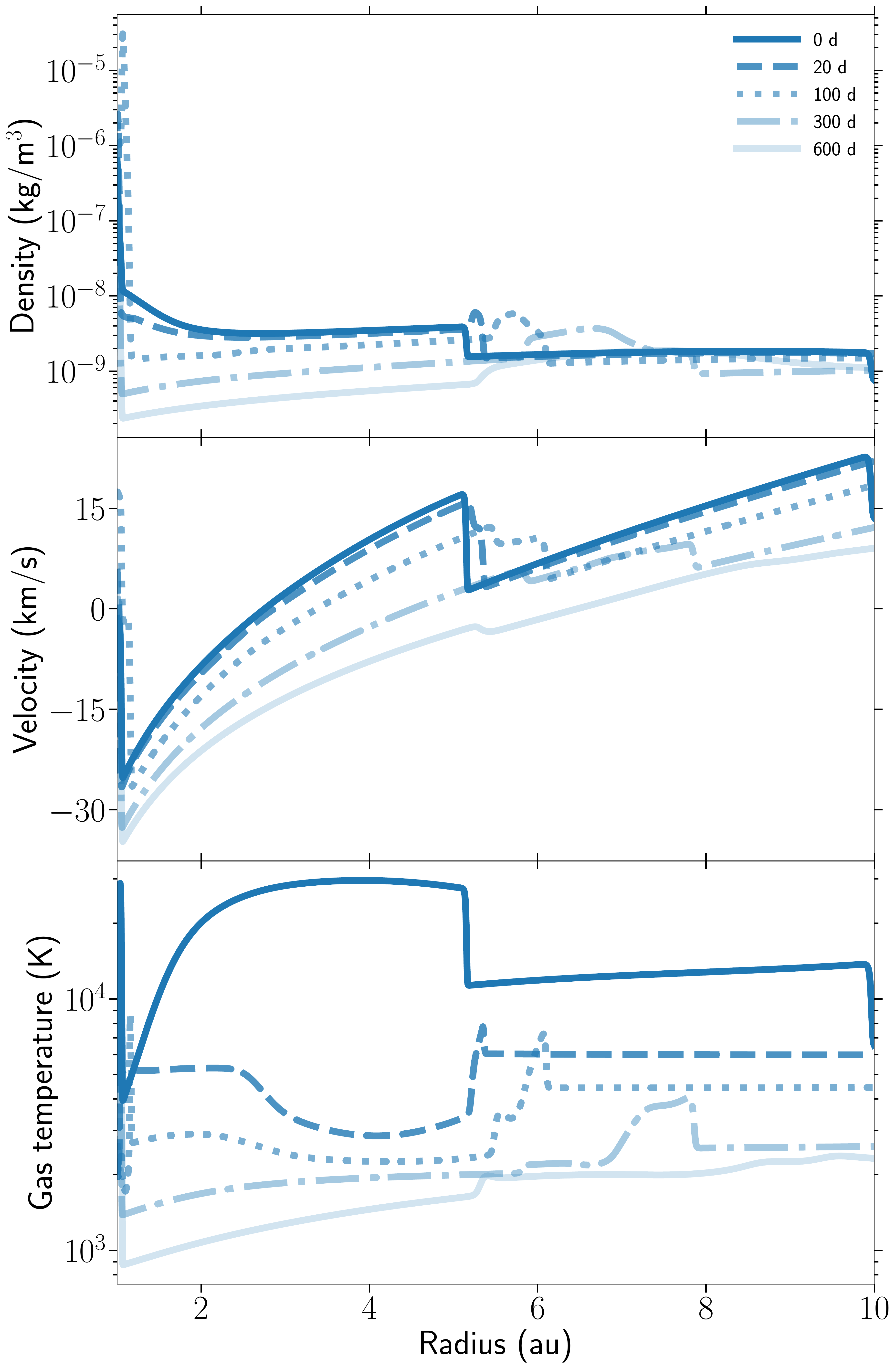}
    \caption{In a hydrochemical framework with $\Delta v = 20$ \kms, the sustainable wind gets destroyed when switching on the chemical and thermal evolution. This happens because of the immense loss of internal energy due to efficient cooling. After roughly two pulsation periods, there is only the incoming shock which dissipates too quickly to persist throughout the wind. Eventually all gas turns cold and falls back onto the star. However, close to the star, the conditions are ideal for dust formation to occur (i.e. cold and dense). All three panels depict the temporal evolution when switching on the chemical and thermal evolution after a `burn-in' phase of four pulsation periods.}
    \label{fig:structureCooling20k}
\end{figure}

\subsection{Hydrochemical simulation}
\label{sec:hydrochemres}
The hydrochemical model starts by switching on the chemical and thermal evolution after a hydrodynamical `burn-in' phase of four pulsation periods. Overall, the chemical cooling is so efficient that the hot gas in the shocks quickly cools down to a couple of 1000~K in mere days, which is extremely fast compared to the dynamical time scale of the pulsations. Due to this immense loss of energy, the internal pressure of the gas drops quickly. Unable to persist counteracting the stellar gravity, the inner wind quickly collapses and material falls back onto the star. Over time, the entire wind structure breaks down resulting in cold gas falling onto the star. After roughly two hydrochemical pulsation periods, there is only the incoming shock which dissipates too quickly to carry on throughout the wind. This thermodynamical development occurs for both the $\Delta v = 2.5$ \kms and $\Delta v = 20$ \kms cases (Figs.~\ref{fig:structureCooling2k5}--\ref{fig:structureCooling20k}, respectively).\\\\
In terms of dynamical and thermal evolution, both the $\Delta v = 2.5$ \kms and $\Delta v = 20$ \kms cases are qualitatively identical. The latter \ado{is a less realistic inner boundary velocity}, however, \ado{it} more closely resembles an AGB wind by compensating for the absent outwards dust acceleration with a larger pulsation velocity amplitude (Section~\ref{sec:hydrores}). In this case, the spatial extend of wind encompasses the entire numerical grid, compared to roughly the inner 4~au in the $\Delta v = 2.5$ \kms case. This makes the $\Delta v = 20$ \kms case more convenient to infer how the thermal and chemical evolution affect the physical structure of the wind. Therefore we limit an in-depth analysis of the heating and cooling processes to the $\Delta v = 20$ \kms case, bearing in mind that the same processes occur in the $\Delta v = 2.5$ \kms case.\\\\
When switching on the hydrochemistry, the initial chemical composition still needs to adjust itself, both chemically and thermally. For one, this leads to a global tremendous drop in temperature, which happens faster than the dynamical evolution of the system\footnote{This adjustment actually happens faster than the hydrodynamical time step ($\sim$2~h). As an initial condition, such a time-dependent adjustment is preferable over the assumption of chemical equilibrium. If this assumption were valid, then the time-dependent evolution would reach this equilibrium as well. If it were not, then the time-dependent evolution would be more correct.}, meaning that the initial chemical composition was unstable in those local conditions. We limit the analysis to three snapshots in time to describe the complex evolution in space, time, temperature and chemical composition. We opt for snapshots at 20, 100, and 300 days after switching on the chemistry (Fig.~\ref{fig:structureCool20}-\ref{fig:structureCool100}-\ref{fig:structureCool300}, respectively), \ado{because} after 20 days, rapid initial chemical and thermal adjustments have ended; after 300 days, the evolution of the heating and cooling processes \ado{and the chemical abundances become similar to the inner half of the 300 days snapshot until the model breaks down after roughly two pulsation periods; and 100 days for an in-between evolutionary snapshot.}

\subsubsection{Hydrochemical wind after 20 days (Fig.
\ref{fig:structureCool20})}
Metal cooling is one of the main coolants in the entire wind. This cooling rate is roughly constant, which correlates with the nearly constant abundance of the involved metals. However, \ch{Si} completely disappears between 3 and 5 au due to the drop in temperature, forming \ch{Si}-bearing molecules. For the same reason, the \ch{C} abundance drops several orders of magnitude. The fact that the metal cooling rate does not drop in this region means that both elements do not significantly contribute to the cooling. It is \ch{Fe} line cooling which is most dominant in this temperature regime (T > 3000~K), followed by \ch{O} cooling \citep[][fig.~3]{Grassi2014}. The slight decrease in cooling rate between 3 and 5 au is most likely due to the drop in temperature rather than the loss of metals. The ionized metal coolants do not contribute as their abundance is negligible. The second, equally effective, cooling process is chemical \ch{H2} cooling\footnote{Both destruction and formation of \ch{H2} are considered as the same thermal processes. When the rate of destruction is greater than the rate of formation, this leads to a net cooling rate. Vice versa for a greater formation rate. Only the reactions listed in Table~\ref{tab:H2diss} participate.}. As the temperature drops between 3 and 5 au, the \ch{H2} dissociation efficiency decreases but the number of \ch{H2} molecules increases, making dissociation reactions more abundant. The balance between both results in a slight net decrease of cooling. \ch{H} and \ch{He} line cooling is only significantly present at high temperatures and traces the temperature profile nicely. It keeps the inner shock from reaching too high temperatures.
% Continuum cooling is consistently two orders of magnitude weaker than chemical and metal cooling. Therefore, it is not effective enough at any point in the wind. It does follow the temperature profile neatly as is suggested by Eq.~\eqref{contCoolProp}. 
CIE cooling is not effective because the abundance of \ch{H2} is too low, as shown by \citet[fig.~2]{Ripamonti2004}. Cosmic ray heating is negligible as a heat source. Another unimportant processes is \ch{H2} line cooling, due to the low amount of \ch{H2} molecules because of too high temperatures. It does, however, neatly follow the \ch{H2} abundance profile. The low temperature region between 3 and 5 au reveals that \ch{CO} line cooling is negligible in our system even though almost all carbon is lock up in \ch{CO}, maximising the \ch{CO} content.

\subsubsection{Hydrochemical wind after 100 days (Fig. \ref{fig:structureCool100})}
Metal cooling is still the most pronounced coolant throughout the wind. The \ch{Fe} and \ch{O} abundances roughly stay constant, whereas, below 6 au, all \ch{C} and \ch{Si} transform into molecular species. As the temperature drops gradually, more \ch{H2} gets produced and eventually chemical heating takes over from chemical cooling. Even though, there is a smooth temperature transition, the switch from cooling to heating is brisk. This turning point reveals itself between 2 to 3 au. Below the critical temperature at this turning point, heating is more dominant (see 3--6 au). This dominance also manifests itself in the \ch{H2} abundance, which grows to mass fractions of about one per cent. \ch{H} and \ch{He} line cooling is still only relevant in the highest temperature regions where the shocks still haven't cooled sufficiently. 
% Continuum cooling is still not efficient enough to compete with metal and chemical cooling. The overall rate is also lower as the density has slightly dropped as compared to the previous snapshot.
The \ch{H2} density is too low for CIE cooling to become relevant. Heating by cosmic rays is overall unimportant. \ch{H2} line cooling increases overall as the production of \ch{H2} goes up due to the decreasing temperature. Even so, it is still not effective enough to matter on a global scale. Between 2 and 5 au, all \ch{C} is locked up in \ch{CO} hereby maximising \ch{CO} cooling capability. Yet it only marginally affects the temperature structure.

\subsubsection{Hydrochemical wind after 300 days (Fig. \ref{fig:structureCool300})}
As before, the metal cooling stays important, yet becomes slightly weaker. The gas has cooled down sufficiently for \ch{H2} formation reactions to be dominant, thereby heating the gas rather than cooling it. Yet, the abrupt cooling spike around 6 au demonstrates the delicate balance between temperature and availability of species, driving the thermal evolution. The density of \ch{H2} is high enough for CIE cooling to be non-zero, but is negligible compared to other cooling processes. Cosmic ray heating is not relevant in the wind. Again, \ch{H2} line cooling follows the \ch{H2} abundance profile but is still too weak to matter. Since the shocks have cooled sufficiently, \ch{H} and \ch{He} line cooling have become irrelevant. \ch{CO} cooling is still not important.\\\\
Note that the wind mainly consists of atomic \ch{H} rather than molecular \ch{H2}. This might look surprising compared to other AGB wind models, which conclude that the wind is mainly molecular. However, such wind models are based on the assumption of chemical equilibrium, which predicts that below $\sim2000$~K all \ch{H} should be molecular \citep[e.g.][fig. 1]{Schirrmacher2003a}. The validity that this conversion happens fast has been questioned but never pursued because a time dependent, kinetic treatment of \ch{H2} formation is needed to provided an answer \citep[e.g.][footnote 11]{Schirrmacher2003a}. This is exactly what is included in our model and it shows that this conversion processes is less efficient than previously assumed. The time to form \ch{H2} is larger than the dynamical time scale of the AGB wind. Note that extending the simulation time will lead to more \ch{H2} because as the temperature continues to decrease, the formation efficiency will increase. Whether at some point the abundance of \ch{H2} will dominate over the \ch{H} abundance is currently unclear because our model breaks down after roughly two pulsation periods due to the absence of dust acceleration (see above).

\begin{figure}
	\includegraphics[width=\columnwidth]{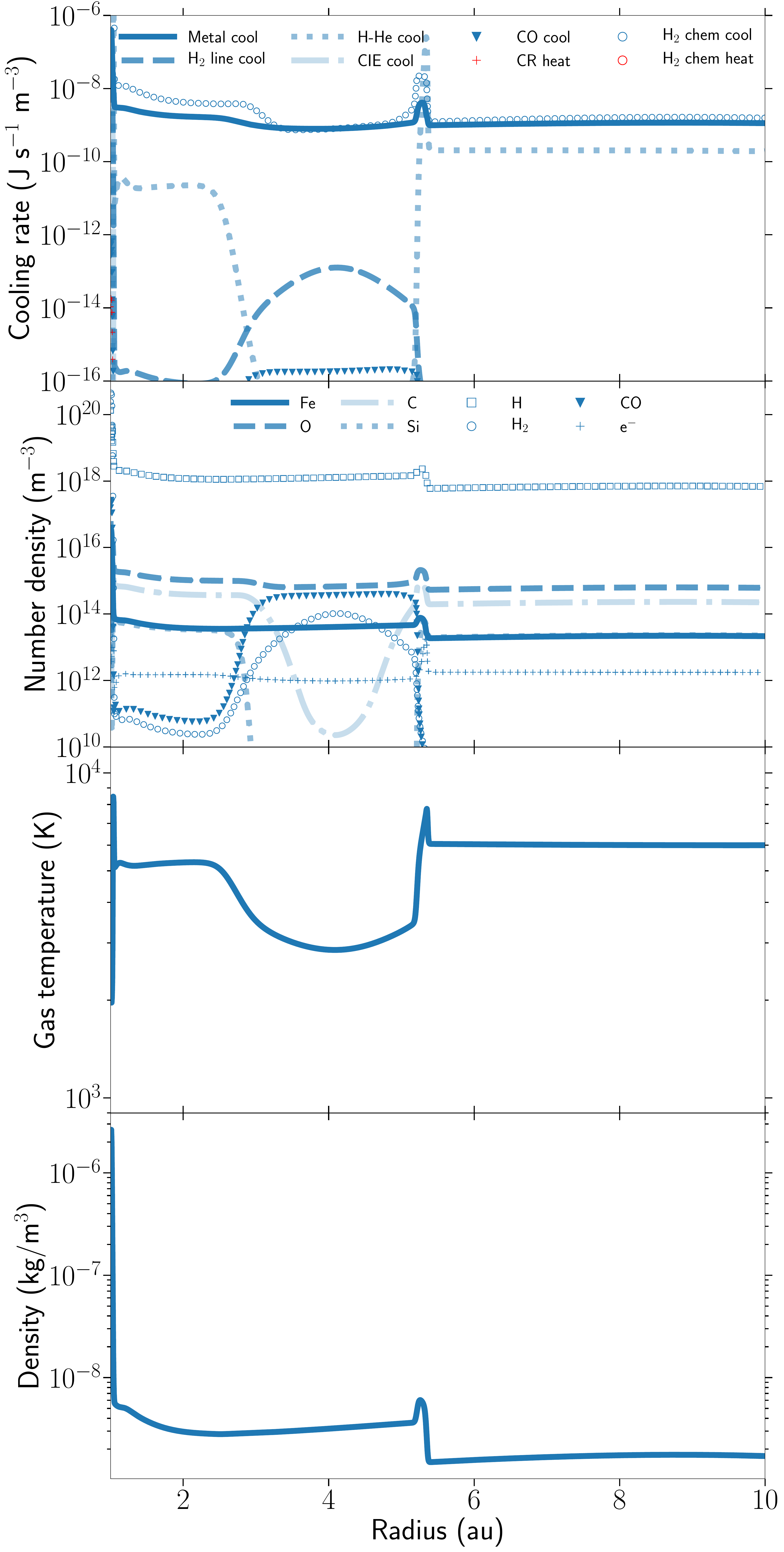}
    \caption{A snapshot of the wind structure 20 days after a `burn-in' phase of four pulsation periods for a hydrochemical model with $\Delta v = 20$~\kms. \textbf{First:} All heating and cooling processes with cut-offs at $10^{-6}$ and $10^{-16}$~J~\si{\per\s\per\m\cubed}. \textbf{Second:} Number densities of the most important species involved in the heating and cooling processes with a lower cut-off at $10^{10}$~\si{\per\m\cubed}. \ado{More species are shown in Appendix~\ref{app:extra_mol}}. \textbf{Third:} Temperature structure of the gas. \textbf{Fourth:} Density structure of the gas.}
    \label{fig:structureCool20}
\end{figure}

\begin{figure}
	\includegraphics[width=\columnwidth]{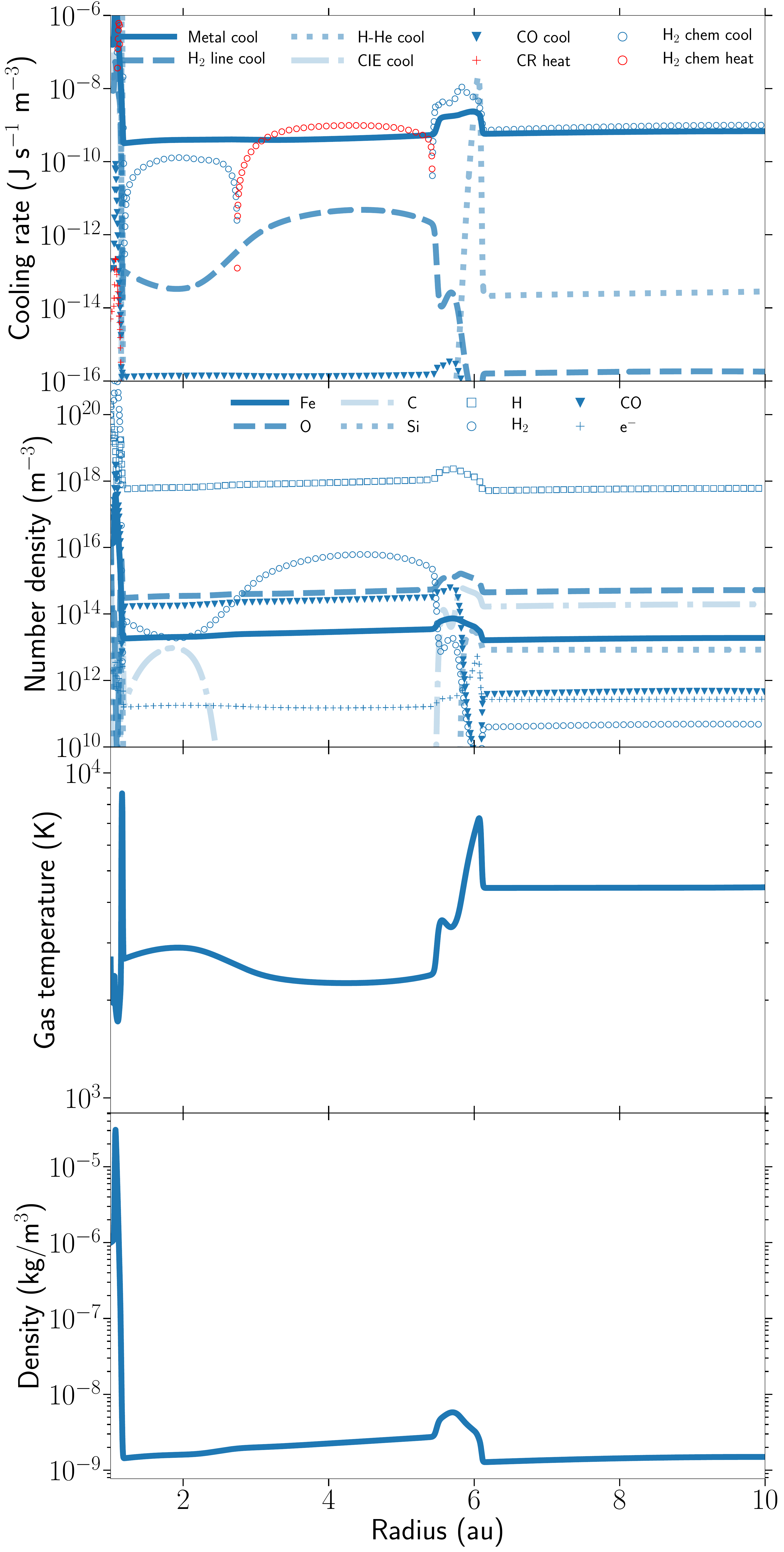}
    \caption{A snapshot of the wind structure 100 days after a `burn-in' phase of four pulsation periods for a hydrochemical model with $\Delta v = 20$~\kms. \textbf{First:} All heating and cooling processes with cut-offs at $10^{-6}$ and $10^{-16}$~J~\si{\per\s\per\m\cubed}. \textbf{Second:} Number densities of the most important species involved in the heating and cooling processes with a lower cut-off at $10^{10}$~\si{\per\m\cubed}. \ado{More species are shown in Appendix~\ref{app:extra_mol}}. \textbf{Third:} Temperature structure of the gas. \textbf{Fourth:} Density structure of the gas.}
    \label{fig:structureCool100}
\end{figure}

\begin{figure}
	\includegraphics[width=\columnwidth]{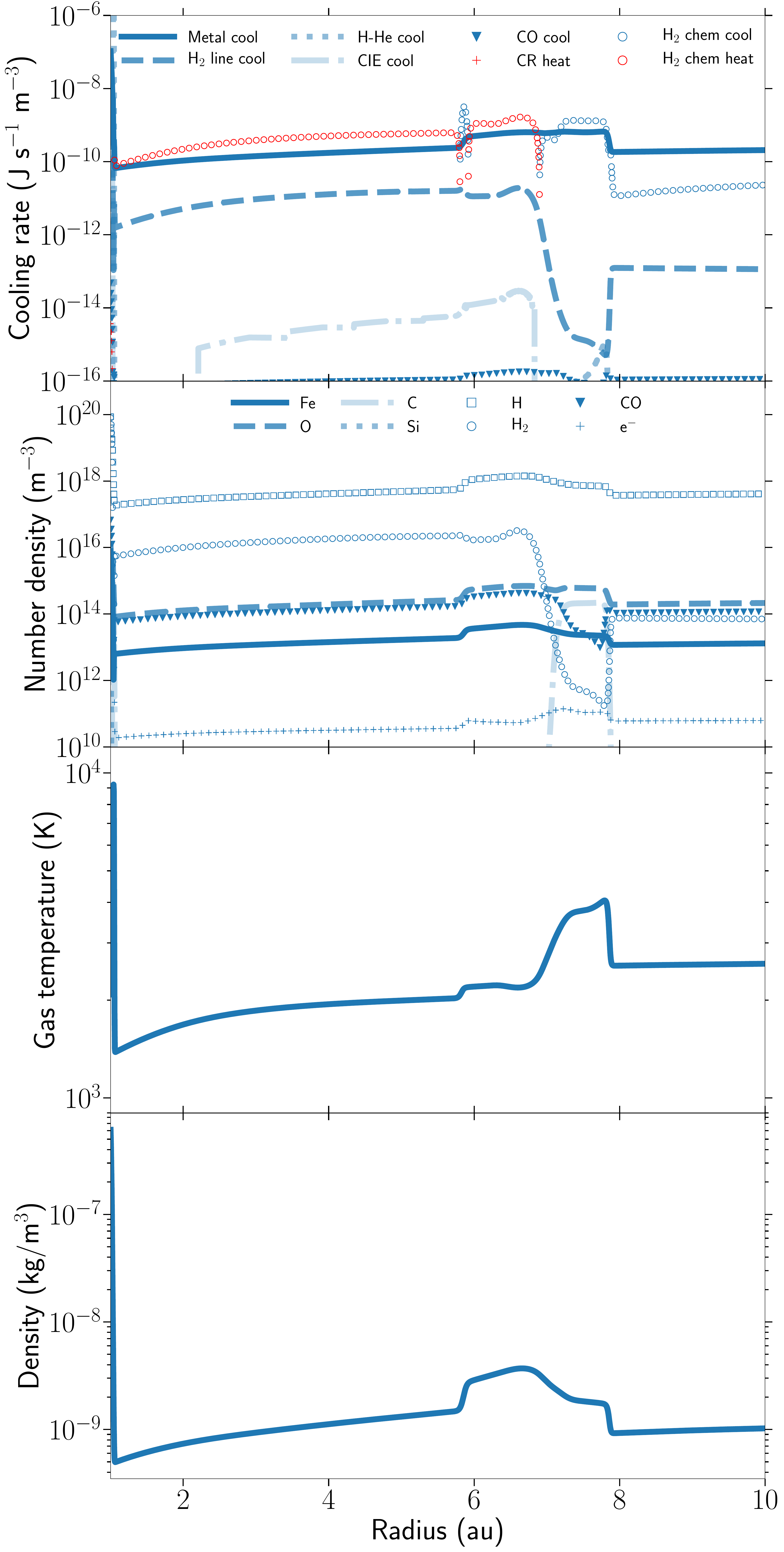}
    \caption{A snapshot of the wind structure 300 days after a `burn-in' phase of four pulsation periods for a hydrochemical model with $\Delta v = 20$~\kms. \textbf{First:} All heating and cooling processes with cut-offs at $10^{-6}$ and $10^{-16}$~J~\si{\per\s\per\m\cubed}. \textbf{Second:} Number densities of the most important species involved in the heating and cooling processes with a lower cut-off at $10^{10}$~\si{\per\m\cubed}. \ado{More species are shown in Appendix~\ref{app:extra_mol}}. \textbf{Third:} Temperature structure of the gas. \textbf{Fourth:} Density structure of the gas.}
    \label{fig:structureCool300}
\end{figure}

\section{Limitations}\label{sec:discussion}
This section will discuss some limitations of our methods and provide suggestions for improvement. The first part will elaborate on the chemical network construction. The second part will address the microphyscial heating and cooling processes. The last part will briefly address the impact of higher spatial dimensions.

\subsection{Chemical network}
We performed an in depth analysis for constructing a reduced chemical network that yields the same results as an extensive network, and that is computationally feasible when combined with a hydrodynamical framework code. However, this network is not complete in terms of number of species, amount of reactions, and the prescription of the reaction rate coefficients, yet sufficient for this proof-of-concept paper. Addition of more complex species and more reactions to the comprehensive network will give a more complete view on the chemical composition of an AGB wind. Furthermore, chemical reaction coefficients are adopted from a database and take the simplified form of a modified Arrhenius' equation. In reality, the prescription of a reaction is much more complicated and susceptible to small temperature changes, and often are a temperature polynomial. Additionally, reaction coefficients are determined either experimentally or theoretically. Experiments are usually only performed at room temperature (300~K) whereas theoretical derivations are regularly based on statistical equilibrium with a reversed reaction. This latter can then be described by a Maxwell-Boltzmann distribution, and the rate then depends on temperature via the partition function of the species. Partition functions can be a rather complex finite sum of exponentials, therefore, they are frequently approximated by a single temperature power law that is only valid in a limited temperature range. \citet{Forrey2013} points out the effects of such assumptions on the rate of the three-body reaction \ch{H + H + H -> H2 + H}. In this particular case, the rate can differ up to two orders of magnitude. As this is an essential reaction for \ch{H2} formation in high density regimes and \ch{H2} is an omnipresent collision partner, it can have far reaching effects on the temperature regulation and on the overall chemistry. Furthermore, due to limited information of databases, we extrapolate reaction coefficients in temperature space, even though they are explicitly valid within a specific range. Lastly, we do not take any dust into account. Yet, dust can have profound repercussions on both chemical and thermal evolution. On one hand, it acts as a catalyst for \ch{H2} formation making this a more efficient path than three-body reactions. On the other hand, the dust can act as a heating and cooling source for the gas by transfer of energy during gas-grain collisions or \ch{H2} formation on its surface.\\\\
We aspire to extend our chemical network with updated reaction rates from literature (e.g. \krome reaction networks, \citealt{Glover2007,Glover2010,Cairnie2017}) and add valuable reactions which are still missing e.g. from combustion, atmospheric, and exoplanetary chemistry. Finally, we will include species relevant for dust formation and gas-grain reactions.

\subsection{Heating and cooling processes}
As a first step, we have limited ourselves to thermal processes provided by \krome. However, \krome is developed as a package to be embedded in any astrophysical simulation and is not specifically tuned for AGB stars. Therefore, some thermal processes can be missing\ado{, e.g. \ch{H2O} line cooling}. The thermal processes predominantly supported by \krome are often taken from studies that have collected cooling rates from older research. E.g.,  \citet{Cen1992} take collisional excitation, collisional ionization, and recombination coefficients from \citet{Black1981}, who in turn collects these rate coefficients from even older papers. Additionally, these coefficients are often determined for specific astrophysical environments with approximations which may not be valid in a different region of interest. For example, our collisional ionization rate of \ch{He} depends on the abundance of \ch{He+} because this rate is determined by a steady state of certain \ch{He} formation and destruction reactions in low density, primordial intergalactic clouds \citep{Black1981}. But, in our model, we know the abundance of \ch{He} due to the chemical evolution of our user provided network. \ado{Note that we assume all are cooling radiation to escape from the model. Adding an escape probability applicable to an AGB wind would increase its accuracy, e.g. using Sobolev's approximation \citep{Sobolev1960}. The escape probability incorporated in the \ch{CO} line cooling tables might also not be valid in an AGB wind \citep{Omukai2000, Omukai2010}.}\adr{\\\\
Note that adopting heating/cooling rates from literature papers can lead to less self-consistency than might be expected. Heating/cooling by collisional processes are determined by collisional reaction rates of the species involved. These rates might be different from the ones used in the chemical network, hence not fully-consistent. For example, our \ch{H2} line cooling function is determined by collision with \ch{H}, \ch{H+}, \ch{H2}, \ch{He}, and \ch{e-}. However, \citet{Glover2008} calculated this cooling rate with a set of chemical reaction rates that is different from ours.}\\\\\ado{
Thus far, any stellar radiation has been ignored. Yet, the gas can absorb this light, hereby heating up. In RHD simulations, this process corresponds to an extra term in the energy equation depending on the (frequency integrated) mean intensity and absorption coefficient of the gas \citep[e.g. appendix A][]{Hofner2016}. The former can be computed via the radiation transfer equation or with an approximated geometrical dilution factor added to the stellar radiation. The latter depends on the abundance of chemical species, and can be calculated given their energy level populations and (de)excitation coefficients. In the past, such calculations have assumed chemical equilibrium, however now, the absorption coefficients can be weighted with the appropriate chemical abundances provided by the hydrochemical evolution. Although, without actually implementing this additional heating term, it is difficult to gauge its importance.} \\\\
We aspire to extend our set of thermal processes with more atomic and molecular line heating/cooling present in AGB winds \citep{Woitke1996a,Schirrmacher2003a}, improve on the internal consistency of already incorporated rates\ado{, and add a heating term due to stellar radiation absorption by gas.}

\subsection{Spatial dimension}
All our simulations are limited to 1D, assuming spherical symmetry. However, departure from this assumption will trigger instabilities \citep{Woitke2006, Freytag2008, Freytag2017}. These instabilities will lead to variations in density and temperature, and consequently alter the chemical composition. This will result in a chemically inhomogeneous wind which is more complex than the ones we obtain from our 1D simulations.

\section{Summary and perspectives}\label{sec:summary}
In this paper, we have, to our knowledge, developed the first self-consistent hydrochemical model for simulating the onset of an AGB wind. We have extended the multi-dimensional MHD code, \mpiamrvac \citep{Keppens2012} such that it can handle accurate chemical evolution by incorporating \krome \citep{Grassi2014}, and by implementing a consistent multi-fluid advection to ensure conservation of chemical species (Section~\ref{sec:method}).\\\\
We opted for a slightly different hydrodynamical setup as compared to literature. Firstly, we drop the hydrostatic equilibrium solution for the initial density, as this is degenerate with the choice of temperature profile and mean molecular weight. Secondly, we use an open inner boundary applied with a simplified sinusoidal velocity variation, in line with 3D~RHD simulations of AGB stars \citep{Freytag2008,Freytag2017}. This in contrast to the more commonly used solid piston approximation \citep[e.g.][]{Bowen1988,Fleischer1992,Winters2000,Schirrmacher2003a,Freytag2008,Liljegren2016,Hofner2016}. Our hydrodynamical setup corresponds to a fixed spatial slice of an AGB winds, where the inner boundary varies according to the pulsational behaviour of the star (Section~\ref{sec:hydromod}).\\\\
We have constructed the first reduced chemical network applicable to an AGB wind. Firstly, we constructed a large, but simplified, chemical network that comprises triatomic molecule reactions from the UMIST database \citep{McElroy2013} extended with some collisional \ch{H}, \ch{He}, \ch{H2} reactions, and three-body \ch{H2} formation reactions. Secondly, we developed a flux-limited reduction algorithm included with a validation procedure to reduce the large network. This reduction method is more rigorous than constructing a network based on intuitions of which reactions are relevant. A reduced network is needed when including time-dependent chemical evolution in a hydrodynamical framework code because a large network is computationally too time consuming as chemical evolution calculations are slower than dynamical evolution calculations. Our reduced network consists of 255 reactions and 70 species (Section~\ref{sec:network}).\\\\
We have included a number of microphysical heating and cooling processes for regulating the temperature of the gas. We have restricted ourselves to a subset of the thermal processes provided by \krome, thought to be important in an AGB wind (Section~\ref{sec:thermalproc}).\\\\
Note that any form of dust or radiation is excluded from this work. This work serves as a proof-of-concept, and using a bottom-up approach for including more physics will enable us to more easily disentangle the physical effects that, together, generate AGB winds.\\\\
To ascertain the impact of chemical and thermal evolution, we have run a purely hydrodynamical, and a hydrochemical simulation. The hydrochemical one switches on chemical and thermal evolution after a hydrodynamical `burn-in' phase. We conclude that a pure hydrodynamical model cannot achieve a sustainable AGB wind by using a realistic inner boundary velocity variation with an amplitude of $\Delta v =2.5$ \si{\km\per\s} \citep[][fig.~6]{Freytag2017}. The gravitational pull of the star is too strong to overcome, and material falls back onto the star. However, a sustainable wind, and therefore mass loss, can be realised by increasing the velocity amplitude by roughly an order of magnitude. Such strong velocity variation is unlikely \citep{Nowotny2010}, therefore, we presume dust acceleration can enhance the outwards motion in the more realistic case. Unfortunately, gas temperatures in this simulation remain too high for dust to form, ranging from 2000 to several \num{10000}~K (Section~\ref{sec:hydrores}). In the hydrochemical model, the sustainable wind structure collapses due to the shear loss of energy by efficient cooling. The whole system evolves towards cold, dense gas that falls back to the star with incoming shocks of maximally \num{10000}~K that quickly dissipate. \ch{H} and \ch{He} line cooling prevents the gas from exceeding this threshold. Below \num{10000}~K, the gas mainly cools by fine-structure lines of \ch{Fe} and \ch{O}, and collisional dissociation of \ch{H2}. When the temperature gets low enough for \ch{H2} formation to be more prevailing than \ch{H2} destruction, the gas will heat instead of cool. Other processes like \ch{H2} rovibration line cooling, CIE cooling, CO rotation line cooling, and cosmic ray heating are not effective enough to affect the temperature structure (Section~\ref{sec:hydrochemres}).\\\\
In conclusion, the hydrochemical evolution of an AGB wind presented here cannot initiate nor sustain a wind. Nevertheless, the results are promising as a cold, dense gas region forms close to the star, which is ideal for dust formation to happen. We expect, once dust has been incorporated into our model, it will gradually accelerate outwards by momentum transfer of stellar photons,  dragging the gas along. This way, dynamical evolution gets reintroduced into the system and will allow for a more meaningful hydrochemical evolution of an AGB wind. It is believed, that this extra outward force can, under the right conditions, be enough to overcome the gravitational pull of the star \citep{Hofner2016}. The inclusion of dust will have repercussion on the dynamics and chemistry of the system. Firstly, the extra acceleration might recreate shocks to temperatures high enough to destroy the newly formed dust \citep[][fig. 4, for temperature-pressure stability limits of different kinds of dust]{Gail2013}. Accordingly, so will the outward motion disappear, leading to the same result as our dust-free model. However, as this gas cools down again, dust might reform and restart the cycle. It might be that averaged over time, material gets lost into the interstellar medium. Secondly, dust also heats the gas by catalysing \ch{H2} formation, and gas-grain collision of warm dust. Heating up the gas can inhibit the sustainable wind structure from breaking down, yet it can also hamper dust formation. Thirdly, dust can act as a cooling source by collisional energy transfer from gas to dust followed by efficient infrared emission, for which the gas is translucent.\\\\
This paper has laid the basis for more accurate modelling of time-dependent interaction between gas-grain chemistry, thermal processes, and dynamics in AGB winds. It is the first in a series where we strive for increased self-consistency regarding chemistry, dust creation, and dynamics. Currently, the results of this work are not intended to be used for direct comparison with observations, because the model does not yet represent a realistic AGB wind, primarily because of the absence of the dynamical force exerted through dust acceleration. We intend to include comparisons with observations once the model has reached a more complete stage.

\section*{Acknowledgements}
J.B would like to thank T. Grassi for the useful discussions on reducing the chemical network and answering questions about \krome. J.B., N.C., and L.D. acknowledge support from the ERC consolidator grant 646758 AEROSOL. All plots were produced using the community open-source Python packages Matplotlib \citep{Hunter2007} and NumPy \citep{Oliphant2006}.

%%%%%%%%%%%%%%%%%%%%%%%%%%%%%%%%%%%%%%%%%%%%%%%%%%



\bibliographystyle{mnras}
\bibliography{Myref} % if your bibtex file is called example.bib

\begin{thebibliography}{}
\makeatletter
\relax
\def\mn@urlcharsother{\let\do\@makeother \do\$\do\&\do\#\do\^\do\_\do\%\do\~}
\def\mn@doi{\begingroup\mn@urlcharsother \@ifnextchar [ {\mn@doi@}
  {\mn@doi@[]}}
\def\mn@doi@[#1]#2{\def\@tempa{#1}\ifx\@tempa\@empty \href
  {http://dx.doi.org/#2} {doi:#2}\else \href {http://dx.doi.org/#2} {#1}\fi
  \endgroup}
\def\mn@eprint#1#2{\mn@eprint@#1:#2::\@nil}
\def\mn@eprint@arXiv#1{\href {http://arxiv.org/abs/#1} {{\tt arXiv:#1}}}
\def\mn@eprint@dblp#1{\href {http://dblp.uni-trier.de/rec/bibtex/#1.xml}
  {dblp:#1}}
\def\mn@eprint@#1:#2:#3:#4\@nil{\def\@tempa {#1}\def\@tempb {#2}\def\@tempc
  {#3}\ifx \@tempc \@empty \let \@tempc \@tempb \let \@tempb \@tempa \fi \ifx
  \@tempb \@empty \def\@tempb {arXiv}\fi \@ifundefined
  {mn@eprint@\@tempb}{\@tempb:\@tempc}{\expandafter \expandafter \csname
  mn@eprint@\@tempb\endcsname \expandafter{\@tempc}}}

\bibitem[\protect\citeauthoryear{Abel, Anninos, Zhang  \& Norman}{Abel
  et~al.}{1997}]{Abel1997}
Abel T.,  Anninos P.,  Zhang Y.,   Norman M.~L.,  1997, \mn@doi [New Astron.]
  {10.1016/S1384-1076(97)00010-9}, 2, 181

\bibitem[\protect\citeauthoryear{Bertschinger \& Chevalier}{Bertschinger \&
  Chevalier}{1985}]{Bertschinger1985}
Bertschinger E.,  Chevalier R.~A.,  1985, \mn@doi [ApJ] {10.1086/163690}, 299,
  167

\bibitem[\protect\citeauthoryear{Black}{Black}{1981}]{Black1981}
Black J.~H.,  1981, MNRAS, 197, 553

\bibitem[\protect\citeauthoryear{Borysow}{Borysow}{2002}]{Borysow2002}
Borysow A.,  2002, \mn@doi [A{\&}A] {10.1051/0004-6361:20020555}, 390, 779

\bibitem[\protect\citeauthoryear{Borysow, J{\o}rgensen  \& Fu}{Borysow
  et~al.}{2001}]{Borysow2001}
Borysow A.,  J{\o}rgensen U.~G.,   Fu Y.,  2001, \mn@doi [J. Quant. Spectrosc.
  Radiat. Transf.] {10.1016/S0022-4073(00)00023-6}, 68, 235

\bibitem[\protect\citeauthoryear{Bowen}{Bowen}{1988}]{Bowen1988}
Bowen G.~H.,  1988, \mn@doi [ApJ] {10.1086/166378}, 329, 299

\bibitem[\protect\citeauthoryear{Cairnie, Forrey, Babb, Stancil  \&
  McLaughlin}{Cairnie et~al.}{2017}]{Cairnie2017}
Cairnie M.,  Forrey R.~C.,  Babb J.~F.,  Stancil P.~C.,   McLaughlin B.~M.,
  2017, \mn@doi [MNRAS] {10.1093/mnras/stx1715}, 11, 1

\bibitem[\protect\citeauthoryear{Capitelli, Coppola, Diomede  \&
  Longo}{Capitelli et~al.}{2007}]{Capitelli2007}
Capitelli M.,  Coppola C.~M.,  Diomede P.,   Longo S.,  2007, \mn@doi [A{\&}A]
  {10.1051/0004-6361:20077600}, 470, 811

\bibitem[\protect\citeauthoryear{Cazaux \& Spaans}{Cazaux \&
  Spaans}{2009}]{Cazaux2009}
Cazaux S.,  Spaans M.,  2009, \mn@doi [A{\&}A] {10.1051/0004-6361:200811302},
  496, 365

\bibitem[\protect\citeauthoryear{Cen \& Renyue}{Cen \& Renyue}{1992}]{Cen1992}
Cen R.,  Renyue 1992, \mn@doi [ApJS] {10.1086/191630}, 78, 341

\bibitem[\protect\citeauthoryear{Cernicharo et~al.,}{Cernicharo
  et~al.}{2010}]{Cernicharo2010}
Cernicharo J.,  et~al., 2010, \mn@doi [A{\&}A] {10.1051/0004-6361/201015150},
  521, L8

\bibitem[\protect\citeauthoryear{Cherchneff}{Cherchneff}{2006}]{Cherchneff2006a}
Cherchneff I.,  2006, \mn@doi [A{\&}A] {10.1051/0004-6361:20064827}, 456, 1001

\bibitem[\protect\citeauthoryear{Cherchneff}{Cherchneff}{2012}]{Cherchneff2012}
Cherchneff I.,  2012, \mn@doi [A{\&}A] {10.1051/0004-6361/201118542}, 545, A12

\bibitem[\protect\citeauthoryear{Cherchneff, Barker  \& Tielens}{Cherchneff
  et~al.}{1992}]{Cherchneff1992}
Cherchneff I.,  Barker J.~R.,   Tielens A. G. G.~M.,  1992, \mn@doi [ApJ]
  {10.1086/173012}, 401, 269

\bibitem[\protect\citeauthoryear{Dalgarno}{Dalgarno}{2006}]{Dalgarno2006}
Dalgarno A.,  2006, \mn@doi [Proc. Natl. Acad. Sci.] {10.1073/pnas.0602117103},
  103, 12269

\bibitem[\protect\citeauthoryear{Dalgarno, Yan  \& Liu3}{Dalgarno
  et~al.}{1999}]{Dalgarno1999}
Dalgarno A.,  Yan M.,   Liu3 W.,  1999, ApJS, 125, 237

\bibitem[\protect\citeauthoryear{Danilovich, {De Beck}, Black, Olofsson  \&
  Justtanont}{Danilovich et~al.}{2016}]{Danilovich2016}
Danilovich T.,  {De Beck} E.,  Black J.~H.,  Olofsson H.,   Justtanont K.,
  2016, \mn@doi [A{\&}A] {10.1051/0004-6361/201527943}, 588, A119

\bibitem[\protect\citeauthoryear{Decin et~al.,}{Decin et~al.}{2010}]{Decin2010}
Decin L.,  et~al., 2010, \mn@doi [A{\&}A] {10.1051/0004-6361/201015069}, 521,
  L4

\bibitem[\protect\citeauthoryear{Feuchtinger, Dorfi  \& Hofner}{Feuchtinger
  et~al.}{1993}]{Feuchtinger1993}
Feuchtinger M.~U.,  Dorfi E.~A.,   Hofner S.,  1993, A{\&}A, 273, 513

\bibitem[\protect\citeauthoryear{Fleischer, Gauger  \& Sedlmayr}{Fleischer
  et~al.}{1992}]{Fleischer1992}
Fleischer A.~J.,  Gauger A.,   Sedlmayr E.,  1992, A{\&}A, 266, 321

\bibitem[\protect\citeauthoryear{Fleischer, Gauger  \& Sedlmayr}{Fleischer
  et~al.}{1995}]{Fleischer1995}
Fleischer A.~J.,  Gauger A.,   Sedlmayr E.,  1995, A{\&}A, 297, 543

\bibitem[\protect\citeauthoryear{Fonfria, Cernicharo, Richter  \& Lacy}{Fonfria
  et~al.}{2008}]{Fonfria2008}
Fonfria J.~P.,  Cernicharo J.,  Richter M.~J.,   Lacy J.~H.,  2008, \mn@doi
  [ApJ] {10.1086/523882}, 673, 445

\bibitem[\protect\citeauthoryear{Forrey}{Forrey}{2013}]{Forrey2013}
Forrey R.~C.,  2013, \mn@doi [ApJ] {10.1088/2041-8205/773/2/L25}, 773, L25

\bibitem[\protect\citeauthoryear{Freytag \& H{\"{o}}fner}{Freytag \&
  H{\"{o}}fner}{2008}]{Freytag2008}
Freytag B.,  H{\"{o}}fner S.,  2008, \mn@doi [A{\&}A]
  {10.1051/0004-6361:20078096}, 483, 571

\bibitem[\protect\citeauthoryear{Freytag, Liljegren  \& H{\"{o}}fner}{Freytag
  et~al.}{2017}]{Freytag2017}
Freytag B.,  Liljegren S.,   H{\"{o}}fner S.,  2017, \mn@doi [A{\&}A]
  {10.1051/0004-6361/201629594}, 600, A137

\bibitem[\protect\citeauthoryear{Gail, Wetzel, Pucci  \& Tamanai}{Gail
  et~al.}{2013}]{Gail2013}
Gail H.-P.,  Wetzel S.,  Pucci A.,   Tamanai A.,  2013, \mn@doi [A{\&}A]
  {10.1051/0004-6361/201321807}, 555, A119

\bibitem[\protect\citeauthoryear{Galli \& Padovani}{Galli \&
  Padovani}{2015}]{Galli2015}
Galli D.,  Padovani M.,  2015, arXiv

\bibitem[\protect\citeauthoryear{Galli \& Palla}{Galli \&
  Palla}{1998}]{Galli1998}
Galli D.,  Palla F.,  1998, A{\&}A, 335, 403

\bibitem[\protect\citeauthoryear{Glassgold, Galli  \& Padovani}{Glassgold
  et~al.}{2012}]{Glassgold2012}
Glassgold A.~E.,  Galli D.,   Padovani M.,  2012, \mn@doi [ApJ]
  {10.1088/0004-637X/756/2/157}, 756, 157

\bibitem[\protect\citeauthoryear{Glover}{Glover}{2015}]{Glover2015}
Glover S. C.~O.,  2015, \mn@doi [MNRAS] {10.1093/mnras/stv1781}, 453, 2901

\bibitem[\protect\citeauthoryear{Glover \& Abel}{Glover \&
  Abel}{2008}]{Glover2008}
Glover S.~C.,  Abel T.,  2008, \mn@doi [MNRAS]
  {10.1111/j.1365-2966.2008.13224.x}, 388, 1627

\bibitem[\protect\citeauthoryear{Glover \& Jappsen}{Glover \&
  Jappsen}{2007}]{Glover2007}
Glover S. C.~O.,  Jappsen A.,  2007, \mn@doi [ApJ] {10.1086/519445}, 666, 1

\bibitem[\protect\citeauthoryear{Glover, Federrath, Low  \& Klessen}{Glover
  et~al.}{2010}]{Glover2010}
Glover S.~C.,  Federrath C.,  Low M.~M.,   Klessen R.~S.,  2010, \mn@doi
  [MNRAS] {10.1111/j.1365-2966.2009.15718.x}, 404, 2

\bibitem[\protect\citeauthoryear{Gobrecht, Cherchneff, Sarangi, Plane  \&
  Bromley}{Gobrecht et~al.}{2016}]{Gobrecht2016}
Gobrecht D.,  Cherchneff I.,  Sarangi A.,  Plane J. M.~C.,   Bromley S.~T.,
  2016, \mn@doi [A{\&}A] {10.1051/0004-6361/201425363}, 585, A6

\bibitem[\protect\citeauthoryear{Goldsmith \& Langer}{Goldsmith \&
  Langer}{1978}]{Goldsmith1978}
Goldsmith P.~F.,  Langer W.~D.,  1978, \mn@doi [ApJ] {10.1086/156206}, 222, 881

\bibitem[\protect\citeauthoryear{Gould \& Salpeter}{Gould \&
  Salpeter}{1963}]{Gould1963}
Gould R.~J.,  Salpeter E.~E.,  1963, \mn@doi [ApJ] {10.1086/147654}, 138, 393

\bibitem[\protect\citeauthoryear{Grassi, Bovino, Gianturco, Baiocchi  \&
  Merlin}{Grassi et~al.}{2012}]{Grassi2012}
Grassi T.,  Bovino S.,  Gianturco F.~A.,  Baiocchi P.,   Merlin E.,  2012,
  \mn@doi [MNRAS] {10.1111/j.1365-2966.2012.21537.x}, 425, 1332

\bibitem[\protect\citeauthoryear{Grassi, Bovino, Schleicher  \&
  Gianturco}{Grassi et~al.}{2013}]{Grassi2013}
Grassi T.,  Bovino S.,  Schleicher D.,   Gianturco F.~A.,  2013, \mn@doi
  [MNRAS] {10.1093/mnras/stt284}, 431, 1659

\bibitem[\protect\citeauthoryear{Grassi, Bovino, Schleicher, Prieto, Seifried,
  Simoncini  \& Gianturco}{Grassi et~al.}{2014}]{Grassi2014}
Grassi T.,  Bovino S.,  Schleicher D.~R.,  Prieto J.,  Seifried D.,  Simoncini
  E.,   Gianturco F.~A.,  2014, \mn@doi [MNRAS] {10.1093/mnras/stu114}, 439,
  2386

\bibitem[\protect\citeauthoryear{Grassi, Bovino, Haugb{\o}lle  \&
  Schleicher}{Grassi et~al.}{2017}]{Grassi2017}
Grassi T.,  Bovino S.,  Haugb{\o}lle T.,   Schleicher D. R.~G.,  2017, \mn@doi
  [Mnras] {10.1093/mnras/stw2871}, 466, 1259

\bibitem[\protect\citeauthoryear{Gredel, Lepp  \& Dalgarno}{Gredel
  et~al.}{1987}]{Gredel1987}
Gredel R.,  Lepp S.,   Dalgarno A.,  1987, \mn@doi [ApJ] {10.1086/185073}, 323,
  L137

\bibitem[\protect\citeauthoryear{Gredel, Lepp, Dalgarno  \& Herbst}{Gredel
  et~al.}{1989}]{Gredel1989}
Gredel R.,  Lepp S.,  Dalgarno A.,   Herbst E.,  1989, \mn@doi [ApJ]
  {10.1086/168117}, 347, 289

\bibitem[\protect\citeauthoryear{Groenewegen, Sevenster, Spoon  \&
  P{\'{e}}rez}{Groenewegen et~al.}{2002}]{Groenewegen2002}
Groenewegen M. A.~T.,  Sevenster M.,  Spoon H. W.~W.,   P{\'{e}}rez I.,  2002,
  \mn@doi [A{\&}A] {10.1051/0004-6361:20020728}, 390, 511

\bibitem[\protect\citeauthoryear{Habing \& Olofsson}{Habing \&
  Olofsson}{2003}]{Habing2003}
Habing H.,  Olofsson H.,  2003, in Habing H.,  Olofsson H.,  eds, Asymptot.
  giant branch Stars. by Harm J. Habing Hans Olofsson. Astron. Astrophys. Libr.
  New York, Berlin Springer, 2003.

\bibitem[\protect\citeauthoryear{Hindmarsh}{Hindmarsh}{1983}]{Hindmarsh1983}
Hindmarsh A.~C.,  1983, IMACS Trans. Sci. Comput., 1, 55

\bibitem[\protect\citeauthoryear{Hindmarsh, Brown, Grant, Lee, Serban, Shumaker
   \& Woodward}{Hindmarsh et~al.}{2005}]{Hindmarsh}
Hindmarsh A.~C.,  Brown P.~N.,  Grant K.~E.,  Lee S.~L.,  Serban R.,  Shumaker
  D.~E.,   Woodward C.~S.,  2005, ACM Trans. Math. Softw., 31, 363

\bibitem[\protect\citeauthoryear{Hirano \& Yoshida}{Hirano \&
  Yoshida}{2013}]{Hirano2013}
Hirano S.,  Yoshida N.,  2013, \mn@doi [ApJ] {10.1088/0004-637X/763/1/52}, 763,
  10

\bibitem[\protect\citeauthoryear{H{\"{o}}fner, Feuchtinger  \&
  Dorfi}{H{\"{o}}fner et~al.}{1995}]{Hofner1995}
H{\"{o}}fner S.,  Feuchtinger M.~U.,   Dorfi E.~A.,  1995, A{\&}A, 297, 815

\bibitem[\protect\citeauthoryear{H{\"{o}}fner, Gautschy, Loidl, Aringer,
  J{\o}rgensen, Loidl, Aringer  \& J{\o}rgensen}{H{\"{o}}fner
  et~al.}{2003}]{Hofner2003}
H{\"{o}}fner S.,  Gautschy R.,  Loidl â.,  Aringer B.,  J{\o}rgensen U.~G.,
  Loidl R.,  Aringer B.,   J{\o}rgensen U.~G.,  2003, \mn@doi [A{\&}A]
  {10.1051/0004-6361:20021757}, 399, 589

\bibitem[\protect\citeauthoryear{H{\"{o}}fner, Bladh, Aringer  \&
  Ahuja}{H{\"{o}}fner et~al.}{2016}]{Hofner2016}
H{\"{o}}fner S.,  Bladh S.,  Aringer B.,   Ahuja R.,  2016, \mn@doi [A{\&}A]
  {10.1051/0004-6361/201628424}, 594, A108

\bibitem[\protect\citeauthoryear{Hollenbach \& McKee}{Hollenbach \&
  McKee}{1979}]{Hollenbach1979}
Hollenbach D.,  McKee C.~F.,  1979, \mn@doi [ApJS] {10.1086/190631}, 41, 555

\bibitem[\protect\citeauthoryear{Hollenbach, McKee, Hollenbach  \&
  McKee}{Hollenbach et~al.}{1989}]{Hollenbach1989}
Hollenbach D.,  McKee C.~F.,  Hollenbach D.,   McKee C.~F.,  1989, \mn@doi
  [ApJ] {10.1086/167595}, 342, 306

\bibitem[\protect\citeauthoryear{Hunter}{Hunter}{2007}]{Hunter2007}
Hunter J.~D.,  2007, \mn@doi [Comput. Sci. Eng.] {10.1109/MCSE.2007.55}, 9, 90

\bibitem[\protect\citeauthoryear{Irikura}{Irikura}{2007}]{Irikura2007}
Irikura K.~K.,  2007, J. Phys. Chem. Ref. Data, 36, 389

\bibitem[\protect\citeauthoryear{Janev, Langer  \& Evans}{Janev
  et~al.}{1987}]{Janev1987}
Janev R.,  Langer W.,   Evans K.,  1987, {Elementary processes in
  Hydrogen-Helium plasmas - Cross sections and reaction rate coefficients}.
Springer, Berlin

\bibitem[\protect\citeauthoryear{J{\o}rgensen, Hammer, Borysow  \&
  Falkesgaard}{J{\o}rgensen et~al.}{2000}]{Jørgensen2000}
J{\o}rgensen U.~G.,  Hammer D.,  Borysow A.,   Falkesgaard J.,  2000, A{\&}A,
  361, 283

\bibitem[\protect\citeauthoryear{Karakas}{Karakas}{2010}]{Karakas2010}
Karakas A.~I.,  2010, \mn@doi [MNRAS] {10.1111/j.1365-2966.2009.16198.x}, 403,
  1413

\bibitem[\protect\citeauthoryear{Karakas \& Lattanzio}{Karakas \&
  Lattanzio}{2007}]{Karakas2007}
Karakas A.,  Lattanzio J.~C.,  2007, \mn@doi [Publ. Astron. Soc. Aust.]
  {10.1071/AS07021}, 24, 103

\bibitem[\protect\citeauthoryear{Keppens, Meliani, van Marle, Delmont, Vlasis
  \& van~der Holst}{Keppens et~al.}{2012}]{Keppens2012}
Keppens R.,  Meliani Z.,  van Marle A.~J.,  Delmont P.,  Vlasis A.,   van~der
  Holst B.,  2012, \mn@doi [J. Comput. Phys.] {10.1016/j.jcp.2011.01.020}, 231,
  718

\bibitem[\protect\citeauthoryear{Kuzmin}{Kuzmin}{2006}]{Kuzmin2006}
Kuzmin D.,  2006, \mn@doi [J. Comput. Phys.] {10.1016/j.jcp.2006.03.034}, 219,
  513

\bibitem[\protect\citeauthoryear{Liljegren, H{\"{o}}fner, Nowotny  \&
  Eriksson}{Liljegren et~al.}{2016}]{Liljegren2016}
Liljegren S.,  H{\"{o}}fner S.,  Nowotny W.,   Eriksson K.,  2016, \mn@doi
  [A{\&}A] {10.1051/0004-6361/201527885}, 589, A130

\bibitem[\protect\citeauthoryear{Lombaert et~al.,}{Lombaert
  et~al.}{2016}]{Lombaert2016}
Lombaert R.,  et~al., 2016, \mn@doi [A{\&}A] {10.1051/0004-6361/201527049},
  588, A124

\bibitem[\protect\citeauthoryear{Maercker, Danilovich, Olofsson, {De Beck},
  Justtanont, Lombaert  \& Royer}{Maercker et~al.}{2016}]{Maercker2016}
Maercker M.,  Danilovich T.,  Olofsson H.,  {De Beck} E.,  Justtanont K.,
  Lombaert R.,   Royer P.,  2016, \mn@doi [A{\&}A]
  {10.1051/0004-6361/201628310}, 591, A44

\bibitem[\protect\citeauthoryear{Maio, Dolag, Ciardi  \& Tornatore}{Maio
  et~al.}{2007}]{Maio2007}
Maio U.,  Dolag K.,  Ciardi B.,   Tornatore L.,  2007, \mn@doi [MNRAS]
  {10.1111/j.1365-2966.2007.12016.x}, 379, 963

\bibitem[\protect\citeauthoryear{Marigo, Bressan, Nanni, Girardi  \&
  Pumo}{Marigo et~al.}{2013}]{Marigo2013}
Marigo P.,  Bressan A.,  Nanni A.,  Girardi L.~L.,   Pumo M.~L.,  2013, \mn@doi
  [MNRAS] {10.1093/mnras/stt1034}, 434, 488

\bibitem[\protect\citeauthoryear{Marigo, Ripamonti, Nanni, Bressan  \&
  Girardi}{Marigo et~al.}{2016}]{Marigo2016}
Marigo P.,  Ripamonti E.,  Nanni A.,  Bressan A.,   Girardi L.,  2016, \mn@doi
  [MNRAS] {10.1093/mnras/stv2547}, 456, 23

\bibitem[\protect\citeauthoryear{McElroy, Walsh, Markwick, Cordiner, Smith  \&
  Millar}{McElroy et~al.}{2013}]{McElroy2013}
McElroy D.,  Walsh C.,  Markwick A.~J.,  Cordiner M.~A.,  Smith K.,   Millar
  T.~J.,  2013, \mn@doi [A{\&}A] {10.1051/0004-6361/201220465}, 550, A36

\bibitem[\protect\citeauthoryear{McQuarrie \& Simon}{McQuarrie \&
  Simon}{1999}]{McQuarrie1999}
McQuarrie D.~A.,  Simon J.~D.,  1999, {Molecular thermodynamics},
  \mn@doi{10.1016/0022-2860(80)80339-5.
}, \url {http://linkinghub.elsevier.com/retrieve/pii/0022286080803395}

\bibitem[\protect\citeauthoryear{Men'shchikov, Balega, Bl{\"{o}}cker, Osterbart
   \& Weigelt}{Men'shchikov et~al.}{2001}]{Menshchikov2001}
Men'shchikov A.~B.,  Balega Y.,  Bl{\"{o}}cker T.,  Osterbart R.,   Weigelt G.,
   2001, \mn@doi [A{\&}A] {10.1051/0004-6361:20000554}, 368, 497

\bibitem[\protect\citeauthoryear{Nejad}{Nejad}{2005}]{Nejad2005}
Nejad L. A.~M.,  2005, \mn@doi [Ap{\&}SS] {10.1007/s10509-005-2100-z}, 299, 1

\bibitem[\protect\citeauthoryear{Neufeld \& Kaufman}{Neufeld \&
  Kaufman}{1993}]{Neufeld1993}
Neufeld D.~A.,  Kaufman M.~J.,  1993, \mn@doi [ApJ] {10.1086/173388}, 418, 263

\bibitem[\protect\citeauthoryear{Nowotny, H{\"{o}}fner  \& Aringer}{Nowotny
  et~al.}{2010}]{Nowotny2010}
Nowotny W.,  H{\"{o}}fner S.,   Aringer B.,  2010, \mn@doi [A{\&}A]
  {10.1051/0004-6361/200911899}, 514

\bibitem[\protect\citeauthoryear{Oliphant}{Oliphant}{2006}]{Oliphant2006}
Oliphant T.~E.,  2006, {A guide to NumPy}

\bibitem[\protect\citeauthoryear{Olofsson, {Gonz{\'{a}}lez Delgado}, Kerschbaum
   \& Sch{\"{o}}ier}{Olofsson et~al.}{2002}]{Olofsson2002}
Olofsson H.,  {Gonz{\'{a}}lez Delgado} D.,  Kerschbaum F.,   Sch{\"{o}}ier
  F.~L.,  2002, \mn@doi [A{\&}A] {10.1051/0004-6361:20020841}, 391, 1053

\bibitem[\protect\citeauthoryear{Omukai}{Omukai}{2000}]{Omukai2000}
Omukai K.,  2000, \mn@doi [ApJ] {10.1086/308776}, 534, 809

\bibitem[\protect\citeauthoryear{Omukai, Hosokawa  \& Yoshida}{Omukai
  et~al.}{2010}]{Omukai2010}
Omukai K.,  Hosokawa T.,   Yoshida N.,  2010, \mn@doi [ApJ]
  {10.1088/0004-637X/722/2/1793}, 722, 1793

\bibitem[\protect\citeauthoryear{Palla, Salpeter  \& Stahler}{Palla
  et~al.}{1983}]{Palla1983}
Palla F.,  Salpeter E.~E.,   Stahler S.~W.,  1983, \mn@doi [ApJ]
  {10.1086/161231}, 271, 632

\bibitem[\protect\citeauthoryear{Plewa \& M{\"{u}}ller}{Plewa \&
  M{\"{u}}ller}{1999}]{Plewa1999}
Plewa T.,  M{\"{u}}ller E.,  1999, A{\&}A, 342, 179

\bibitem[\protect\citeauthoryear{Poulaert, Brouillard, Claeys, McGowan  \&
  Wassenhove}{Poulaert et~al.}{1978}]{Poulaert1978}
Poulaert G.,  Brouillard F.,  Claeys W.,  McGowan J.~W.,   Wassenhove G.~V.,
  1978, \mn@doi [J. Phys. B At. Mol. Phys.] {10.1088/0022-3700/11/21/006}, 11,
  L671

\bibitem[\protect\citeauthoryear{Ripamonti \& Abel}{Ripamonti \&
  Abel}{2004}]{Ripamonti2004}
Ripamonti E.,  Abel T.,  2004, \mn@doi [MNRAS]
  {10.1111/j.1365-2966.2004.07422.x}, 348, 1019

\bibitem[\protect\citeauthoryear{Ruuth \& Spiteri}{Ruuth \&
  Spiteri}{2002}]{Ruuth2002}
Ruuth S.~J.,  Spiteri R.~J.,  2002, \mn@doi [J. Sci. Comput.]
  {10.1023/A:1015156832269}, 17, 211

\bibitem[\protect\citeauthoryear{Santoro \& Shull}{Santoro \&
  Shull}{2006}]{Santoro2006}
Santoro F.,  Shull J.~M.,  2006, \mn@doi [ApJ] {10.1086/501518}, 643, 26

\bibitem[\protect\citeauthoryear{Savin, Krsti, Haiman  \& Stancil}{Savin
  et~al.}{2004}]{Savin2004}
Savin D.~W.,  Krsti P.~S.,  Haiman Z.,   Stancil P.~C.,  2004, \mn@doi [ApJ]
  {10.1086/421108}, 606, L167

\bibitem[\protect\citeauthoryear{Schirrmacher, Woitke  \&
  Sedlmayr}{Schirrmacher et~al.}{2003}]{Schirrmacher2003a}
Schirrmacher V.,  Woitke P.,   Sedlmayr E.,  2003, \mn@doi [A{\&}A]
  {10.1051/0004-6361:20030444}, 404, 267

\bibitem[\protect\citeauthoryear{Sch{\"{o}}ier \& Olofsson}{Sch{\"{o}}ier \&
  Olofsson}{2001}]{Schoier2001}
Sch{\"{o}}ier F.~L.,  Olofsson H.,  2001, \mn@doi [A{\&}A]
  {10.1051/0004-6361:20010072}, 368, 969

\bibitem[\protect\citeauthoryear{Sch{\"{o}}ier, Ramstedt, Olofsson, Lindqvist,
  Bieging  \& Marvel}{Sch{\"{o}}ier et~al.}{2013}]{Schoier2013}
Sch{\"{o}}ier F.~L.,  Ramstedt S.,  Olofsson H.,  Lindqvist M.,  Bieging J.~H.,
    Marvel K.~B.,  2013, \mn@doi [A{\&}A] {10.1051/0004-6361/201220400}, 550,
  A78

\bibitem[\protect\citeauthoryear{Sobolev}{Sobolev}{1960}]{Sobolev1960}
Sobolev V.~V.,  1960, Cambridge Harvard Univ. Press

\bibitem[\protect\citeauthoryear{Spitzer}{Spitzer}{1978}]{Spitzer1978}
Spitzer L.,  1978, {Physical Processes in the Interstellar Medium}.
Wiley-VCH Verlag GmbH, Weinheim, Germany, \mn@doi{10.1002/9783527617722}, \url
  {http://doi.wiley.com/10.1002/9783527617722}

\bibitem[\protect\citeauthoryear{T{\'{o}}th \& Odstr{\v{c}}il}{T{\'{o}}th \&
  Odstr{\v{c}}il}{1996}]{Toth1996}
T{\'{o}}th G.,  Odstr{\v{c}}il D.,  1996, \mn@doi [J. Comput. Phys.]
  {10.1006/jcph.1996.0197}, 128, 82

\bibitem[\protect\citeauthoryear{Tupper}{Tupper}{2002}]{Tupper2002}
Tupper P.~F.,  2002, BIT Numer. Math., 42, 447

\bibitem[\protect\citeauthoryear{Wakelam et~al.,}{Wakelam
  et~al.}{2012}]{Wakelam2012}
Wakelam V.,  et~al., 2012, \mn@doi [ApJS] {10.1088/0067-0049/199/1/21}, 199, 21

\bibitem[\protect\citeauthoryear{Willson}{Willson}{2000}]{Willson2000}
Willson A.~L.,  2000, ARA{\&}A, 38, 573

\bibitem[\protect\citeauthoryear{Willson \& Bowen}{Willson \&
  Bowen}{1984}]{Willson1984}
Willson L.~A.,  Bowen G.~W.,  1984, in Stalio R.,  ed., Relat. Chromoshperic
  Coronal Heat Mass Loss Stars. p.~127, \url
  {https://ui.adsabs.harvard.edu/{\#}abs/1984rcch.conf..127W/abstract}

\bibitem[\protect\citeauthoryear{Winters, {Le Bertre}, Jeong, Helling  \&
  Sedlmayr}{Winters et~al.}{2000}]{Winters2000}
Winters J.~M.,  {Le Bertre} T.,  Jeong K.~S.,  Helling C.,   Sedlmayr E.,
  2000, A{\&}A, 361, 641

\bibitem[\protect\citeauthoryear{Woitke}{Woitke}{2006}]{Woitke2006}
Woitke P.,  2006, \mn@doi [A{\&}A] {10.1051/0004-6361:20054202}, 452, 537

\bibitem[\protect\citeauthoryear{Woitke, Krueger, Sedlmayr, Woitke, Krueger  \&
  Sedlmayr}{Woitke et~al.}{1996}]{Woitke1996a}
Woitke P.,  Krueger D.,  Sedlmayr E.,  Woitke P.,  Krueger D.,   Sedlmayr E.,
  1996, A{\&}A, 311, 927

\bibitem[\protect\citeauthoryear{Wood}{Wood}{1979}]{Wood1979a}
Wood P.~R.,  1979, \mn@doi [ApJ] {10.1086/156721}, 227, 220

\bibitem[\protect\citeauthoryear{Yee \& C.}{Yee \& C.}{1989}]{Yee1989}
Yee C. H.,  1989, Technical report, {A class of high resolution explicit and
  implicit shock-capturing methods}, \url
  {https://ntrs.nasa.gov/search.jsp?R=19890016281}.
NASA Ames Research Center, \url
  {https://ntrs.nasa.gov/search.jsp?R=19890016281}

\makeatother
\end{thebibliography}

% \begin{thebibliography}{99}
% \bibitem[\protect\citeauthoryear{Author}{2012}]{Author2012}
% Author A.~N., 2013, Journal of Improbable Astronomy, 1, 1
% \bibitem[\protect\citeauthoryear{Others}{2013}]{Others2013}
% Others S., 2012, Journal of Interesting Stuff, 17, 198
% \end{thebibliography}

%%%%%%%%%%%%%%%%%%%%%%%%%%%%%%%%%%%%%%%%%%%%%%%%%%

%%%%%%%%%%%%%%%%% APPENDICES %%%%%%%%%%%%%%%%%%%%%

\appendix
\section{Adiabatic index}\label{app:gamma}
The adiabatic index $\gamma$ is \ado{defined as}
\begin{equation}\label{appA:gamma}
    \gamma \equiv \frac{C_p}{C_V},
\end{equation}
 \ado{where the heat capacity at constant volume is given by}
\begin{equation}\label{cv}
    C_V = \left.\frac{\partial E}{\partial T}\right|_V,
\end{equation}
 \ado{and the heat capacity at constant pressure by}
\begin{equation}\label{cp}
    C_p = \left.\frac{\partial H}{\partial T}\right|_p,
\end{equation}
 \ado{where $E$ is the internal energy and $H=E+pV$ is the enthalpy. Equation \eqref{cp} can be rewritten as}
\begin{align}
    C_p &= \left.\frac{\partial H}{\partial T}\right|_p\nonumber\\
        &= \left.\frac{\partial E}{\partial T}\right|_p + p\left.\frac{\partial V}{\partial T}\right|_p\nonumber\\
        &= \left.\frac{\partial E}{\partial T}\right|_p + p\left.\frac{\partial \frac{Nk_BT}{p}}{\partial T}\right|_p\nonumber\\
        &= \left.\frac{\partial E}{\partial T}\right|_p + Nk_B\nonumber\\
        &= \left.\frac{\partial E}{\partial T}\right|_V + Nk_B\nonumber\\
        &= C_V + Nk_B \label{appA:cv-cp}
\end{align}
\ado{where in the second to last step we stated that} 
\begin{equation}
    \left.\frac{\partial E}{\partial T}\right|_p = \left.\frac{\partial E}{\partial T}\right|_V,
\end{equation}
\ado{which is valid because all internal energy terms are independent of $p$ and $V$. This is the case for any atom or molecule be it monoatomic, diatomic (Eqs. \ref{appA:Etr}, \ref{appA:Evib}, \ref{appA:Erot}), linear polyatomic, or non-linear polyatomic \citep{McQuarrie1999}. When substituting Eq. \eqref{appA:cv-cp} into Eq. \eqref{appA:gamma}, the adiabatic index is prescribed by}
% determined by the internal degrees of freedom and is related to the specific heat capacity $C_V$ by:
\begin{equation}\label{eq:CV}
\frac{N\kb}{\gamma-1} = C_V = \left.\frac{\partial E}{\partial T}\right|_V.
\end{equation}
\ado{The internal energy of a system consisting of $N$ particles is defined as}
\begin{equation}\label{app:E_N}
    E =  \left.-\frac{\partial \ln Z}{\partial \beta}\right|_{V,N},
\end{equation}
% where $\left< E\right>$ is the thermodynamic total energy and $\kb$ the Boltzmann constant. The total energy is calculated via:
% \begin{equation}\label{eq:Eint}
% \left< E\right> = -\frac{\partial \ln Z}{\partial \beta},
% \end{equation}
with $\beta = \frac{1}{\kb T}$ and $Z$ the total partition function of the gas. According to the Born-Oppenheimer approximation, rotational, vibrational, and electronic energies are independent of each other, and the partition function of one particle can be written as the product of separate contributors namely translational, rotational, vibrational, and electronic degrees of freedom, $Z_1=Z\tr Z\rot Z\vib Z\elec$. When dealing with a system of $N$ non-interacting particles, the system's partition function is given by
\begin{equation}\label{app:Z_N}
    Z = \frac{1}{N!}Z_1^N.
\end{equation}
\ado{Substituting this into Eq.~\eqref{app:E_N} results in}
\begin{equation}\label{eq:Eint}
    E = \left.NkT^2\frac{\pd\ln Z_1}{\pd T}\right|_{V} = N E_1,
\end{equation}
\ado{with $E_1$ the internal energy of one particle.} We limit the calculation of $\gamma$ to mono and diatomic molecules. Note that mono atomic molecules do not possess any rotational or vibrational freedom. \ado{All subsections below are representations for one particle ($N=1$).}

\subsection{Translational freedom}
The translational partition function is given by:
\begin{equation}
    Z_{\text{trans}}= \left(\frac{2\pi m \kb T}{h^2}\right)^{3/2} V
\end{equation}
with $m$ the mass of the species, $\kb$ the Boltzmann constant, $T$ the temperature, $h$ the Planck constant, and $V$ the available volume. According to Eq.~\eqref{eq:Eint}, the internal energy is then,
\begin{equation}\label{appA:Etr}
    E_{\text{trans}}=\frac{3}{2}\kb T,
\end{equation}
that via Eq.~\eqref{eq:CV} a heat capacity at constant volume
\begin{equation}
    C_{V_\text{trans}} = \frac{3}{2}\kb,
\end{equation}
gives.

\subsection{Vibrational freedom}
When approximating a diatomic molecule by an harmonic oscillator, the vibrational energy levels relative to the bottom of the potential well are
\begin{equation}
    \varepsilon_n = \left(n+\frac{1}{2}\right) h\nu, \qquad\qquad n=0,1,2,...
\end{equation}
where $n$ is the vibrational quantum number and all energy levels are equally spaced by $\Delta \varepsilon=h\nu$. The vibrational partition function then becomes
\begin{equation}
    Z_{\text{vib}}= \sum_{n=0}^{\infty} e^{-\beta \varepsilon_n} = e^{\frac{-\beta h\nu}{2}} \sum_{n=0}^{\infty} e^{-\beta h\nu} = \frac{e^{\frac{-\beta h\nu}{2}}}{1-e^{-\beta h\nu}},
\end{equation}
where the last equality is a standard geometric series. Following the same procedure as before, the internal energy is given by,
\begin{equation}\label{appA:Evib}
    E_{\text{vib}}=\kb \left( \frac{\theta_v}{2} + \frac{\theta_v}{e^{\theta_v/T}-1}\right),
\end{equation}
where $\theta_v=\frac{h\nu}{\kb}$ is the vibrational temperature. The heat capacity at constant volume is then prescribed by,
\begin{equation}
    C_{V_\text{vib}} = \kb \left(\frac{\theta_v}{T}\right)^2 \frac{e^{\theta_v/T}}{\left(e^{\theta_v/T}-1\right)^2}.
\end{equation}

\subsection{Rotational freedom}
In the rigid rotor approximation, the rotational energy levels of a diatomic molecule can be described as,
\begin{equation}
    \varepsilon_J=\frac{J(J+1)\hbar^2}{2I}, \qquad\qquad J=0,1,2,...
\end{equation}
where $J$ is the rotational quantum number, and $I=\mu R^2$ the moment of inertia of the molecule with $\mu$ the reduced mass and $R$ the distance of each nuclei to the centre of mass. Two consecutive energy levels, $J$ to $J$+1, are then separated by $\Delta \varepsilon(J)=\frac{(J+1)\hbar^2}{2I}$. For hetero nuclear diatomic molecules the rotational partition function is then,
\begin{equation}
    Z_{\text{rot}}=\sum_{J=0}^{\infty} \omega_J e^{-\beta \varepsilon_J}=\sum_{J=0}^{\infty}(2J+1) e^{-\theta_r J(J+1)/T}, 
\end{equation}
where $\theta_r=\frac{\hbar^2}{2I\kb}$ is the rotational temperature and $\omega_J$ the degeneracy factor. This series does not converge to an analytic function but for high temperatures, $T\gg \theta_r$, the series can be approximated by its integral from,
\begin{equation}
    Z_{\text{rot}} \approx \int_{J=0}^{\infty}(2J+1) e^{-\theta_r J(J+1)/T} dJ = \frac{T}{\theta_r}.
\end{equation}
This high temperature approximation is valid during our entire simulation as the temperature is always much larger than the highest rotational temperature, which in this case is that of \ch{H2} ($\theta_r = 85.4$\,K). For homo nuclear diatomic molecules, the calculation of the rotational partition function is less straightforward. This because the total wave function of the molecule, that is, the electronic, vibrational, rotational, translational, and nuclear wave function, must be either symmetric or anti-symmetric under the interchange of the two identical nuclei. Only the nuclear and rotational wave function get coupled by this symmetry requirement. The rotational-nuclear partition function for homo nuclear diatomic molecules is then given by:
\begin{align}
Z_{\text{rot,nucl}} &= I(2I+1)\sum_{J \text{even}}(2J+1) e^{-\theta_r J(J+1)/T} \nonumber \\
&+ (I+1)(2I+1)\sum_{J \text{odd}}(2J+1)e^{-\theta_r J(J+1)/T}.
\end{align}
With $I$ the nuclear spin. Note that this is for Fermionic nuclei (half-integer spin), whereas for Bosonic nuclei (integer spin) the even and odd degeneracy must be reversed.
Once again, we apply the high temperature approximation to write the series as an integral,
\begin{align}
    \sum_{J \text{even}} \approx \sum_{J \text{odd}} \approx \frac{1}{2} \sum_{J \text{all}}  &\approx \frac{1}{2}\int_{J=0}^{\infty}(2J+1) e^{-\theta_r J(J+1)/T} dJ \nonumber \\
    &= \frac{T}{2\theta_r}.
\end{align}
The total rotational-nuclear partition function can then be decoupled and is given by:
\begin{equation}
    Z_{\text{rot,nucl}} = (2I+1)^2 \frac{T}{2\theta_r} =Z_{\text{nucl}}Z_{\text{rot}}.
\end{equation}
Subsequently, the internal energy of the nuclear part is zero, while the rotational part is given by:
\begin{equation}\label{appA:Erot}
    E_{\text{rot}} = \kb T.
\end{equation}
Thus only a rotational contribution to the heat capacity at constant volume,
\begin{equation}
    C_{V_\text{rot}} = \kb.
\end{equation}

\subsection{Electronic freedom}
The excited electronic energy levels are typically at much higher energies than the pure vibrational and rotational energy levels. Therefore, they contribute only a fraction to the partition function of molecules. Because of this, we neglect this contribution to the heat capacity at constant volume.

\subsection{Total heat capacity at constant volume}
The total heat capacity at constant volume of each species can be written as the sum of all its individual constituents. The total heat capacity at constant volume of the gas is then the sum of all atomic and molecular contributions, weighted by their fractional number density:
\begin{equation}\label{eq:cv_tot}
    C_V =  \sum_{\text{mono}} \frac{n_i}{n_\text{tot}} C_{V_\text{trans}} + \sum_{\text{dia}} \frac{n_i}{n_\text{tot}} \left( C_{V_\text{trans}} + C_{V_\text{rot}} + C_{V_\text{vib}} \right),
\end{equation}
where the first sum is over the mono atomic molecules and the second one over the diatomic molecules. The adiabatic index can then be calculated by substituting this in Eq.~\eqref{eq:CV}. We do limit the vibrational contributions to the molecules of which the vibrational temperature is provided by \krome and taken from \citet{Irikura2007} (\ch{H2}, \ch{C2}, \ch{CH}, \ch{CO}, \ch{CO+}, \ch{N2}, \ch{NH}, \ch{NO}, \ch{O2}, \ch{OH}). The calculations described above correspond to the option 'GAMMA=VIB' in \krome.

\subsection{Internal energy pressure relation}\label{appAsect:e-gamma}
\ado{According to Eq. \eqref{eq:CV} the interal energy can be derived from the heat capacity via}
\begin{equation}
    E = \int C_V dT.
\end{equation}
\ado{If $C_V$ is constant over $T$, this can be brought outside the integral and results in}
\begin{equation}\label{eq:e-P}
    E = C_V T = \frac{Nk_B T}{\gamma - 1} = \frac{pV}{\gamma - 1},
\end{equation}
\ado{which describes the pressure in function of the internal energy and closes the system of differential equations that dictate the dynamical behaviour (Eqs. \ref{eq:cons_mass}--\ref{energy}). Note that this is the first term of Eq.~\eqref{eq:energy_density}. However, we know $C_V$ is temperature dependent, so this cannot be taken out of the integral. Yet, if the change in $C_V$ is small over the taken temperature change, it is still a decent approximation. It turns out that $C_V$ does not vary rapidly over temperature. Moreover, the temperature jumps per hydrodynamical time step are small enough for the change in $C_V$ to stay small.\\\\
One can improve the energy determination by summing over all individual energy contributions of the microphysical states (translation, rotation, etc) rather than using the approximation of Eq.~\eqref{eq:e-P}. However, when wanting to relate pressure in function of the internal energy, each temperature term in the internal energy has to be substituted with the ideal gas law. Yet, inverting this function does not yield a unique solution for the internal energy due to its complex temperature dependence and the summation over all microphysical states of all species. The non-uniqueness of this approach renders it unfeasible to be used instead the approximation of Eq.~\eqref{eq:e-P}.\\\\
The summation over all microphysical states of all species can be used as an improvement to update the internal energy after a chemical evolution step performed by \krome. Currently, we use Eq.~\eqref{eq:e-P} with an updated temperature and adiabatic index to determine the energy change, via Eq.~\eqref{eq:temp_change},  and Eqs.~\eqref{eq:CV} and \eqref{eq:cv_tot}, respectively. However this is beyond the current scope, as it implies knowing all microphysical energy states of all considered species either experimentally of via quantum chemical computations.}

\section{Degeneracy of hydrostatic equilibrium}\label{app:HE}

\begin{figure}
	\includegraphics[width=\columnwidth]{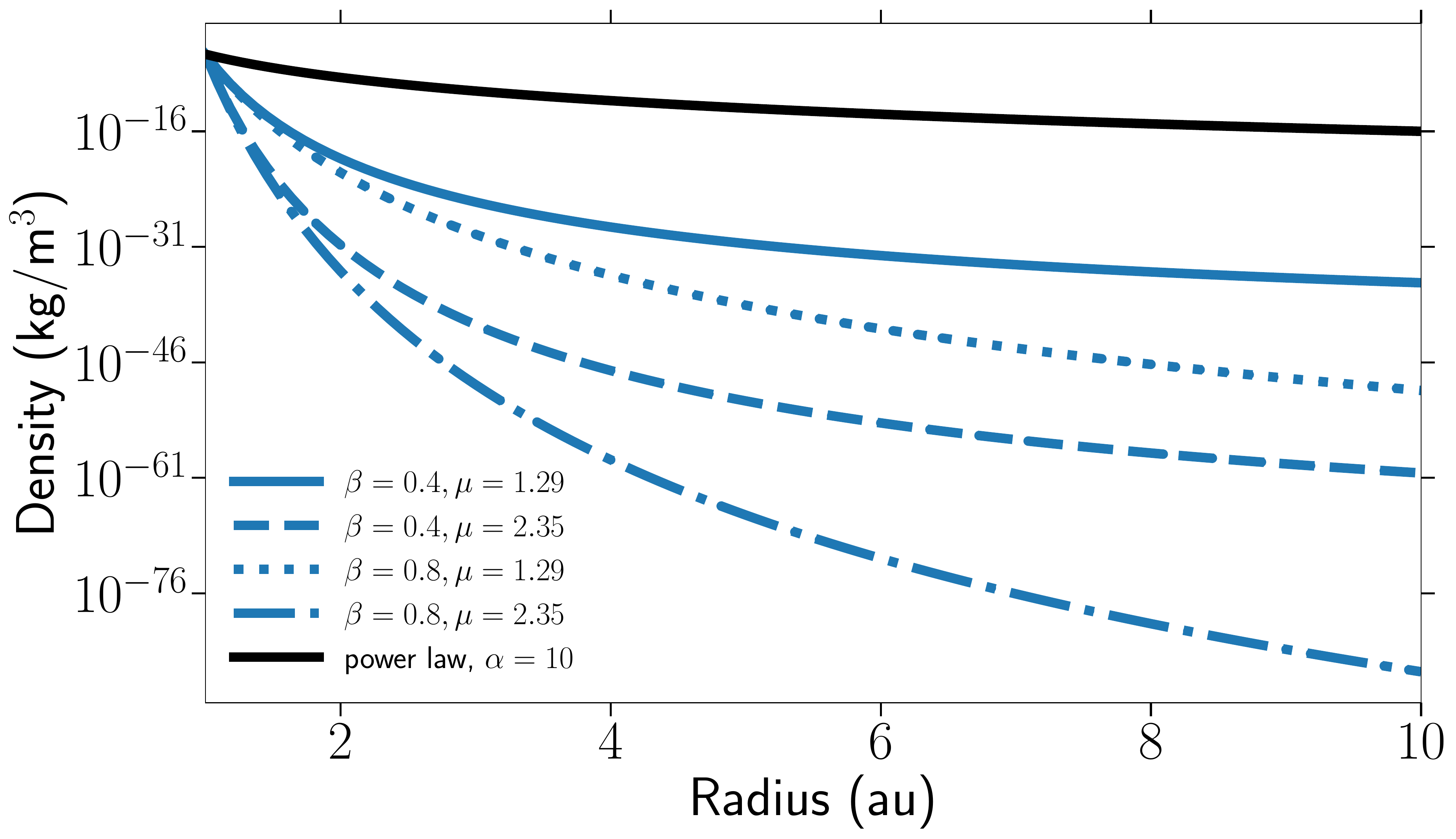}
    \caption{It is quite meaningless to use \textit{the} hydrostatic equilibrium solution of the gas density (Eq.~\ref{rhoStat}) as an initial condition because there exist no single one solution. The solution is degenerate due to parameter freedom of the temperature power law exponent, $\beta$, and the mean molecular weight, $\mu$. For reasonable values of $\beta$ and $\mu$, these solutions differ several orders of magnitude. We therefore choose a density power law (Eq.~\ref{rhoPow}) with $\alpha =10$, consistent with model results of \citet{Hofner2016} and \citet{Freytag2017}.}
    \label{fig:HE}
\end{figure}

In the hydrostatic case and assuming spherical symmetry, the momentum equation (Eq.~\ref{momentum}) reduces to:

\begin{equation}
\label{momStat}
\frac{\partial p}{\partial r} = \frac{-GM_\star}{r^2} \rho(r).
\end{equation}
Using the ideal gas law,

\begin{equation}\label{IGlaw}
p(r)=T(r) \rho(r) \kbmu,
\end{equation}
and the assumption of a power law for the temperature profile,

\begin{equation}
T(r)= T_\star \left( \frac{r}{R_\star} \right)^{-\beta},
\end{equation}
integration of equation (Eq.~\ref{momStat}) results in an expression for the gas density:

\begin{equation}\label{rhoStat}
\rho(r) = \rho_\star \left(\frac{r}{R_\star} \right)^\beta\text{exp}\left\{ \frac{-G M_\star}{T_\star R_\star} \frac{\mu m_\text{H}}{\kb (\beta-1)}  \left[\left(\frac{r}{R_\star}\right)^{\beta-1}-1 \right] \right\},
\end{equation}
with $\rho_\star$ the stellar surface density, $\beta$ the temperature profiles exponent, and $\mu$ the mean molecular weight in units of atomic hydrogen mass, $m_\text{H}$. The hydrostatic solution\footnote{Note the extra radial factor in front of the exponential function which was overlooked in the derivation by \citet{Cherchneff1992, Marigo2016} due to improper derivation when substituting Eq.~\eqref{IGlaw} in Eq.~\eqref{momStat}. Luckily, the exponential function dominates, leading to a difference of maximal 10 per cent.} of the density (Eq. \ref{rhoStat}) has two degrees of freedom, namely the temperature exponent $\beta$ and the mean molecular weight $\mu$. The former typically has a value between $0.4$ and $0.8$ but varies radially, with a steeper temperature profile closer to the star \citep[][fig.~10]{Freytag2017}. As a rough guess, the latter can lie between two extreme values of $\mu_\text{H/He} = 1.29$ for a pure H-He mixture and $\mu_\text{\ch{H2}/He} = 2.35$ for a pure \ch{H2}-He mixture, where we take reasonable mass fractions of $X_\text{(H or \ch{H2})}=0.7$ and $X_\text{He}=0.3$. These four \textit{equilibrium} solutions already differ orders of magnitude (Fig.~\ref{fig:HE}).

\section{Thermal processes}
Here we present the details on the metal cooling (Table~\ref{tab:Zcool}) and \ch{H2} chemical heating and cooling (Table~\ref{tab:H2diss}).

\begin{table}
    \center
    \caption{Characteristics adopted for metal cooling \citep[][table~6]{Grassi2014}.}\label{tab:Zcool}
    \begin{tabular}{llll}
        Coolant & Fine-structure levels    & Transitions   & Partners \\ \hline
        C       & 3         & 3              & H, \ch{H+}, \ch{H2}, \ch{e-}          \\
        \ch{C+}     & 2          &   1            &  H, \ch{e-}          \\
        O       & 3          & 3               & H, \ch{H+}, \ch{e-}          \\
        \ch{O+}     &   3        &   3            &  \ch{e-}         \\
        Si      & 3          & 3              & H, \ch{H+}          \\
        \ch{Si+}    & 2           & 1              &   H, \ch{e-}         \\
        Fe      & 5          & 6              & H, \ch{e-}          \\
        \ch{Fe+}    &  5         &   5            &  H, \ch{e-}          \\ 
    \end{tabular}
\end{table}

\begin{table}
    \center
    \caption{\ch{H2} chemical heating and cooling reactions (subset of the ones provided by \citet{Grassi2014}).}\label{tab:H2diss}
    \begin{tabular}{lr}

        Cooling & Energy (eV)\\ 
        \hline
        \ch{H2 + H -> H + H + H} & 4.48\\
        \ch{H2 + e- -> H + H + e-}  & 4.48\\\\
        Heating & Energy (eV) $/ f_{\text{crit}}^a$ \\\hline
        \ch{H- + H -> H2  + e-}      & 3.53    \\
        \ch{H + H2+ -> H2 + H+} & 1.83 \\
        \ch{H + H + H -> H2 + H}   &  4.48\\
        \ch{H2 + H + H -> H2 + H2} & 4.48\\\\
        \hline
        \multicolumn{2}{p{\columnwidth}}{$^a$ Critical density factor (\citealt{Hollenbach1979}, Eq.~6.45; \citealt{Omukai2000})}
    \end{tabular}
\end{table}

\section{Additional chemical evolution}\label{app:extra_mol}
\begin{figure}
	\includegraphics[width=\columnwidth]{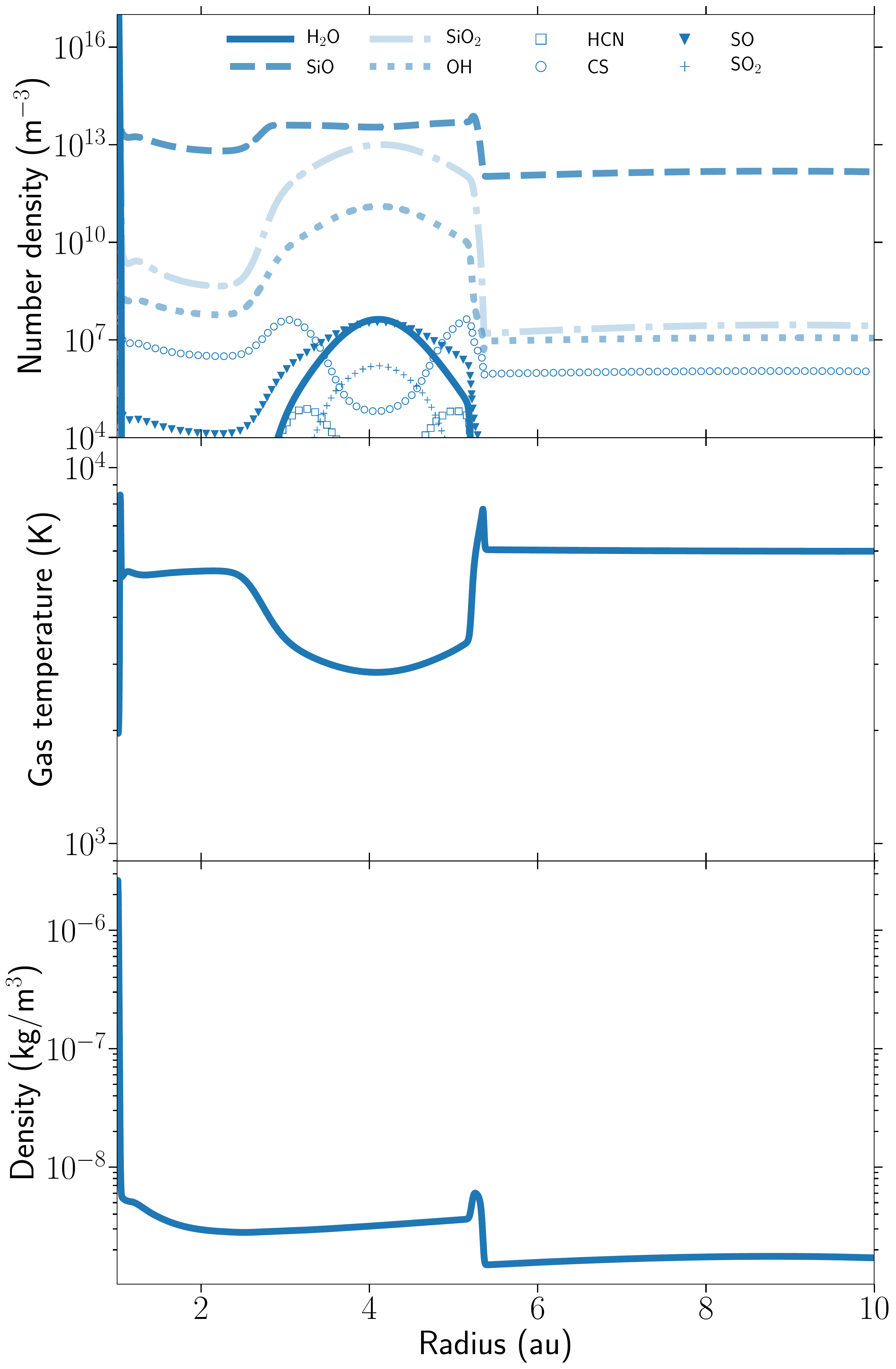}
    \caption{A snapshot of the wind structure 20 days after a `burn-in' phase of four pulsation periods for a hydrochemical model with $\Delta v = 20$~\kms. \textbf{First:} Number densities of relevant species in AGB winds with a lower cut-off at $10^{4}$~\si{\per\m\cubed}. Beware that none of the abundances should be used as a comparison reference because our AGB wind model is far from complete (Section~\ref{sec:summary}). \textbf{Second:} Temperature structure of the gas. \textbf{Third:} Density structure of the gas.}
    \label{fig:structureExtra20}
\end{figure}

\begin{figure}
	\includegraphics[width=\columnwidth]{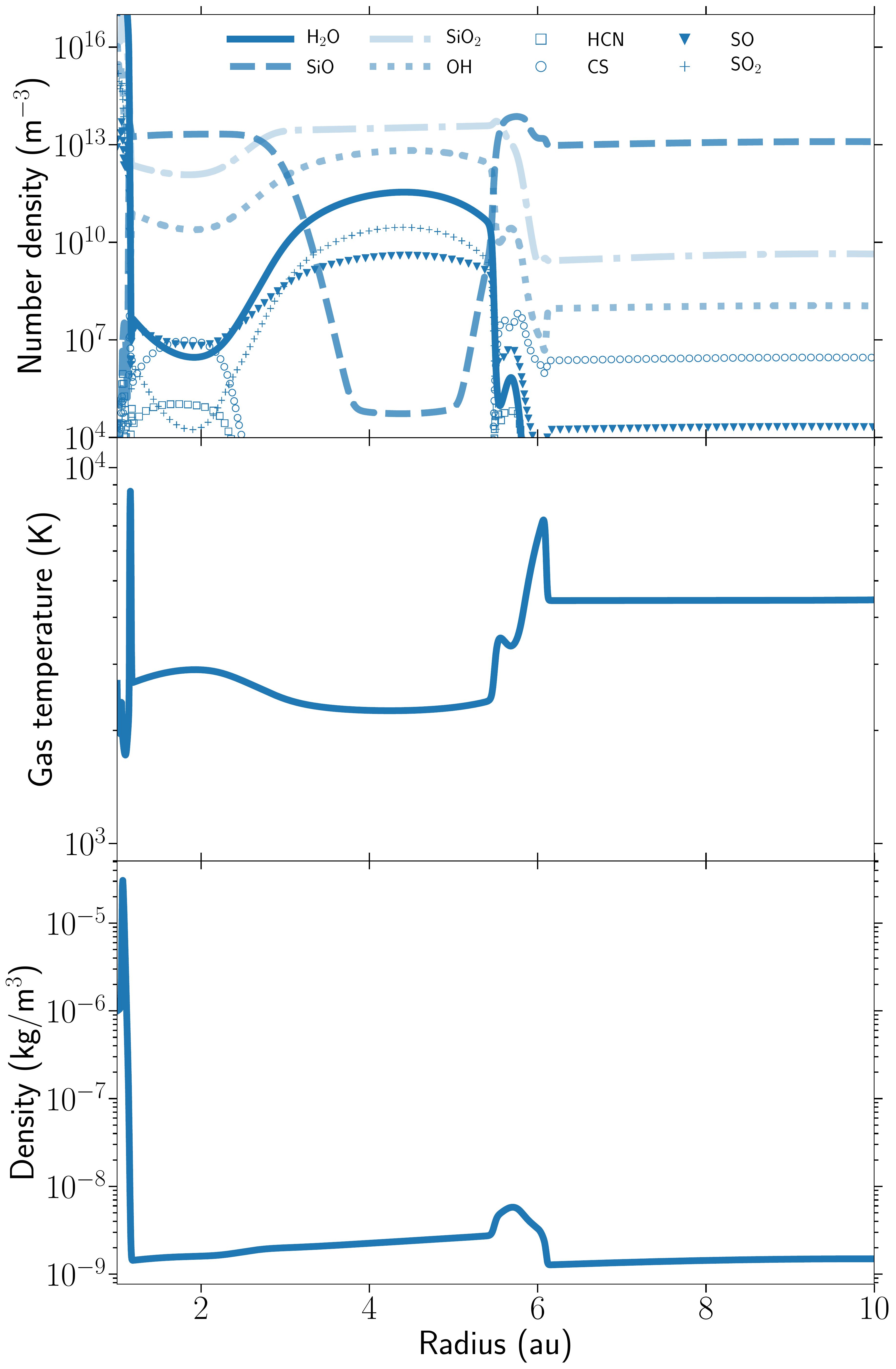}
    \caption{A snapshot of the wind structure 100 days after a `burn-in' phase of four pulsation periods for a hydrochemical model with $\Delta v = 20$~\kms. \textbf{First:} Number densities of relevant species in AGB winds with a lower cut-off at $10^{4}$~\si{\per\m\cubed}. Beware that none of the abundances should be used as a comparison reference because our AGB wind model is far from complete (Section~\ref{sec:summary}). \textbf{Second:} Temperature structure of the gas. \textbf{Third:} Density structure of the gas.}
    \label{fig:structureExtra100}
\end{figure}

\begin{figure}
	\includegraphics[width=\columnwidth]{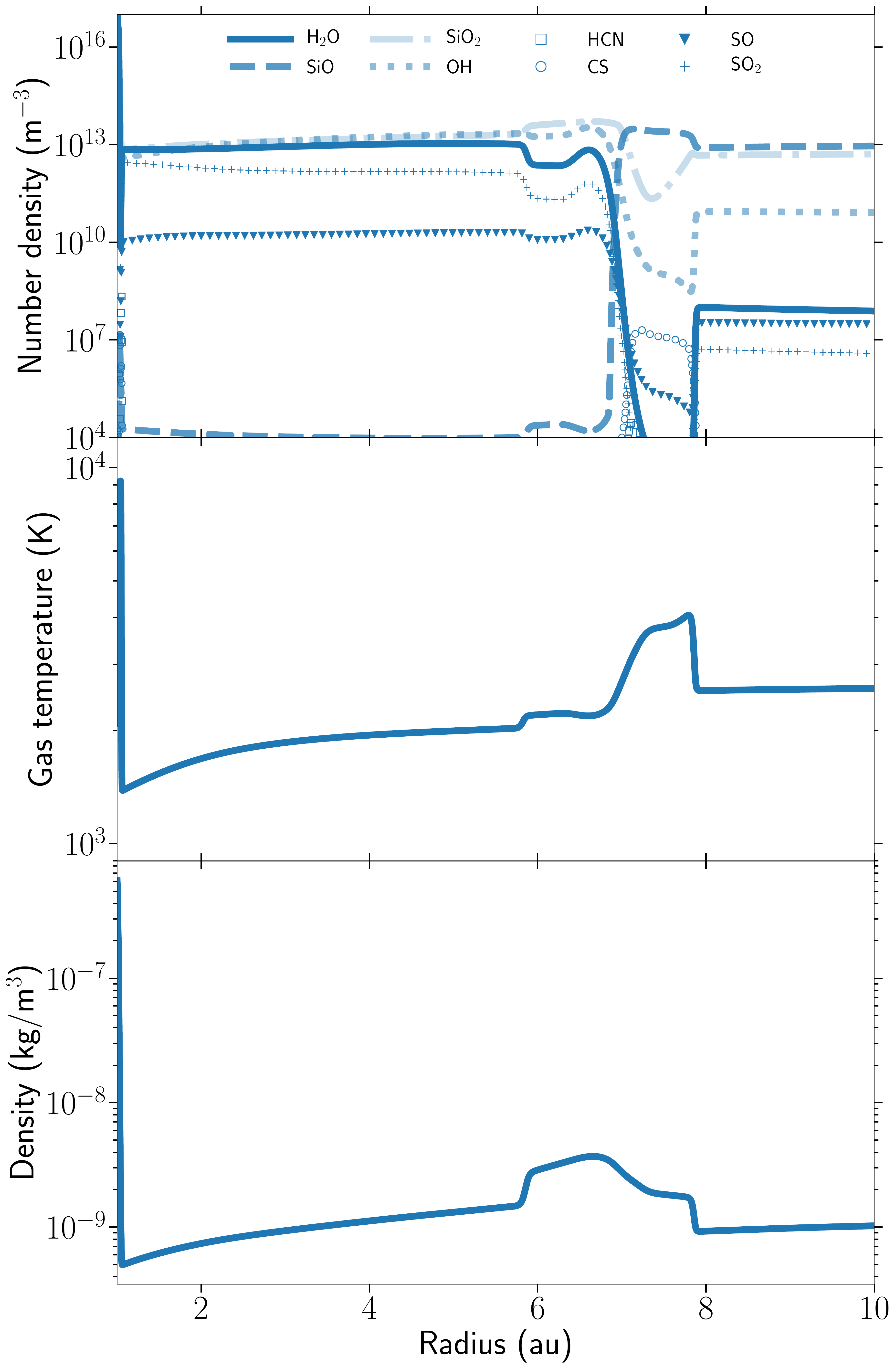}
    \caption{A snapshot of the wind structure 300 days after a `burn-in' phase of four pulsation periods for a hydrochemical model with $\Delta v = 20$~\kms. \textbf{First:} Number densities of relevant species in AGB winds with a lower cut-off at $10^{4}$~\si{\per\m\cubed}. Beware that none of the abundances should be used as a comparison reference because our AGB wind model is far from complete (Section~\ref{sec:summary}). \textbf{Second:} Temperature structure of the gas. \textbf{Third:} Density structure of the gas.}
    \label{fig:structureExtra300}
\end{figure}

\ado{In addition to the evolution and abundance of the chemical species relevant to heating and cooling processes (Figs.~\ref{fig:structureCool20}--\ref{fig:structureCool300}), we show extra ones that are relevant in AGB winds (Figs.~\ref{fig:structureExtra20}--\ref{fig:structureExtra300}). Beware that none of these results should be used as a comparison reference because our AGB wind model is still incomplete (Section~\ref{sec:summary}). These results are merely shown as a proof that our chemical model does not yield highly unrealistic values. For example, the abundance of \ch{HCN} is negligible as expected in an \ch{O}-rich environment. Another example is that as the temperature decreases, more complex molecules start to form. Note the gradual rise of \ch{H2O}, \ch{SiO2}, and \ch{SO2} abundances as the temperature drops, starting between 3 and 5~au but spreading to the inner half.}

\section{Reduced network}\label{app:network}
Here we present the reduced chemical network where $T$ represents the gas temperature, $T_e = T/11604.525$~eV\,K$^{-1}$ is the gas temperature in electron volt, and $\zeta = 1.36 \cdot 10^{-17}$s$^{-1}$ is the chosen cosmic ray (CR) flux \citep[falls within the typical range, ][and references therein]{Dalgarno2006}. A digital, computer readable version of the network can be found as Supplementary material of the original paper (online). This serves as a \krome input file and is also made available in \krome as a subset of the network \textit{react\_AGBwind\_nucleation}\footnote{\url{https://bitbucket.org/tgrassi/krome}}, (without the nucleation part, which is addressed in an upcoming paper).
\onecolumn
\renewcommand{\arraystretch}{1.5}
\begin{longtable}{lllhp{4cm}}
\hline\rule[0mm]{-1mm}{4mm}
No. & Reaction & Rate coefficient (cm$^{3(N-1)}$ s$^{-1}$) with $N$ number of reactants& Temperature limit & Reference \\ \hline
%********************************
% Include \usepackage{chemformula}
% Chemical reaction network (LaTeX format)
% Table columns format {lllll}
% Set "h" as table-spec for the column to hide it
%%\newcolumntype{h}{>{\setbox0=\hbox\bgroup}c<{\egroup}@{}}
1 & \ch{C + C -> C2 + $\gamma$} & k$_{1} = 4.36 \cdot 10^{-18} (T/300)^{0.35} \operatorname{exp}{\left (- \frac{161.3}{T} \right )}$ &  & UMIST \\
2 & \ch{C + CH -> C2 + H} & k$_{2} = 6.59\cdot 10^{-11}$ &  & UMIST \\
3 & \ch{C + CN -> C2 + N} & k$_{3} = 4.98 \cdot 10^{-10} \operatorname{exp}{\left (- \frac{18116}{T} \right )}$ &  & UMIST \\
4 & \ch{C + CO -> C2 + O} & k$_{4} = 2.94 \cdot 10^{-11} (T/300)^{0.5} \operatorname{exp}{\left (- \frac{58025}{T} \right )}$ &  & UMIST \\
5 & \ch{C ->[CR] C+ + e-} & k$_{5} = 1.69117647059 \cdot \zeta$ & 10 $ < $ T $ < $ 41000 K & UMIST \\
6 & \ch{C + CS -> S + C2} & k$_{6} = 1.44 \cdot 10^{-11} (T/300)^{0.5} \operatorname{exp}{\left (- \frac{20435}{T} \right )}$ &  & UMIST \\
7 & \ch{C + e- -> C- + $\gamma$} & k$_{7} = 2.25\cdot 10^{-15}$ &  & UMIST \\
8 & \ch{C + HCO+ -> CO + CH+} & k$_{8} = 1.1\cdot 10^{-9}$ &  & UMIST \\
9 & \ch{C + HS -> CS + H} & k$_{9} = 1\cdot 10^{-10}$ &  & UMIST \\
10 & \ch{C + HS -> S + CH} & k$_{10} = 1.2 \cdot 10^{-11} (T/300)^{0.58} \operatorname{exp}{\left (- \frac{5880}{T} \right )}$ &  & UMIST \\
11 & \ch{C + N -> CN + $\gamma$} & k$_{11} = 5.72 \cdot 10^{-19} (T/300)^{0.37} \operatorname{exp}{\left (- \frac{51}{T} \right )}$ &  & UMIST \\
12 & \ch{C + N2 -> CN + N} & k$_{12} = 8.69 \cdot 10^{-11} \operatorname{exp}{\left (- \frac{22600}{T} \right )}$ &  & UMIST \\
13 & \ch{C + NH -> N + CH} & k$_{13} = 1.73 \cdot 10^{-11} (T/300)^{0.5} \operatorname{exp}{\left (- \frac{4000}{T} \right )}$ &  & UMIST \\
14 & \ch{C + NH -> CN + H} & k$_{14} = 1.2\cdot 10^{-10}$ &  & UMIST \\
15 & \ch{C + NO -> CO + N} & k$_{15} =  9 \cdot 10^{-11}(T/300)^{-0.16}$ &  & UMIST \\
16 & \ch{C + NO -> CN + O} & k$_{16} =  6 \cdot 10^{-11}(T/300)^{-0.16}$ &  & UMIST \\
17 & \ch{C + NS -> S + CN} & k$_{17} =  1.5 \cdot 10^{-10}(T/300)^{-0.16}$ &  & UMIST \\
18 & \ch{C + O -> CO + $\gamma$} & k$_{18} = 4.69 \cdot 10^{-19} (T/300)^{1.52} \operatorname{exp}{\left (\frac{50.5}{T} \right )}$ &  & UMIST \\
19 & \ch{C + O2 -> CO + O} & k$_{19} = 5.56 \cdot 10^{-11} (T/300)^{0.41} \operatorname{exp}{\left (\frac{26.9}{T} \right )}$ &  & UMIST \\
20 & \ch{C + OH -> O + CH} & k$_{20} = 2.25 \cdot 10^{-11} (T/300)^{0.5} \operatorname{exp}{\left (- \frac{14800}{T} \right )}$ &  & UMIST \\
21 & \ch{C + OH -> CO + H} & k$_{21} = 1\cdot 10^{-10}$ &  & UMIST \\
22 & \ch{C + S -> CS + $\gamma$} & k$_{22} = 4.36 \cdot 10^{-19} (T/300)^{0.22}$ &  & UMIST \\
23 & \ch{C + SO -> CS + O} & k$_{23} = 3.5\cdot 10^{-11}$ &  & UMIST \\
24 & \ch{C + SO -> S + CO} & k$_{24} = 3.5\cdot 10^{-11}$ &  & UMIST \\
25 & \ch{C + SO2 -> CO + SO} & k$_{25} = 7\cdot 10^{-11}$ &  & UMIST \\
26 & \ch{C + SiO+ -> Si+ + CO} & k$_{26} = 1\cdot 10^{-9}$ &  & UMIST \\
27 & \ch{C+ + e- -> C + $\gamma$} & k$_{27} =  2.36 \cdot 10^{-12} (T/300)^{-0.29}\operatorname{exp}{\left (\frac{17.6}{T} \right )}$ &  & UMIST \\
28 & \ch{C+ + Fe -> Fe+ + C} & k$_{28} = 2.6\cdot 10^{-9}$ &  & UMIST \\
29 & \ch{C+ + Mg -> Mg+ + C} & k$_{29} = 1.1\cdot 10^{-9}$ &  & UMIST \\
30 & \ch{C+ + Si -> Si+ + C} & k$_{30} = 2.1\cdot 10^{-9}$ &  & UMIST \\
31 & \ch{C- + H+ -> C + H} & k$_{31} =  7.51 \cdot 10^{-8}(T/300)^{-0.5}$ &  & UMIST \\
32 & \ch{C2 + S -> CS + C} & k$_{32} = 1\cdot 10^{-10}$ &  & UMIST \\
33 & \ch{CH + N -> NH + C} & k$_{33} = 3.3 \cdot 10^{-11} (T/300)^{0.65} \operatorname{exp}{\left (- \frac{1207}{T} \right )}$ &  & UMIST \\
34 & \ch{CH + N -> CN + H} & k$_{34} =  1.66 \cdot 10^{-10}(T/300)^{-0.9}$ &  & UMIST \\
35 & \ch{CH + O -> OH + C} & k$_{35} = 2.52 \cdot 10^{-11} \operatorname{exp}{\left (- \frac{2381}{T} \right )}$ &  & UMIST \\
36 & \ch{CH + O -> CO + H} & k$_{36} = 6.2 \cdot 10^{-11} (T/300)^{0.1} \operatorname{exp}{\left (\frac{4.5}{T} \right )}$ &  & UMIST \\
37 & \ch{CH + O -> HCO+ + e-} & k$_{37} =  1.9 \cdot 10^{-11} (T/300)^{-2.19}\operatorname{exp}{\left (- \frac{165.1}{T} \right )}$ &  & UMIST \\
38 & \ch{CH + S -> CS + H} & k$_{38} = 5\cdot 10^{-11}$ &  & UMIST \\
39 & \ch{CH + S -> HS + C} & k$_{39} = 1.73 \cdot 10^{-11} (T/300)^{0.5} \operatorname{exp}{\left (- \frac{4000}{T} \right )}$ &  & UMIST \\
40 & \ch{CH+ + e- -> C + H} & k$_{40} =  1.5 \cdot 10^{-7}(T/300)^{-0.42}$ &  & UMIST \\
41 & \ch{CN + O2 -> OCN + O} & k$_{41} =  2.2 \cdot 10^{-11} (T/300)^{-0.19}\operatorname{exp}{\left (\frac{31.9}{T} \right )}$ &  & UMIST \\
42 & \ch{CN + S -> NS + C} & k$_{42} = 5.71 \cdot 10^{-11} (T/300)^{0.5} \operatorname{exp}{\left (- \frac{32010}{T} \right )}$ &  & UMIST \\
43 & \ch{CO ->[CR] CO+ + e-} & k$_{43} = 2.86764705882 \cdot \zeta$ &  & UMIST \\
44 & \ch{CO ->[CR] C + O} & k$_{44} = 5 \cdot \zeta$ &  & KIDA \\
45 & \ch{CO+ + e- -> O + C} & k$_{45} =  2 \cdot 10^{-7}(T/300)^{-0.48}$ &  & UMIST \\
46 & \ch{Fe+ + e- -> Fe + $\gamma$} & k$_{46} =  2.55 \cdot 10^{-12}(T/300)^{-0.69}$ &  & UMIST \\
47 & \ch{H + C -> CH + $\gamma$} & k$_{47} = 1\cdot 10^{-17}$ &  & UMIST \\
48 & \ch{H + C- -> CH + e-} & k$_{48} = 5\cdot 10^{-10}$ &  & UMIST \\
49 & \ch{H + C2 -> CH + C} & k$_{49} = 4.67 \cdot 10^{-10} (T/300)^{0.5} \operatorname{exp}{\left (- \frac{30450}{T} \right )}$ &  & UMIST \\
50 & \ch{H + CH -> C + H2} & k$_{50} = 1.31 \cdot 10^{-10} \operatorname{exp}{\left (- \frac{80}{T} \right )}$ &  & UMIST \\
51 & \ch{H + CH -> C + H + H} & k$_{51} = 6 \cdot 10^{-9} \operatorname{exp}{\left (- \frac{40200}{T} \right )}$ &  & UMIST \\
52 & \ch{H + CH+ -> C+ + H2} & k$_{52} =  9.6 \cdot 10^{-10} (T/300)^{-0.37}\operatorname{exp}{\left (- \frac{29.1}{T} \right )}$ &  & UMIST \\
53 & \ch{H + CH2 -> CH + H2} & k$_{53} = 2.2\cdot 10^{-10}$ &  & UMIST \\
54 & \ch{H + CO -> OH + C} & k$_{54} = 1.1 \cdot 10^{-10} (T/300)^{0.5} \operatorname{exp}{\left (- \frac{77700}{T} \right )}$ &  & UMIST \\
55 & \ch{H + CO2 -> CO + OH} & k$_{55} = 3.38 \cdot 10^{-10} \operatorname{exp}{\left (- \frac{13163}{T} \right )}$ &  & UMIST \\
56 & \ch{H ->[CR] H+ + e-} & k$_{56} = 0.439705882353 \cdot \zeta$ & 10 $ < $ T $ < $ 41000 K & UMIST \\
57 & \ch{H + e- -> H- + $\gamma$} & k$_{57} = 3.37 \cdot 10^{-16} (T/300)^{0.64} \operatorname{exp}{\left (- \frac{9.2}{T} \right )}$ &  & UMIST \\
%%*********************
%% Manually add \left. \right. or close bracktes for needed lines.
58 & \ch{H + e- -> H+ + e- + e-} & k$_{58}\begin{aligned}[t] & = \operatorname{exp}\Big (- 2.3914985 \cdot 10^{-6} \ln^{8}{\left (T_{e} \right )} \\ 
& + 0.111954395 \ln^{7}{\left (T_{e} \right )} \\ 
& - 0.263197617 \ln^{6}{\left (T_{e} \right )} \\ 
& + 0.348255977 \ln^{5}{\left (T_{e} \right )} \\ 
& - 0.2877056 \ln^{4}{\left (T_{e} \right )} \\ 
& + 1.56315498 \ln^{3}{\left (T_{e} \right )} \\ 
& - 5.73932875 \ln^{2}{\left (T_{e} \right )} \\ 
& + 13.536556 \ln{\left (T_{e} \right )} \\ 
& - 32.71396786 \Big )\end{aligned}$ &  & \citet{Janev1987} \\
59 & \ch{H + H + H -> H2 + H} & k$_{59} =  6 \cdot 10^{-32}T^{-0.25} + 2 \cdot 10^{-31}T^{-0.5}$ &  & \citet{Forrey2013} \\
60 & \ch{H + H + He -> H2 + He} & k$_{60} =  6.9 \cdot 10^{-32}T^{-0.4}$ &  &  \citet{Glover2008} \\
61 & \ch{H + H2 -> H + H + H} & k$_{61} =  4.67 \cdot 10^{-7} (T/300)^{-1}\operatorname{exp}{\left (- \frac{55000}{T} \right )}$ &  & UMIST \\
62 & \ch{H + H2+ -> H2 + H+} & k$_{62} = 6.4\cdot 10^{-10}$ &  & UMIST \\
63 & \ch{H + H2O -> OH + H + H} & k$_{63} = 5.8 \cdot 10^{-9} \operatorname{exp}{\left (- \frac{52900}{T} \right )}$ &  & UMIST \\
64 & \ch{H + H2O -> OH + H2} & k$_{64} = 1.59 \cdot 10^{-11} (T/300)^{1.2} \operatorname{exp}{\left (- \frac{9610}{T} \right )}$ &  & UMIST \\
65 & \ch{H + H2S -> HS + H2} & k$_{65} = 3.71 \cdot 10^{-12} (T/300)^{1.94} \operatorname{exp}{\left (- \frac{455}{T} \right )}$ &  & UMIST \\
66 & \ch{H + HCN -> CN + H2} & k$_{66} = 6.2 \cdot 10^{-10} \operatorname{exp}{\left (- \frac{12500}{T} \right )}$ &  & UMIST \\
67 & \ch{H + HCO -> CO + H2} & k$_{67} = 1.5\cdot 10^{-10}$ &  & UMIST \\
68 & \ch{H + HS -> S + H2} & k$_{68} = 2.5\cdot 10^{-11}$ &  & UMIST \\
69 & \ch{H + HS+ -> S+ + H2} & k$_{69} = 1.1\cdot 10^{-10}$ &  & UMIST \\
70 & \ch{H + He+ -> He + H+} & k$_{70} = 1.2 \cdot 10^{-15} (T/300)^{0.25}$ &  & UMIST \\
71 & \ch{H + HeH+ -> He + H2+} & k$_{71} = 9.1\cdot 10^{-10}$ &  & UMIST \\
72 & \ch{H + NH -> N + H2} & k$_{72} = 1.73 \cdot 10^{-11} (T/300)^{0.5} \operatorname{exp}{\left (- \frac{2400}{T} \right )}$ &  & UMIST \\
73 & \ch{H + NH2 -> NH + H2} & k$_{73} = 4.56 \cdot 10^{-12} (T/300)^{1.2} \operatorname{exp}{\left (- \frac{2161}{T} \right )}$ &  & UMIST \\
74 & \ch{H + NO -> OH + N} & k$_{74} = 3.6 \cdot 10^{-10} \operatorname{exp}{\left (- \frac{24910}{T} \right )}$ &  & UMIST \\
75 & \ch{H + NO -> O + NH} & k$_{75} =  9.29 \cdot 10^{-10} (T/300)^{-0.1}\operatorname{exp}{\left (- \frac{35220}{T} \right )}$ &  & UMIST \\
76 & \ch{H + NS -> HS + N} & k$_{76} = 7.27 \cdot 10^{-11} (T/300)^{0.5} \operatorname{exp}{\left (- \frac{15700}{T} \right )}$ &  & UMIST \\
77 & \ch{H + NS -> S + NH} & k$_{77} = 7.27 \cdot 10^{-11} (T/300)^{0.5} \operatorname{exp}{\left (- \frac{20735}{T} \right )}$ &  & UMIST \\
78 & \ch{H + O -> OH + $\gamma$} & k$_{78} =  9.9 \cdot 10^{-19}(T/300)^{-0.38}$ &  & UMIST \\
79 & \ch{H + O+ -> O + H+} & k$_{79} = 5.66 \cdot 10^{-10} (T/300)^{0.36} \operatorname{exp}{\left (\frac{8.6}{T} \right )}$ &  & UMIST \\
80 & \ch{H + O- -> OH + e-} & k$_{80} = 5\cdot 10^{-10}$ &  & UMIST \\
81 & \ch{H + O2 -> O + O + H} & k$_{81} = 6 \cdot 10^{-9} \operatorname{exp}{\left (- \frac{52300}{T} \right )}$ &  & UMIST \\
82 & \ch{H + O2 -> OH + O} & k$_{82} = 2.61 \cdot 10^{-10} \operatorname{exp}{\left (- \frac{8156}{T} \right )}$ &  & UMIST \\
83 & \ch{H + OCN -> OH + CN} & k$_{83} = 1\cdot 10^{-10}$ &  & UMIST \\
84 & \ch{H + OCN -> HCN + O} & k$_{84} = 1.87 \cdot 10^{-11} (T/300)^{0.9} \operatorname{exp}{\left (- \frac{2924}{T} \right )}$ &  & UMIST \\
85 & \ch{H + OCN -> NH + CO} & k$_{85} = 1.26 \cdot 10^{-10} \operatorname{exp}{\left (- \frac{515}{T} \right )}$ &  & UMIST \\
86 & \ch{H + OCS -> HS + CO} & k$_{86} = 1.23 \cdot 10^{-11} \operatorname{exp}{\left (- \frac{1949}{T} \right )}$ &  & UMIST \\
87 & \ch{H + OH -> O + H + H} & k$_{87} = 6 \cdot 10^{-9} \operatorname{exp}{\left (- \frac{50900}{T} \right )}$ &  & UMIST \\
88 & \ch{H + OH -> O + H2} & k$_{88} = 6.99 \cdot 10^{-14} (T/300)^{2.8} \operatorname{exp}{\left (- \frac{1950}{T} \right )}$ &  & UMIST \\
89 & \ch{H + S- -> HS + e-} & k$_{89} = 1\cdot 10^{-10}$ &  & UMIST \\
90 & \ch{H + S2 -> HS + S} & k$_{90} = 2.25 \cdot 10^{-10} (T/300)^{0.5} \operatorname{exp}{\left (- \frac{8355}{T} \right )}$ &  & UMIST \\
91 & \ch{H + SO -> S + OH} & k$_{91} =  5.9 \cdot 10^{-10} (T/300)^{-0.31}\operatorname{exp}{\left (- \frac{11100}{T} \right )}$ &  & UMIST \\
92 & \ch{H + SO -> HS + O} & k$_{92} = 1.73 \cdot 10^{-11} (T/300)^{0.5} \operatorname{exp}{\left (- \frac{19930}{T} \right )}$ &  & UMIST \\
93 & \ch{H + Si+ -> SiH+ + $\gamma$} & k$_{93} =  1.17 \cdot 10^{-17}(T/300)^{-0.14}$ &  & UMIST \\
94 & \ch{H + SiH+ -> Si+ + H2} & k$_{94} = 1.9\cdot 10^{-9}$ &  & UMIST \\
95 & \ch{H+ + e- -> H + $\gamma$} & k$_{95} =  3.5 \cdot 10^{-12}(T/300)^{-0.75}$ &  & UMIST \\
96 & \ch{H+ + Fe -> Fe+ + H} & k$_{96} = 7.4\cdot 10^{-9}$ &  & UMIST \\
97 & \ch{H+ + H -> H2+ + $\gamma$} & k$_{97} = 1.15 \cdot 10^{-18} (T/300)^{1.49} \operatorname{exp}{\left (- \frac{228}{T} \right )}$ &  & UMIST \\
98 & \ch{H+ + Mg -> Mg+ + H} & k$_{98} = 1.1\cdot 10^{-9}$ &  & UMIST \\
99 & \ch{H+ + NH -> NH+ + H} & k$_{99} =  2.1 \cdot 10^{-9}(T/300)^{-0.5}$ &  & UMIST \\
100 & \ch{H+ + Na -> Na+ + H} & k$_{100} = 1.2\cdot 10^{-9}$ & 10 $ < $ T $ < $ 10000 K & KIDA \\
101 & \ch{H+ + O -> O+ + H} & k$_{101} = 6.86 \cdot 10^{-10} (T/300)^{0.26} \operatorname{exp}{\left (- \frac{224.3}{T} \right )}$ &  & UMIST \\
102 & \ch{H+ + OH -> OH+ + H} & k$_{102} =  2.1 \cdot 10^{-9}(T/300)^{-0.5}$ &  & UMIST \\
103 & \ch{H+ + P -> P+ + H} & k$_{103} = 1\cdot 10^{-9}$ &  & UMIST \\
104 & \ch{H+ + S -> S+ + H} & k$_{104} = 1.3\cdot 10^{-9}$ &  & UMIST \\
105 & \ch{H+ + Si -> Si+ + H} & k$_{105} = 9.9\cdot 10^{-10}$ &  & UMIST \\
106 & \ch{H+ + SiO -> SiO+ + H} & k$_{106} =  3.3 \cdot 10^{-9}(T/300)^{-0.5}$ &  & UMIST \\
107 & \ch{H- + C -> CH + e-} & k$_{107} = 1\cdot 10^{-9}$ &  & UMIST \\
%%*********************
%% Manually add \left. \right. or close bracktes for needed lines.
108 & \ch{H- + e- -> H + e- + e-} & k$_{108}\begin{aligned}[t] & = \operatorname{exp}\Big(- 2.631285809207 \cdot 10^{-6} \ln^{8}{\left (T_{e} \right )} \\ 
& + 0.1068275202678 \ln^{7}{\left (T_{e} \right )} \\ 
& - 0.1656194699504 \ln^{6}{\left (T_{e} \right )} \\ 
& + 0.1178329782711 \ln^{5}{\left (T_{e} \right )} \\ 
& - 0.3365012031363 \ln^{4}{\left (T_{e} \right )} \\ 
& + 0.1623316639567 \ln^{3}{\left (T_{e} \right )} \\ 
& - 0.2827443061704 \ln^{2}{\left (T_{e} \right )} \\ 
& + 2.360852208681 \ln{\left (T_{e} \right )} \\ 
& - 18.1849334273 \Big )\end{aligned}$ &  & \citet{Janev1987} \\
109 & \ch{H- + Fe+ -> H + Fe} & k$_{109} =  7.51 \cdot 10^{-8}(T/300)^{-0.5}$ &  & UMIST \\
110 & \ch{H- + H -> H2 + e-} & k$_{110} = 4.82 \cdot 10^{-9} (T/300)^{0.2} \operatorname{exp}{\left (- \frac{4.3}{T} \right )}$ &  & UMIST \\
%%*********************
%% Manually add \left. \right. or close bracktes for needed lines.
111 & \ch{H- + H -> H + H + e-} & k$_{111}\begin{aligned}[t] & = \operatorname{exp}\Big (- 8.6838246118 \cdot 10^{-8} \ln^{9}{\left (T_{e} \right )} \\ 
& + 2.4555011970392 \cdot 10^{-6} \ln^{8}{\left (T_{e} \right )} \\ 
& - 2.585009680264 \cdot 10^{-5} \ln^{7}{\left (T_{e} \right )} \\ 
& + 8.66396324309 \cdot 10^{-5} \ln^{6}{\left (T_{e} \right )} \\ 
& + 0.2012250284791 \ln^{5}{\left (T_{e} \right )} \\ 
& - 0.14327641212992 \ln^{4}{\left (T_{e} \right )} \\ 
& + 0.846445538663 \ln^{3}{\left (T_{e} \right )} \\ 
& - 0.142101352155415 \ln^{2}{\left (T_{e} \right )} \\ 
& + 1.13944933584163 \ln{\left (T_{e} \right )} \\ 
& - 20.3726089653332 \Big)\end{aligned}$ & T $ > $ 1160 K & \citet{Janev1987},\newline \citet{Abel1997} \\
112 & \ch{H- + H+ -> H + H} & k$_{112} =  7.51 \cdot 10^{-8}(T/300)^{-0.5}$ &  & UMIST \\
113 & \ch{H- + H+ -> H2+ + e-} & k$_{113} =  1 \cdot 10^{-8}T^{-0.4}$ &  & \citet{Poulaert1978} \\
114 & \ch{H- + Mg+ -> H + Mg} & k$_{114} =  7.51 \cdot 10^{-8}(T/300)^{-0.5}$ &  & UMIST \\
115 & \ch{H- + N -> NH + e-} & k$_{115} = 1\cdot 10^{-9}$ &  & UMIST \\
116 & \ch{H- + Na+ -> H + Na} & k$_{116} =  7.51 \cdot 10^{-8}(T/300)^{-0.5}$ &  & UMIST \\
117 & \ch{H- + O -> OH + e-} & k$_{117} = 1\cdot 10^{-9}$ &  & UMIST \\
118 & \ch{H- + O+ -> H + O} & k$_{118} =  7.51 \cdot 10^{-8}(T/300)^{-0.5}$ &  & UMIST \\
119 & \ch{H- + S+ -> H + S} & k$_{119} =  7.51 \cdot 10^{-8}(T/300)^{-0.5}$ &  & UMIST \\
120 & \ch{H- + Si+ -> H + Si} & k$_{120} =  7.51 \cdot 10^{-8}(T/300)^{-0.5}$ &  & UMIST \\
121 & \ch{H2 + C -> CH + H} & k$_{121} = 6.64 \cdot 10^{-10} \operatorname{exp}{\left (- \frac{11700}{T} \right )}$ &  & UMIST \\
122 & \ch{H2 + C -> CH2 + $\gamma$} & k$_{122} = 1\cdot 10^{-17}$ &  & UMIST \\
123 & \ch{H2 + CH -> CH2 + H} & k$_{123} = 5.46 \cdot 10^{-10} \operatorname{exp}{\left (- \frac{1943}{T} \right )}$ &  & UMIST \\
124 & \ch{H2 + CN -> HCN + H} & k$_{124} = 4.4 \cdot 10^{-13} (T/300)^{2.87} \operatorname{exp}{\left (- \frac{820}{T} \right )}$ &  & UMIST \\
125 & \ch{H2 ->[CR] H+ + H-} & k$_{125} = 0.286764705882 \cdot \zeta$ &  & UMIST \\
126 & \ch{H2 ->[CR] H+ + H + e-} & k$_{126} = 0.210294117647 \cdot \zeta$ &  & UMIST \\
127 & \ch{H2 ->[CR] H + H} & k$_{127} = 0.955882352941 \cdot \zeta$ &  & UMIST \\
128 & \ch{H2 ->[CR] H2+ + e-} & k$_{128} = 0.882352941176 \cdot \zeta$ &  & UMIST \\
129 & \ch{H2 + e- -> H + H + e-} & k$_{129} = 3.22 \cdot 10^{-9} (T/300)^{0.35} \operatorname{exp}{\left (- \frac{102000}{T} \right )}$ &  & UMIST \\
130 & \ch{H2 + e- -> H + H-} & k$_{130} = 35.5T^{-2.28} \operatorname{exp}{\left (- \frac{46707}{T} \right )}$ &  & \citet{Capitelli2007} \\
131 & \ch{H2 + F -> HF + H} & k$_{131} = 1 \cdot 10^{-10} \operatorname{exp}{\left (- \frac{400}{T} \right )}$ &  & UMIST \\
132 & \ch{H2 + F+ -> H2+ + F} & k$_{132} = 6.24\cdot 10^{-10}$ &  & UMIST \\
133 & \ch{H2 + H + H -> H2 + H2} & k$_{133} =  \frac{1}{8} \left(6 \cdot 10^{-32}T^{-0.25} + 2 \cdot 10^{-31}T^{-0.5}\right)$ &  & \citet{Glover2008} \\
134 & \ch{H2 + H+ -> H2+ + H} & k$_{134}\begin{aligned}[t] & = \Big (3.5311932 \cdot 10^{-13} \ln^{7}{\left (T \right )} \\ 
& - 1.8171411 \cdot 10^{-11} \ln^{6}{\left (T \right )} \\ 
& + 3.9731542 \cdot 10^{-10} \ln^{5}{\left (T \right )} \\ 
& - 4.7813728 \cdot 10^{-9} \ln^{4}{\left (T \right )} \\ 
& + 3.4172805 \cdot 10^{-8} \ln^{3}{\left (T \right )} \\ 
& - 1.4491368 \cdot 10^{-7} \ln^{2}{\left (T \right )} \\ 
& + 3.3735382 \cdot 10^{-7} \ln{\left (T \right )} \\ 
& - 3.3232183 \cdot 10^{-7} \Big) \operatorname{exp}\left (- \frac{21237.15}{T} \right ) \end{aligned}$& 100 $\leqslant$ T $\leqslant$ 30000 K &  \citet{Savin2004}, \newline \citet{Glover2010} \\
135 & \ch{H2 + HS -> H2S + H} & k$_{135} = 6.52 \cdot 10^{-12} (T/300)^{0.9} \operatorname{exp}{\left (- \frac{8050}{T} \right )}$ &  & UMIST \\
136 & \ch{H2 + N -> NH + H} & k$_{136} = 1.69 \cdot 10^{-9} \operatorname{exp}{\left (- \frac{18095}{T} \right )}$ &  & UMIST \\
137 & \ch{H2 + NH -> NH2 + H} & k$_{137} = 5.96 \cdot 10^{-11} \operatorname{exp}{\left (- \frac{7782}{T} \right )}$ &  & UMIST \\
138 & \ch{H2 + O -> OH + H} & k$_{138} = 3.14 \cdot 10^{-13} (T/300)^{2.7} \operatorname{exp}{\left (- \frac{3150}{T} \right )}$ &  & UMIST \\
139 & \ch{H2 + O+ -> OH+ + H} & k$_{139} = 1.7\cdot 10^{-9}$ &  & UMIST \\
140 & \ch{H2 + O2 -> OH + OH} & k$_{140} = 3.16 \cdot 10^{-10} \operatorname{exp}{\left (- \frac{21890}{T} \right )}$ &  & UMIST \\
141 & \ch{H2 + OH -> H2O + H} & k$_{141} = 2.5 \cdot 10^{-12} (T/300)^{1.52} \operatorname{exp}{\left (- \frac{1736}{T} \right )}$ &  & UMIST \\
142 & \ch{H2 + S -> HS + H} & k$_{142} = 1.76 \cdot 10^{-13} (T/300)^{2.88} \operatorname{exp}{\left (- \frac{6126}{T} \right )}$ &  & UMIST \\
143 & \ch{H2 + S+ -> HS+ + H} & k$_{143} = 1.1 \cdot 10^{-10} \operatorname{exp}{\left (- \frac{9860}{T} \right )}$ &  & UMIST \\
144 & \ch{H2+ + C -> CH+ + H} & k$_{144} = 2.4\cdot 10^{-9}$ &  & UMIST \\
145 & \ch{H2+ + e- -> H + H} & k$_{145} =  1.6 \cdot 10^{-8}(T/300)^{-0.43}$ &  & UMIST \\
146 & \ch{H2+ + He -> HeH+ + H} & k$_{146} = 1.3\cdot 10^{-10}$ &  & UMIST \\
147 & \ch{H2+ + O -> OH+ + H} & k$_{147} = 1.5\cdot 10^{-9}$ &  & UMIST \\
148 & \ch{HCO+ + e- -> CO + H} & k$_{148} =  2.4 \cdot 10^{-7}(T/300)^{-0.69}$ &  & UMIST \\
149 & \ch{HCO+ + Fe -> Fe+ + HCO} & k$_{149} = 1.9\cdot 10^{-9}$ &  & UMIST \\
150 & \ch{HF + Si+ -> SiF+ + H} & k$_{150} =  5.7 \cdot 10^{-9}(T/300)^{-0.5}$ &  & UMIST \\
151 & \ch{HS + HS -> H2S + S} & k$_{151} = 1.3\cdot 10^{-11}$ &  & UMIST \\
152 & \ch{HS+ + e- -> S + H} & k$_{152} =  2 \cdot 10^{-7}(T/300)^{-0.5}$ &  & UMIST \\
153 & \ch{He ->[CR] He+ + e-} & k$_{153} = 0.477941176471 \cdot \zeta$ & 10 $ < $ T $ < $ 41000 K & UMIST \\
%%*********************
%% Manually add \left. \right. or close bracktes for needed lines.
154 & \ch{He + e- -> He+ + e- + e-} & k$_{154}\begin{aligned}[t] & = \operatorname{exp}\Big (- 3.64916141 \cdot 10^{-6} \ln^{8}{\left (T_{e} \right )} \\ 
& + 0.206723616 \ln^{7}{\left (T_{e} \right )} \\ 
& - 0.50090561 \ln^{6}{\left (T_{e} \right )} \\ 
& + 0.679539123 \ln^{5}{\left (T_{e} \right )} \\ 
& - 0.56851189 \ln^{4}{\left (T_{e} \right )} \\ 
& + 3.5803875 \ln^{3}{\left (T_{e} \right )} \\ 
& - 10.7532302 \ln^{2}{\left (T_{e} \right )} \\ 
& + 23.91596563 \ln{\left (T_{e} \right )} \\ 
& - 44.9864886 \Big )\end{aligned}$ &  & \citet{Janev1987} \\
155 & \ch{He+ + e- -> He + $\gamma$} & k$_{155} =  5.36 \cdot 10^{-12}(T/300)^{-0.5}$ &  & UMIST \\
156 & \ch{He+ + HF -> F+ + H + He} & k$_{156} =  1.1 \cdot 10^{-8}(T/300)^{-0.5}$ &  & UMIST \\
157 & \ch{He+ + Si -> Si+ + He} & k$_{157} = 3.3\cdot 10^{-9}$ &  & UMIST \\
158 & \ch{He+ + SiO2 -> O2 + Si+ + He} & k$_{158} = 2\cdot 10^{-9}$ &  & UMIST \\
159 & \ch{HeH+ + e- -> He + H} & k$_{159} =  1 \cdot 10^{-8}(T/300)^{-0.6}$ &  & UMIST \\
160 & \ch{Mg + HCO+ -> HCO + Mg+} & k$_{160} = 2.9\cdot 10^{-9}$ &  & UMIST \\
161 & \ch{Mg + S+ -> S + Mg+} & k$_{161} = 2.8\cdot 10^{-10}$ &  & UMIST \\
162 & \ch{Mg + Si+ -> Si + Mg+} & k$_{162} = 2.9\cdot 10^{-9}$ &  & UMIST \\
163 & \ch{Mg + SiO+ -> SiO + Mg+} & k$_{163} = 1\cdot 10^{-9}$ &  & UMIST \\
164 & \ch{Mg+ + e- -> Mg + $\gamma$} & k$_{164} =  2.78 \cdot 10^{-12}(T/300)^{-0.68}$ &  & UMIST \\
165 & \ch{N + C2 -> CN + C} & k$_{165} = 5\cdot 10^{-11}$ &  & UMIST \\
166 & \ch{N + CN -> N2 + C} & k$_{166} = 1 \cdot 10^{-10} (T/300)^{0.18}$ &  & UMIST \\
167 & \ch{N + CO2 -> NO + CO} & k$_{167} = 3.2 \cdot 10^{-13} \operatorname{exp}{\left (- \frac{1710}{T} \right )}$ &  & UMIST \\
168 & \ch{N ->[CR] N+ + e-} & k$_{168} = 1.98529411765 \cdot \zeta$ & 10 $ < $ T $ < $ 41000 K & UMIST \\
169 & \ch{N + CS -> S + CN} & k$_{169} = 3.8 \cdot 10^{-11} (T/300)^{0.5} \operatorname{exp}{\left (- \frac{1160}{T} \right )}$ &  & UMIST \\
170 & \ch{N + HS -> NS + H} & k$_{170} = 1\cdot 10^{-10}$ &  & UMIST \\
171 & \ch{N + HS -> S + NH} & k$_{171} = 1.73 \cdot 10^{-11} (T/300)^{0.5} \operatorname{exp}{\left (- \frac{9060}{T} \right )}$ &  & UMIST \\
172 & \ch{N + NH -> N2 + H} & k$_{172} = 4.98\cdot 10^{-11}$ &  & UMIST \\
173 & \ch{N + NO -> N2 + O} & k$_{173} =  3.38 \cdot 10^{-11} (T/300)^{-0.17}\operatorname{exp}{\left (\frac{2.8}{T} \right )}$ &  & UMIST \\
174 & \ch{N + O2 -> NO + O} & k$_{174} = 2.26 \cdot 10^{-12} (T/300)^{0.86} \operatorname{exp}{\left (- \frac{3134}{T} \right )}$ &  & UMIST \\
175 & \ch{N + OH -> O + NH} & k$_{175} = 1.88 \cdot 10^{-11} (T/300)^{0.1} \operatorname{exp}{\left (- \frac{10700}{T} \right )}$ &  & UMIST \\
176 & \ch{N + OH -> NO + H} & k$_{176} =  6.5 \cdot 10^{-11} (T/300)^{-0.23}\operatorname{exp}{\left (- \frac{14.9}{T} \right )}$ &  & UMIST \\
177 & \ch{N + PN -> P + N2} & k$_{177} = 1\cdot 10^{-18}$ &  & KIDA \\
178 & \ch{N + PO -> PN + O} & k$_{178} =  3 \cdot 10^{-11}(T/300)^{-0.6}$ &  & KIDA \\
179 & \ch{N + PO -> P + NO} & k$_{179} = 2.55\cdot 10^{-12}$ &  & KIDA \\
180 & \ch{N + SO -> NS + O} & k$_{180} = 4.68 \cdot 10^{-11} (T/300)^{0.5} \operatorname{exp}{\left (- \frac{8254}{T} \right )}$ &  & UMIST \\
181 & \ch{N + SO -> S + NO} & k$_{181} = 1.73 \cdot 10^{-11} (T/300)^{0.5} \operatorname{exp}{\left (- \frac{750}{T} \right )}$ &  & UMIST \\
182 & \ch{N + SiO+ -> NO + Si+} & k$_{182} = 2.1\cdot 10^{-10}$ &  & UMIST \\
183 & \ch{N+ + e- -> N + $\gamma$} & k$_{183} =  3.5 \cdot 10^{-12} (T/300)^{-0.53}\operatorname{exp}{\left (\frac{3.2}{T} \right )}$ &  & UMIST \\
184 & \ch{N2 ->[CR] N + N} & k$_{184} = 5 \cdot \zeta$ &  & KIDA \\
185 & \ch{NH + O -> OH + N} & k$_{185} = 1.16\cdot 10^{-11}$ &  & UMIST \\
186 & \ch{NH + O -> NO + H} & k$_{186} = 6.6\cdot 10^{-11}$ &  & UMIST \\
187 & \ch{NH + S -> NS + H} & k$_{187} = 1\cdot 10^{-10}$ &  & UMIST \\
188 & \ch{NH + S -> HS + N} & k$_{188} = 1.73 \cdot 10^{-11} (T/300)^{0.5} \operatorname{exp}{\left (- \frac{4000}{T} \right )}$ &  & UMIST \\
189 & \ch{NH+ + e- -> N + H} & k$_{189} =  4.3 \cdot 10^{-8}(T/300)^{-0.5}$ &  & UMIST \\
190 & \ch{Na + Fe+ -> Fe + Na+} & k$_{190} = 1\cdot 10^{-11}$ &  & UMIST \\
191 & \ch{Na + Mg+ -> Mg + Na+} & k$_{191} = 1\cdot 10^{-11}$ &  & UMIST \\
192 & \ch{Na + S+ -> S + Na+} & k$_{192} = 2.6\cdot 10^{-10}$ &  & UMIST \\
193 & \ch{Na + Si+ -> Si + Na+} & k$_{193} = 2.7\cdot 10^{-9}$ &  & UMIST \\
194 & \ch{Na+ + e- -> Na + $\gamma$} & k$_{194} =  2.76 \cdot 10^{-12}(T/300)^{-0.68}$ &  & UMIST \\
195 & \ch{O + C2 -> CO + C} & k$_{195} =  2 \cdot 10^{-10}(T/300)^{-0.12}$ &  & UMIST \\
196 & \ch{O + CN -> NO + C} & k$_{196} = 5.37 \cdot 10^{-11} \operatorname{exp}{\left (- \frac{13800}{T} \right )}$ &  & UMIST \\
197 & \ch{O + CN -> CO + N} & k$_{197} = 2.54\cdot 10^{-11}$ &  & UMIST \\
198 & \ch{O ->[CR] O+ + e-} & k$_{198} = 2.5 \cdot \zeta$ & 10 $ < $ T $ < $ 41000 K & UMIST \\
199 & \ch{O + CS -> SO + C} & k$_{199} = 4.68 \cdot 10^{-11} (T/300)^{0.5} \operatorname{exp}{\left (- \frac{28940}{T} \right )}$ &  & UMIST \\
200 & \ch{O + CS -> S + CO} & k$_{200} =  2.48 \cdot 10^{-10} (T/300)^{-0.65}\operatorname{exp}{\left (- \frac{783}{T} \right )}$ &  & UMIST \\
201 & \ch{O + e- -> O- + $\gamma$} & k$_{201} = 1.5\cdot 10^{-15}$ &  & UMIST \\
202 & \ch{O + H2O -> OH + OH} & k$_{202} = 1.85 \cdot 10^{-11} (T/300)^{0.95} \operatorname{exp}{\left (- \frac{8571}{T} \right )}$ &  & UMIST \\
203 & \ch{O + HCN -> CO + NH} & k$_{203} = 7.3 \cdot 10^{-13} (T/300)^{1.14} \operatorname{exp}{\left (- \frac{3742}{T} \right )}$ &  & UMIST \\
204 & \ch{O + HCN -> CN + OH} & k$_{204} = 6.21 \cdot 10^{-10} \operatorname{exp}{\left (- \frac{12439}{T} \right )}$ &  & UMIST \\
205 & \ch{O + HCN -> OCN + H} & k$_{205} = 1.36 \cdot 10^{-12} (T/300)^{1.38} \operatorname{exp}{\left (- \frac{3693}{T} \right )}$ &  & UMIST \\
206 & \ch{O + HS -> S + OH} & k$_{206} = 1.74 \cdot 10^{-11} (T/300)^{0.67} \operatorname{exp}{\left (- \frac{956}{T} \right )}$ &  & UMIST \\
207 & \ch{O + HS -> SO + H} & k$_{207} =  1.74 \cdot 10^{-10} (T/300)^{-0.2}\operatorname{exp}{\left (- \frac{5.7}{T} \right )}$ &  & UMIST \\
208 & \ch{O + N2 -> NO + N} & k$_{208} = 2.51 \cdot 10^{-10} \operatorname{exp}{\left (- \frac{38602}{T} \right )}$ &  & UMIST \\
209 & \ch{O + NS -> S + NO} & k$_{209} = 1\cdot 10^{-10}$ &  & UMIST \\
210 & \ch{O + O -> O2 + $\gamma$} & k$_{210} = 4.9 \cdot 10^{-20} (T/300)^{1.58}$ &  & UMIST \\
211 & \ch{O + OH -> O2 + H} & k$_{211} =  3.69 \cdot 10^{-11} (T/300)^{-0.27}\operatorname{exp}{\left (- \frac{12.9}{T} \right )}$ &  & UMIST \\
212 & \ch{O + SO -> S + O2} & k$_{212} = 6.6 \cdot 10^{-13} \operatorname{exp}{\left (- \frac{2760}{T} \right )}$ &  & UMIST \\
213 & \ch{O + SO2 -> SO + O2} & k$_{213} = 9.1 \cdot 10^{-12} \operatorname{exp}{\left (- \frac{9837}{T} \right )}$ &  & UMIST \\
214 & \ch{O + Si -> SiO + $\gamma$} & k$_{214} = 5.52 \cdot 10^{-18} (T/300)^{0.31}$ &  & UMIST \\
215 & \ch{O + SiO+ -> O2 + Si+} & k$_{215} = 2\cdot 10^{-10}$ &  & UMIST \\
216 & \ch{O+ + e- -> O + $\gamma$} & k$_{216} =  3.24 \cdot 10^{-12}(T/300)^{-0.66}$ &  & UMIST \\
217 & \ch{O+ + Fe -> Fe+ + O} & k$_{217} = 2.9\cdot 10^{-9}$ &  & UMIST \\
218 & \ch{O- + Fe+ -> O + Fe} & k$_{218} =  7.51 \cdot 10^{-8}(T/300)^{-0.5}$ &  & UMIST \\
219 & \ch{O- + H+ -> O + H} & k$_{219} =  7.51 \cdot 10^{-8}(T/300)^{-0.5}$ &  & UMIST \\
220 & \ch{O- + Mg+ -> O + Mg} & k$_{220} =  7.51 \cdot 10^{-8}(T/300)^{-0.5}$ &  & UMIST \\
221 & \ch{O2 + S -> SO + O} & k$_{221} = 1.76 \cdot 10^{-12} (T/300)^{0.81} \operatorname{exp}{\left (\frac{30.8}{T} \right )}$ &  & UMIST \\
222 & \ch{OH + CN -> HCN + O} & k$_{222} = 1 \cdot 10^{-11} \operatorname{exp}{\left (- \frac{1000}{T} \right )}$ &  & UMIST \\
223 & \ch{OH + CN -> OCN + H} & k$_{223} = 7\cdot 10^{-11}$ &  & UMIST \\
224 & \ch{OH + CO -> CO2 + H} & k$_{224} = 2.81 \cdot 10^{-13} \operatorname{exp}{\left (- \frac{176}{T} \right )}$ &  & UMIST \\
225 & \ch{OH + CS -> CO + HS} & k$_{225} = 3\cdot 10^{-11}$ &  & UMIST \\
226 & \ch{OH + CS -> H + OCS} & k$_{226} = 1.7\cdot 10^{-10}$ &  & UMIST \\
227 & \ch{OH + F -> HF + O} & k$_{227} = 1.6\cdot 10^{-10}$ &  & UMIST \\
228 & \ch{OH + H2S -> HS + H2O} & k$_{228} = 6.3 \cdot 10^{-12} \operatorname{exp}{\left (- \frac{80}{T} \right )}$ &  & UMIST \\
229 & \ch{OH + OH -> H2O + O} & k$_{229} = 1.65 \cdot 10^{-12} (T/300)^{1.14} \operatorname{exp}{\left (- \frac{50}{T} \right )}$ &  & UMIST \\
230 & \ch{OH + S -> SO + H} & k$_{230} = 6.6\cdot 10^{-11}$ &  & UMIST \\
231 & \ch{OH + SO -> SO2 + H} & k$_{231} = 8.6\cdot 10^{-11}$ &  & UMIST \\
232 & \ch{OH + Si -> SiO + H} & k$_{232} = 1\cdot 10^{-10}$ &  & UMIST \\
233 & \ch{OH + Si+ -> SiO+ + H} & k$_{233} =  6.3 \cdot 10^{-10}(T/300)^{-0.5}$ &  & UMIST \\
234 & \ch{OH + SiO -> SiO2 + H} & k$_{234} = 2\cdot 10^{-12}$ &  & UMIST \\
235 & \ch{OH+ + e- -> O + H} & k$_{235} =  3.75 \cdot 10^{-8}(T/300)^{-0.5}$ &  & UMIST \\
236 & \ch{P + O2 -> PO + O} & k$_{236} = 1\cdot 10^{-13}$ &  & KIDA \\
237 & \ch{P+ + e- -> P + $\gamma$} & k$_{237} =  3.41 \cdot 10^{-12}(T/300)^{-0.65}$ &  & UMIST \\
238 & \ch{S + e- -> S- + $\gamma$} & k$_{238} = 5\cdot 10^{-15}$ &  & UMIST \\
239 & \ch{S + HS -> S2 + H} & k$_{239} = 4.5\cdot 10^{-11}$ &  & UMIST \\
240 & \ch{S + SO2 -> SO + SO} & k$_{240} = 9.76 \cdot 10^{-12} \operatorname{exp}{\left (- \frac{4545}{T} \right )}$ &  & UMIST \\
241 & \ch{S+ + e- -> S + $\gamma$} & k$_{241} =  5.49 \cdot 10^{-12}(T/300)^{-0.59}$ &  & UMIST \\
242 & \ch{S+ + Fe -> Fe+ + S} & k$_{242} = 1.8\cdot 10^{-10}$ &  & UMIST \\
243 & \ch{Si + CO -> SiO + C} & k$_{243} = 1.3 \cdot 10^{-9} \operatorname{exp}{\left (- \frac{34513}{T} \right )}$ &  & UMIST \\
244 & \ch{Si + CO2 -> SiO + CO} & k$_{244} = 2.72 \cdot 10^{-11} \operatorname{exp}{\left (- \frac{282}{T} \right )}$ &  & UMIST \\
245 & \ch{Si + HCO+ -> SiH+ + CO} & k$_{245} = 1.6\cdot 10^{-9}$ &  & UMIST \\
246 & \ch{Si + NO -> SiO + N} & k$_{246} =  9 \cdot 10^{-11} (T/300)^{-0.96}\operatorname{exp}{\left (- \frac{28}{T} \right )}$ &  & UMIST \\
247 & \ch{Si + O2 -> SiO + O} & k$_{247} =  1.72 \cdot 10^{-10} (T/300)^{-0.53}\operatorname{exp}{\left (- \frac{17}{T} \right )}$ &  & UMIST \\
248 & \ch{Si + P+ -> P + Si+} & k$_{248} = 1\cdot 10^{-9}$ &  & UMIST \\
249 & \ch{Si + S+ -> S + Si+} & k$_{249} = 1.6\cdot 10^{-9}$ &  & UMIST \\
250 & \ch{Si+ + e- -> Si + $\gamma$} & k$_{250} =  4.26 \cdot 10^{-12}(T/300)^{-0.62}$ &  & UMIST \\
251 & \ch{Si+ + Fe -> Fe+ + Si} & k$_{251} = 1.9\cdot 10^{-9}$ &  & UMIST \\
252 & \ch{SiF+ + e- -> Si + F} & k$_{252} =  2 \cdot 10^{-7}(T/300)^{-0.5}$ &  & UMIST \\
253 & \ch{SiH+ + e- -> Si + H} & k$_{253} =  2 \cdot 10^{-7}(T/300)^{-0.5}$ &  & UMIST \\
254 & \ch{SiO+ + e- -> Si + O} & k$_{254} =  2 \cdot 10^{-7}(T/300)^{-0.5}$ &  & UMIST \\
255 & \ch{SiO+ + Fe -> Fe+ + SiO} & k$_{255} = 1\cdot 10^{-9}$ &  & UMIST \\\\\hline
\multicolumn{5}{p{\columnwidth}}{$T_e = T/11604.525$~eV\,K$^{-1}$ is the gas temperature in electron -- $\zeta = 1.36 \cdot 10^{-17}$s$^{-1}$ is the cosmic ray (CR) flux}
\end{longtable}

% Don't change these lines
% \bsp	% typesetting comment
\label{lastpage}
\end{document}